\documentclass[fleqn,usenatbib]{mnras}

\usepackage[T1]{fontenc}

\DeclareRobustCommand{\VAN}[3]{#2}
\let\VANthebibliography\thebibliography
\def\thebibliography{\DeclareRobustCommand{\VAN}[3]{##3}\VANthebibliography}

\usepackage{graphicx}	
\usepackage{amsmath}	
\usepackage{amssymb}	

\newcommand{\oii}{\mbox{O\,{\scshape ii}}}
\newcommand{\oiii}{\mbox{O\,{\scshape iii}}}

\newcommand{\ciii}{\mbox{C\,{\scshape iii}}}
\newcommand{\civ}{\mbox{C\,{\scshape iv}}}

\title[Pilot-WINGS]{Pilot-WINGS: An extended MUSE view of the structure of Abell 370}

\author[D.\ J.\ Lagattuta et al.]{David J.\ Lagattuta,$^{1,2}$\thanks{E-mail: david.j.lagattuta@durham.ac.uk}
Johan Richard,$^{3}$
Franz Erik Bauer,$^{4,5,6}$ 
Catherine Cerny,$^{1,2}$
Ad\'ela\"\i de Claeyssens,$^{3}$
\newauthor
Lucia Guaita$^{4,7}$,
Mathilde Jauzac,$^{1,2,8,9}$
Alexandre Jeanneau,$^{3}$
Anton M.\ Koekemoer,$^{10}$
Guillaume Mahler,$^{1,2}$ 
\newauthor
Gonzalo Prieto Lyon,$^{4,5}$
Matteo Bianconi,$^{11}$
Thomas Connor,$^{12}$
Renyue Cen,$^{13}$
Alastair Edge,$^{1}$
\newauthor
Andreas L.\ Faisst,$^{14}$
Marceau Limousin,$^{15}$
Richard Massey,$^{1,2}$
Mauro Sereno,$^{16,17}$
Keren Sharon,$^{18}$
\newauthor
and John R.\ Weaver$^{19,20}$
\\
$^{1}$Centre for Extragalactic Astronomy, Durham University, South Road, Durham DH1 3LE, UK\\
$^{2}$Institute for Computational Cosmology, Durham University, South Road, Durham DH1 3LE, UK\\
$^{3}$Univ Lyon, Univ Lyon1, Ens de Lyon, CNRS, Centre de Recherche Astrophysique de Lyon UMR5574, 69230, Saint-Genis-Laval, France\\
$^{4}$Instituto de Astrof{\'{\i}}sica and Centro de Astroingenier{\'{\i}}a, Facultad de F{\'{i}}sica, Pontificia Universidad Cat{\'{o}}lica de Chile, Casilla 306, Santiago 22, Chile\\
$^{5}$Millennium Institute of Astrophysics (MAS), Nuncio Monse{\~{n}}or S{\'{o}}tero Sanz 100, Providencia, Santiago, Chile\\
$^{6}$Space Science Institute, 4750 Walnut Street, Suite 205, Boulder, Colorado 80301\\
$^{7}$N\'ucleo de Astronom\'ia, Facultad de Ingenier\'ia, Universidad Diego Portales, Av. Ej\'ercito 441, Santiago, Chile\\ 
$^{8}$Astrophysics Research Centre, University of KwaZulu-Natal, Westville Campus, Durban 4041, South Africa\\
$^{9}$School of Mathematics, Statistics \& Computer Science, University of KwaZulu-Natal, Westville Campus, Durban 4041, South Africa\\
$^{10}$Space Telescope Science Institute, 3700 San Martin Dr., Baltimore, MD 21218, USA\\
$^{11}$School of Physics \& Astronomy, University of Birmingham, Birmingham, B15 2TT, UK\\
$^{12}$Jet Propulsion Laboratory, California Institute of Technology, 4800 Oak Grove Drive, Pasadena, CA 91109, USA\\
$^{13}$Princeton University Observatory, Princeton, NJ 08544, USA\\
$^{14}$IPAC, M/C 314-6, California Institute of Technology, 1200 East California Boulevard, Pasadena, CA 91125, USA\\
$^{15}$Aix Marseille Univ, CNRS, CNES, LAM, Marseille, France\\
$^{16}$INAF - Osservatorio di Astrofisica e Scienza dello Spazio di Bologna, via Piero Gobetti 93/3, I-40129 Bologna, Italy\\
$^{17}$INFN, Sezione di Bologna, viale Berti Pichat 6/2, I-40127 Bologna, Italy\\
$^{18}$Department of Astronomy, University of Michigan, 1085 S. University Ave, Ann Arbor, MI 48109, USA\\
$^{19}$Cosmic Dawn Center (DAWN), Denmark\\
$^{20}$Niels Bohr Institute, University of Copenhagen, Lyngbyvej 2, Copenhagen \O\ DK-2100, Denmark\\
}

\date{Accepted XXX. Received YYY; in original form ZZZ}

\pubyear{2022}

\usepackage{newtxtext,newtxmath}
\begin{document}

\label{firstpage}
\pagerange{\pageref{firstpage}--\pageref{lastpage}}
\maketitle

\begin{abstract}
We investigate the strong-lensing cluster Abell 370 (A370) using a wide Integral Field Unit (IFU) spectroscopic mosaic from the Multi-Unit Spectroscopic Explorer (MUSE). IFU spectroscopy provides significant insight into the structure and mass content of galaxy clusters, yet IFU-based cluster studies focus almost exclusively on the central Einstein-radius region. Covering over 14 arcmin$^2$, the new MUSE mosaic extends significantly beyond the A370 Einstein radius, providing, for the first time, a detailed look at the cluster outskirts. Combining these data with wide-field, multi-band \emph{Hubble} Space Telescope (\emph{HST}) imaging from the BUFFALO project, we analyse the distribution of objects within the cluster and along the line of sight. Identifying 416 cluster galaxies, we use kinematics to trace the radial mass profile of the halo, providing a mass estimate independent from the lens model.  We also measure radially-averaged properties of the cluster members, tracking their evolution as a function of infall. Thanks to the high spatial resolution of our data, we identify six cluster members acting as galaxy-galaxy lenses, which constrain localized mass distributions beyond the Einstein radius. Finally, taking advantage of MUSE's 3D capabilities, we detect and analyse multiple spatially extended overdensities outside of the cluster that influence lensing-derived halo mass estimates. We stress that much of this work is only possible thanks to the robust, extended IFU coverage, highlighting its importance even in less optically dense cluster regions. Overall, this work showcases the power of combining \emph{HST}+MUSE, and serves as the initial step towards a larger and wider program targeting several clusters.

\end{abstract}




\section{Introduction}

Mass distributions in galaxy clusters reveal valuable information about the nature of the Universe, providing evidence for the existence of dark matter \citep[e.g.][]{clo06,bra08,har15,bas16}, tracing the growth of total (dark + baryonic) matter over a range of physical scales \citep[e.g.][]{gon07,tes11,mcd17,mul19}, and mapping nodes of the Universe’s large-scale structure \citep[e.g.][]{mas07,lee08,ser18}, all of which constrain important aspects of cosmology \citep[e.g.][]{car02,fu08,vik14,kaf19}. 

Studies involving gravitational lensing \citep[e.g.][]{bra05,mas10,mer11,zit12,smi16,men17,bon18,ace20} are at the forefront of these efforts, as unlike other techniques, lensing provides a direct mass estimate. It is derived from observed distortions of distant background galaxies, without needing to make assumptions about the dynamical state or baryon astrophysics of the cluster, or its morphological symmetry.  Lensing estimates also provide a highly precise picture of dark matter mass fraction and substructure distribution, and as a result the popularity of lensing has steadily increased since its early usage in the 1980s.  The past two decades have, in particular, seen a tremendous increase in the number of lensing-based mass results, driven largely by the advent of wide-field, high resolution imaging arrays -- such as the Advanced Camera for Surveys \citep[ACS;][]{for03} and Wide-Field Camera 3 \citep[WFC3;][]{kim08} cameras on the \emph{Hubble Space Telescope} (\emph{HST}) -- which have significantly improved our ability to detect and characterize background galaxies. 

However, imaging alone provides an incomplete picture of the lensing signal, which is dependent on the 3-dimensional separation between foreground (lens) and background (source) objects. To derive physical mass measurements for the system, we must therefore determine line-of-sight distances to the lens and source objects, which can be achieved with accurate redshift measurements. While large spectroscopic campaigns have targeted lensing clusters in the past, these efforts have often relied on multi-object spectrographs (MOS), such as VIMOS \citep{lef03} on the Very Large Telescope (VLT) and Hectospec \citep{fab05} at the MMT observatory, which are not optimally designed for the crowded nature of galaxy clusters. Furthermore, MOS instruments still largely target objects that have been optically pre-selected, potentially biasing mass results by missing lensing constraints without strong continuum emission or cluster members that do not fall on the typical red sequence.

By contrast, integral field unit (IFU) spectroscopy has dramatically changed the nature of cluster redshift campaigns, as IFUs are particularly efficient at handling crowded fields and provide redshift information for all objects in a field simultaneously, regardless of visual appearance.  The Multi-Unit Spectroscopic Explorer (MUSE; \citealt{bac10}) has been especially valuable in these efforts. With its wide field of view (1\arcmin $\times$ 1\arcmin) and high sensitivity (upwards of 40\% throughput) at moderate resolution (R$\sim$3000 spectrally, 0\farcs2/pixel spatially), MUSE is a powerful resource for our redshifting efforts in cluster fields. Indeed, lensing studies combining MUSE spectroscopy with \emph{HST} imaging have produced some of the highest quality lens models to date, with the MUSE data significantly enhancing the model both on the source-side -- by increasing the number of spectroscopically-confirmed background galaxies \citep[e.g.,][]{mah19,jau19,cam19,ric20} -- and the lens-side, by identifying additional cluster members and measuring their central velocity dispersions, allowing for a more accurate description of small-scale mass distributions \citep[e.g.,][]{berg19,berg21}.

Despite these benefits, MUSE observations of clusters have up to now largely focused on their very central regions, with data usually extending only to the line which separates the singly-imaged weak-lensing regime from the multiply-imaged strong-lensing regime. In some ways this is understandable: as the optically densest region, the core benefits the most from IFU coverage, while the multiply-imaged galaxies found there place the tightest constraints on the mass model.  Nevertheless, the more extended ``outskirts'' regions can also benefit from MUSE spectroscopy. Specifically, cluster members lying in these regions live in relatively less dense environments when compared to those in the core, and many have not been exposed as long to the harsh intra-cluster medium. This allows them to retain a larger fraction of their gas \citep[e.g.,][]{von10,hai15,oma16} and maintain greater star formation rates (SFR).  With MUSE data extending from the core to the outskirts, changes in these parameters can be tracked directly as a function of position and infall radius, revealing a clearer picture of baryonic and dark matter mass evolution as galaxies traverse through the cluster. Since an increased SFR also tends to produce bluer galaxies that can be missed by a standard \citep[e.g.,][]{gla00} cluster member colour cut, the MUSE data provide an additional benefit of capturing these galaxies automatically, without needing to invoke a colour evolution model to target them specifically.  At the same time, identifying additional (bluer) cluster members places tighter constraints on the substructure mass distribution within the cluster, and with a robust spectroscopic sample of cluster members at larger radii it becomes more feasible to employ additional analysis (including the use of cluster kinematics) to complement the lensing data alone. 

Therefore, in this paper we take the first steps at exploring extended cluster regions with IFU spectroscopy, using a panoramic (14 arcmin$^2$) mosaic of MUSE data, in conjunction with multiband \emph{HST} imaging, to investigate the initial outskirts region of the first-known lensing cluster: Abell 370 (A370; \citealt{sou87,ham87}).  These efforts are a further continuation of two previous studies, \citet[][hereafter L17]{lag17}  and \citet[][hereafter L19]{lag19}, which targeted A370 with narrower MUSE fields, respectively covering 1 arcmin$^2$ and 4 arcmin$^2$ areas around the established centre. With this wider data set we are, for the first time, able to diversify our analysis by characterizing the extended cluster structure and probing colour variations in cluster members, all while continuing to investigate the total mass profile and -- thanks to the 3D capabilities of MUSE -- search for additional mass components along the expanded line of sight. This work, which we are calling Pilot-WINGS (the Pilot \textcolor{red}{W}ide-area \textcolor{red}{IN}tegral-field \textcolor{red}{G}alaxy \textcolor{red}{S}urvey) also sets the stage for a larger proposed study (BUFFALO-WINGS), targeting several clusters out to still greater radii.

The manuscript is organized as follows: in Section \ref{sec:data} we highlight the data used for our analysis including the acquisition and reduction techniques used. In Section \ref{sec:redshifting} we present the MUSE redshift catalogue, providing overall statistics as well as a specific breakdown of foreground, cluster, and background galaxies. We use these redshifts to investigate mass distributions within the cluster and along the line of sight in Section \ref{sec:analysis}. We summarize and conclude in Section \ref{sec:conclusions}, where we also lay out future plans for the BUFFALO-WINGS survey.

Throughout this work we assume a flat cosmological model with $\Omega_{\rm \Lambda} = 0.7$, $\Omega_{\rm M} = 0.3$, and $H_{0} = 70$ km s$^{-1}$ kpc$^{-1}$. Assuming these parameters, 1 arcminute spans 309.7 kpc at the systemic cluster redshift ($z = 0.375$). Unless otherwise specified, all magnitudes are presented in the AB system \citep{oke74}. 

\section{Data Acquisition and Reduction}
\label{sec:data}

\subsection{HST}
\label{subsec:hst}
For high-resolution imaging of A370, we use multi-band \emph{HST} data observed in the \emph{Hubble} Frontier Fields (HFF) survey \citep{lot17} and its successor, the Beyond Ultra-deep Frontier Fields And Legacy Observations (BUFFALO; \citealt{ste20}) project. Both programs target the same six lensing clusters, though their observation strategies differ in scope: The A370 HFF data consist of 160 \emph{HST} orbits in 7 broad bands (F435W, F606W, F814W, F105W, F125W, F140W and F160W) and are largely stacked over the field of view of a single instrumental pointing  (ACS for the optical bands and WFC3 for the IR bands). This provides incredibly deep imaging (a 5$\sigma$ point-source limiting magnitude of 29 in all bands) over the crowded central core. Conversely, the BUFFALO data are shallower (16 orbits in total; limiting magnitude $\sim$27.1) and cover a subset of the HFF bands (F606W, F814W, F105W, F125W and F160W), but expand the area of the observational footprint by a factor of 3, providing a nearly 6\arcmin $\times$6\arcmin\ window on the core and outskirts (Fig.\,\ref{fig:footprint}).  

In this work we take advantage of both configurations by using images that combine the HFF and BUFFALO exposures into a single mosaic, using the highest available image resolution (0\farcs03/pixel).  This combination provides robust, multi-band imaging over the entire region covered by MUSE, maximizing our ability to combine photometric and spectroscopic information to advance our science goals. A full description of the methods used to create these images can be found in \citet[][Section 2]{ste20}, but we briefly summarize the procedure here. Individual exposures (of both HFF and BUFFALO data) are reduced using a process based on the standard STScI method, but which includes a number of improvements. In particular, the astrometric precision of each frame is refined to an uncertainty of less than one milliarcsecond and geometric distortions are mapped and rectified.  In addition an updated estimate of background flux is measured and removed from each frame and an improved cosmic-ray rejection algorithm is applied \citep{koe11,koe13}. Finally, all frames in a given filter are combined to form full-depth mosaics, and each mosaic is registered to the GAIA-DR2 absolute coordinate system. While not yet publicly available, the final mosaics are expected to appear on the BUFFALO data products page\footnote{\url{https://doi.org/10.17909/t9-w6tj-wp63}} in the near future.

\subsection{MUSE}
\label{subsec:muse}
\subsubsection{Acquisition}
\label{subsubsec:acquisition}
The MUSE data used in this work were obtained in ESO program 0102.A-0533(A) (PIs: F.~Bauer and D.~Lagattuta) and designed to complement and expand the previous deep observations of the A370 core (2-8 hours exposure time per pointing) used in L17 and L19: programs 094.A-0115(A) (PI: J.~Richard) and 096.A-0710(A) (PI: F.~Bauer).  In contrast to the core data, these new fields are shallower; each position is covered by a single 1-hour Observation Block (OB) with an effective exposure time of 45 minutes when accounting for overheads. Every OB consists of four $\sim$11-minute exposures, the first of which has a PA of 29 degrees (east of north) to match the orientation of the deep observations. Subsequent exposures are rotated 90 degrees clockwise relative to their predecessors and include a small 0.5\arcsec\ dither to mitigate the effects of detector systematics and maximize sky coverage. During acquisition, we operate MUSE in wide-field adaptive-optics (WFM-AO-N) mode, exposing under high-quality observing conditions: $<0.7\arcsec$ seeing, $<10\%$ cloud cover, target airmass $<1.5$ and moon distance $>90$ degrees. This ensures that, even though our exposure times are shallow, we are still able to effectively detect and identify a significant number of faint foreground and background galaxies, developing a robust 3D image of the cluster field. 

\begin{figure}
    \centering
    \includegraphics[width=0.49\textwidth]{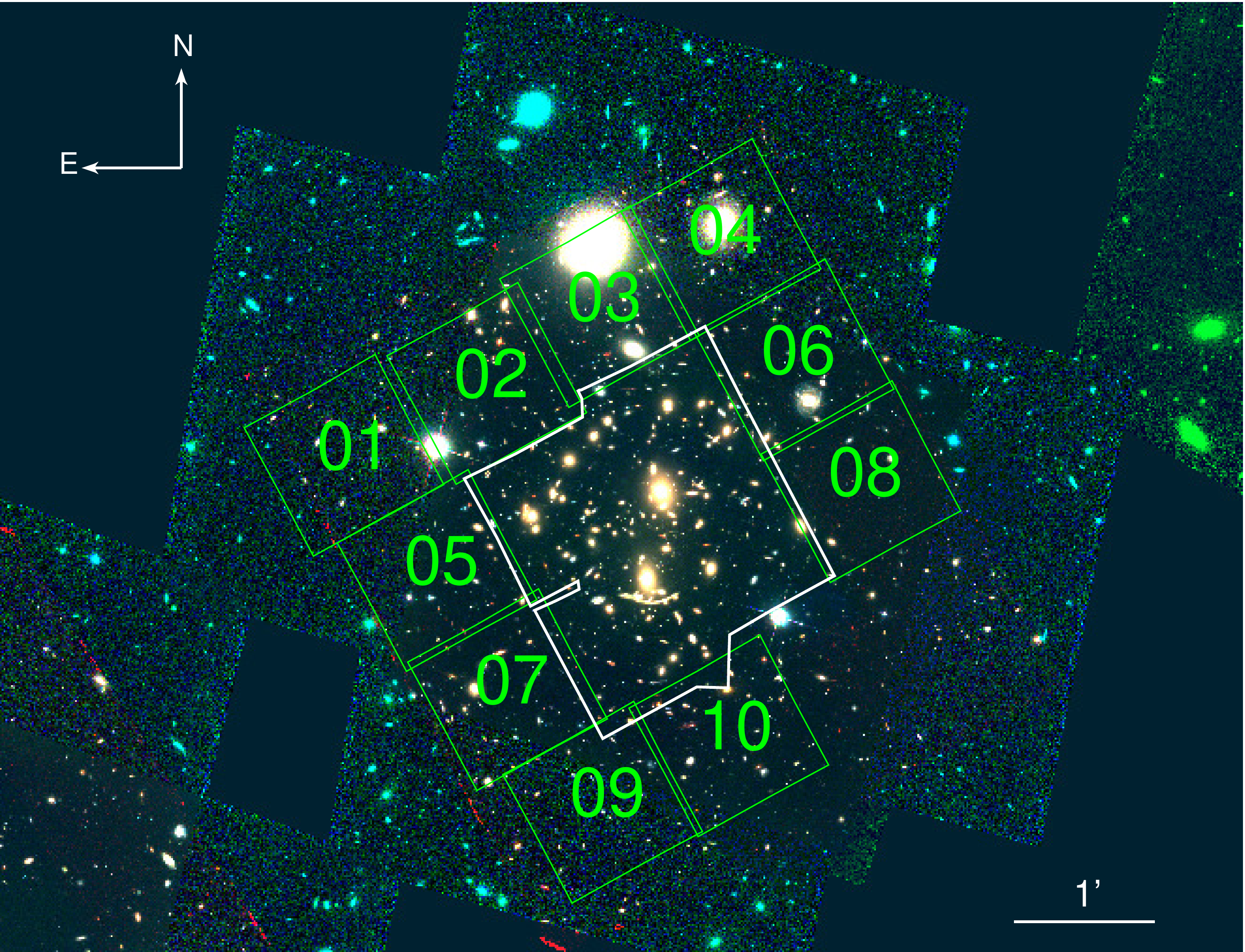}
    \caption{Data of the Abell 370 (A370) cluster used in this work. Newly-acquired shallow (1-hour depth) MUSE data appear as numbered green boxes, nearly surrounding the cluster centre. The footprint of previously-obtained deep MUSE data (2-8 hour depth) is shown as a white contour. We subsequently combine these two data sets to create a contiguous region of spectroscopic coverage spanning 14 arcmin$^2$. The coloured image shows the combined \emph{Hubble Space Telescope} coverage of A370, using F606W/F814W/F160W filters. Like the spectroscopic footprint, the imaging region also shows a deep core (the \emph{Hubble} Frontier Field region) and shallower outskirts (the BUFFALO region).}
    \label{fig:footprint}
\end{figure}

All frames were taken between 30 July and 24 December 2019. Although we requested 15 OBs in total, only 10 were successfully completed before the end of the observing semester, forming an almost complete ring surrounding the central cluster region. When combined with the core data this created a nearly 4\arcmin $\times$ 4\arcmin\ footprint covering the cluster and its surroundings (Fig.\,\ref{fig:footprint}), serving as a spectroscopic counterpart of BUFFALO and the original \emph{HST} Frontier Fields program.

\subsubsection{Data Reduction}
\label{subsubsec:reduction}
To reduce the data we largely follow the technique described in \citet{ric20}, which is based on recipes in the public MUSE Data Reduction Pipeline User Manual\footnote{\url{https://www.eso.org/sci/software/pipelines/muse/}} and \citet{wei20}. Full details can be found in those sources (see, e.g., Section 2.5 in \citealt{ric20}), but for clarity we briefly summarize the process here. 

First, we use calibration files associated with a given OB to characterize systematic effects in the raw data: we create bias- and flatfield-correction frames, generate a trace table (to map the position of the instrument slitlets on the detector), determine a unique wavelength solution, measure the line-spread function, and correct for additional illumination effects with twilight and night-time lamp flat exposures. 

After applying these corrections, we reformat the calibrated data for each exposure into non-interpolated \emph{pixel tables} (rather than fully-formed data cubes) to undergo additional calibrations. This is done to reduce memory and processing costs, but does not affect the final output product. While in the \emph{pixel table} format, we apply flux calibration with a standard star exposure taken shortly before or after the science frames, correct for telluric effects and differential atmospheric dispersion, perform a basic sky subtraction, and transform any residual velocity offsets to the Solar System's barycentric rest frame. Additionally, since we use AO corrections, we also fit and subtract residual Raman-line flux induced by the laser guide star.

During this step we also align individual exposures to a common world coordinate system (WCS) -- in this case the WCS defined in the A370 BUFFALO imaging. This is done separately from the standard ESO reduction pipeline, using a stand-alone python script (\texttt{imphot}; \citealt{bac17}) which cross-correlates the MUSE data with the high-resolution \emph{HST} imaging, resampling the \emph{HST} data to match the MUSE pixel scale and Point-Spread Function (PSF). As an added benefit, \texttt{imphot} also normalizes the MUSE flux to match the \emph{HST} photometry in the F606W and F814W bands (i.e., the bands that overlap the MUSE wavelength range), providing a second opportunity for flux calibration during final combination. 

We then process the \emph{pixel table} of each exposure into a data cube using the ESO pipeline. As part of this process, we correct any remaining low-level flat-fielding systematics using a self-calibration routine in the \texttt{muse\_scipost} recipe in the pipeline. In order to remove the effects of diffuse light, which can strongly bias the self-calibration procedure, we generate a mask to ignore regions of sky close to bright cluster members and foreground galaxies (especially the two large, low-redshift galaxies seen in tiles 3 and 4 in Fig.\,\ref{fig:footprint}.) Through empirical testing, we have determined that the optimal threshold for ``clean'' sky selection is to mask all pixels with a flux value greater than 0.3 electron s$^{-1}$ in the \texttt{imphot}-smoothed \emph{HST} data.

Once the final cubes are created, we combine all exposures of a given OB, scaled by their \texttt{imphot}-derived flux normalizations, into a final ``master'' cube. This is done using another external program, \texttt{mpdaf} \citep{piq19}. Each master cube consists of both the reduced data and its variance, and is regularly sampled at 0\farcs2 per pixel in the two spatial directions and 1.25 \AA\ per pixel in the wavelength direction. Individual 3D (spatial + wavelength) pixels are known as spatial pixels, or ``spaxels''. All cubes cover the wavelengths between 4750 and 9350 \AA, and span an area of 1 arcmin$^2$. To match the final ESO pipeline output format as closely as possible, \texttt{mpdaf} employs the same drizzle method as the standard pipeline when resampling the data, and also performs a cosmic-ray rejection to remove unwanted bright defects.  Finally we run the Zurich Atmospheric Purge \citep[ZAP;][]{sot16} on each cube in order to remove any final sky-substraction residuals. ZAP requires a mask to operate, so for consistency we use the same mask that we employed in the self-calibration step. After running ZAP, the cubes are considered fully reduced, and ready to be analysed for redshifting.

Unlike the data in the core, we do not combine all cubes into a mosaic, as the resulting data structure would be too memory-intensive to open. Due to slight overlaps between neighboring cubes (and also between individual cubes and the core mosaic) this means that some objects -- those lying in overlap regions -- will initially appear multiple times during redshifting and cataloguing (Section \ref{sec:redshifting}). Therefore, when compiling the final catalogue we manually inspect entries to remove any duplicate copies of these galaxies, in order to ensure an accurate census.

\section{Redshift Fitting}
\label{sec:redshifting}

To create redshift catalogues out of the MUSE data, we proceed much like we have in past work (e.g. L19; \citealt{ric20}), namely, we first detect sources in the MUSE cube, then extract their spectra and fit redshift solutions, visually inspect these solutions, and finally compile the results into a master catalogue. Below, we summarize these steps, but again refer the interested reader to Section 3 of \citet{ric20} for a full description of the process.

\subsection{Object detection and extraction}
To maximize the number of redshift candidates we can identify, we use two different techniques to detect sources in the field. Ultimately, both methods utilize Source Extractor (SExtractor; \citealt{ber96}) for object detection, but each targets the data in a different way. 

In the first method, we take advantage of the BUFFALO imaging to select objects that have broad continuum emission (i.e., those that are visible in the \emph{HST} data). Since we aim to collect as many candidates as possible, we first create median-subtracted versions of the BUFFALO images by subtracting the local background flux from each pixel. We take the background to be the median flux in a 1\farcs5 $\times$ 1\farcs5 box surrounding a given pixel. This creates a better contrast around faint galaxies that may otherwise be overwhelmed by the light of their brighter neighbors, without affecting the cores of these bright galaxies, which are still easily detectable.  Afterwards, we combine all the median-subtracted frames into a single, inverse-variance weighted ``detection'' image, to improve the overall SNR. We run SExtractor on this detection image (which has also been cropped to match the MUSE cube footprint) to create a list of detections, which we call \emph{prior sources}. We do note that, in some cases (such as clumpy spiral galaxies) this results in an over-segmentation and many duplicate entries in the map, though we flag and re-combine these entries during inspection. We also note that the sensitivity of the detection image can vary significantly over a given MUSE pointing, since the outer regions of the BUFFALO mosaic are considerably shallower ($\sim$1.9 magnitudes in depth) than the centre, which contain the HFF data. Nevertheless, to be as uniform as possible, we use the same SExtractor parameters for all cubes, and over all parts of the image. To account for the differences in image depth, we also include \emph{HST} weight maps (also cropped to match the MUSE footprint) in our SExtractor analysis.

In the second method, we run SExtractor on the MUSE data directly, using the \texttt{muselet} routine available in \texttt{mpdaf}.  To operate, \texttt{muselet} first creates a copy of the cube, replacing the data in each wavelength slice with its equivalent ``pseudo-narrow-band'': the program averages the flux in a narrow wavelength range (centred on the current slice) to make a ``source'', then subtracts the average flux measured from the combined regions just red-ward and blue-ward of the source to remove the ``continuum''. This effectively identifies excess flux above the local continuum level, and is an efficient way to detect emission-line objects. Both the source and continuum parameters are scalable in \texttt{muselet}, and in this work we set the source wavelength range to be $\delta_{\lambda} \leq 2.5$\AA\ (or five slices in total) and the continuum range to be 25\AA\ (or 20 slices) both blue-ward and red-ward of the source. After creating the narrow-band cube, the software runs SExtractor on each slice to identify emission peaks and, after combining peaks at different wavelength slices that are spatially ``close'' to one another (i.e. within the same seeing disk) into ``multi-line objects'', outputs a list of unique detections in the cube, which we call \emph{muselet sources}.

After generating both \emph{prior} and \emph{muselet} sources, we extract a spectrum for each object based on the SExtractor segmentation map created in the detection step. For \emph{prior} sources we first expand the detection region to account for differences in the MUSE pixel scale and convolve the \emph{HST}-based segmentation map with the MUSE PSF.  Because the PSF varies as a function of wavelength, we actually create different convolution maps at ten ``anchor'' wavelengths (evenly spaced at 500\AA\ intervals throughout the MUSE spectral range) and interpolate the results over the entire data cube. Conversely, for the \emph{muselet} sources we simply take the SExtractor detection map of the object's brightest emission line, but projected over all wavelengths. In both cases, we then sum all of the spaxels covered by the new detection footprint, weighted using the optimal \citet{hor86} algorithm, to create an initial 1D spectrum.  To remove any remaining systematics due to sky variation or the extended flux of bright neighbors, we also estimate a local sky spectrum by summing the 500 nearest ``blank'' spaxels (spaxels that are not associated with any detection object in the field) that fall within a 0\farcs4 - 4\farcs0 circular annulus centered on the target object. After subtracting the sky contribution from the source spectrum, we are left with a final ``science'' extraction spectrum that is suitable for redshifting. 

To aid us in the redshifting process, we run each spectrum though \texttt{Marz} \citep{hin16}, a tool that cross-correlates a spectrum with a series of custom templates to provide an initial redshift guess. For this work we use the same templates that were developed to analyse the MUSE coverage of the \emph{Hubble} Ultra Deep Field \citep{ina17}.  Following that paper, we also generate the top five \texttt{Marz} solutions for each spectrum rather than a single best-fit redshift, to allow for possibility of line mismatches or contamination from neighbours. These are then used in the inspection phase of the analysis.

\subsection{Inspection and catalogue creation}
\label{subsec:inspection}
After extracting spectra from all sources, we visually inspect the results in order to determine redshifts for the final catalogue. Based on past experience, we select all \emph{muselet} sources for this process, but limit \emph{prior} sources to  objects with a SNR$>$1.0 in the combined detection image (corresponding roughly to a magnitude limit of m$_{\rm F814W}$ = 25.5), as objects fainter than this cutoff have largely sky-noise dominated spectra with no discernible features. Applying these limits we are left with 1694 objects to inspect.  

Given the large number of systems to check, we share inspection duties between nine coauthors (``redshifters''), grouped into three teams: Team Durham (DJL, GM, CC), Team Lyon (JR, AC, AJ), and Team Santiago (FEB, LG, GPL). To optimize the process, we divide the initial object list evenly between the three teams.  However, rather than subdivide these lists further, each member of a given team inspects the entire subset individually. This ensures three independent checks of a candidate spectrum, while providing an efficient way to identify and resolve discrepancies within teams (by way of a team-level reconciliation meeting).

To maintain a level of consistency in the analysis, all redshifters use a customized graphical user interface for inspection. This tool (named Source Inspector; Bacon et al. in prep) displays the spectrum in question, along with its ``white-light'' image (the combined 2D MUSE spectrum stacked along the spectral direction) and all \emph{HST} images of the target. In addition the tool also overlays key lines at a given redshift, and the user is able to switch between \texttt{Marz} redshift solutions or input a custom solution to look for alternatives.  Source Inspector also highlights any other extracted sources within a $\sim$ 10\arcsec\ box centred on the current target so the user can easily identify nearby contaminants or cross-match \emph{prior}- and \emph{muselet}-identified sources. When the user is satisfied with a redshift choice for a given object, the entry is saved into a database and the next target object is shown.

Once all members of a team observe their object list, the team meets to compare results and resolve any conflicts. These can include disagreements over the identified redshift, the confidence in the chosen redshift (from low to high), or matches between nearby sources. Final parameters for each object are chosen, and a ``resolved'' redshift list is recorded (again using Source Inspector) and sent for final combination.  In this last step we combine all lists and do a final cross check to eliminate any duplicates that exist between different lists or between the new lists and the original redshift catalogue from L19. This final combination results in the master redshift list, which contains spatial and spectral information for all remaining objects. 

\begin{figure*}
    \centering
    \includegraphics[width=0.49\textwidth]{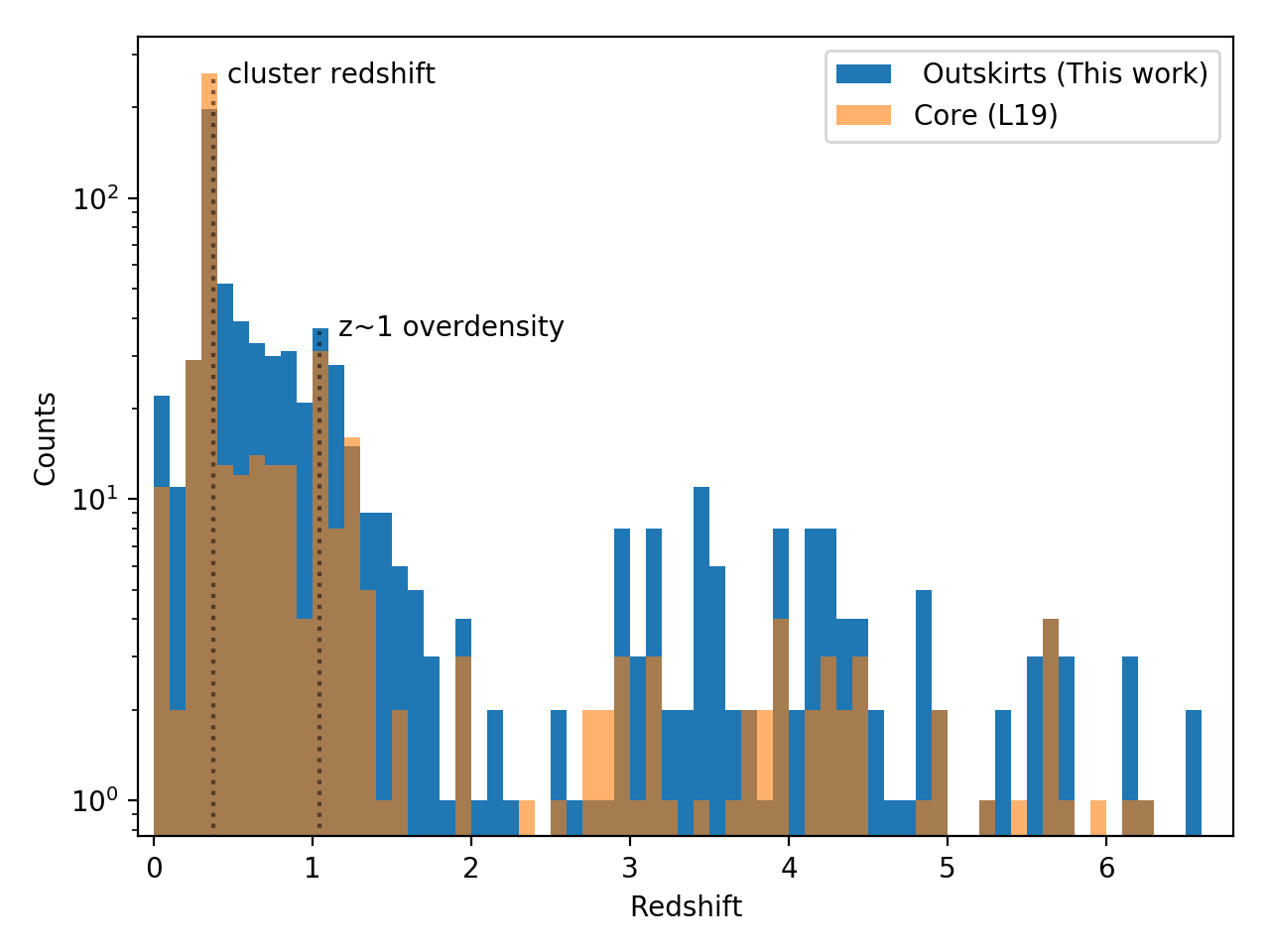}
    \includegraphics[width=0.49\textwidth]{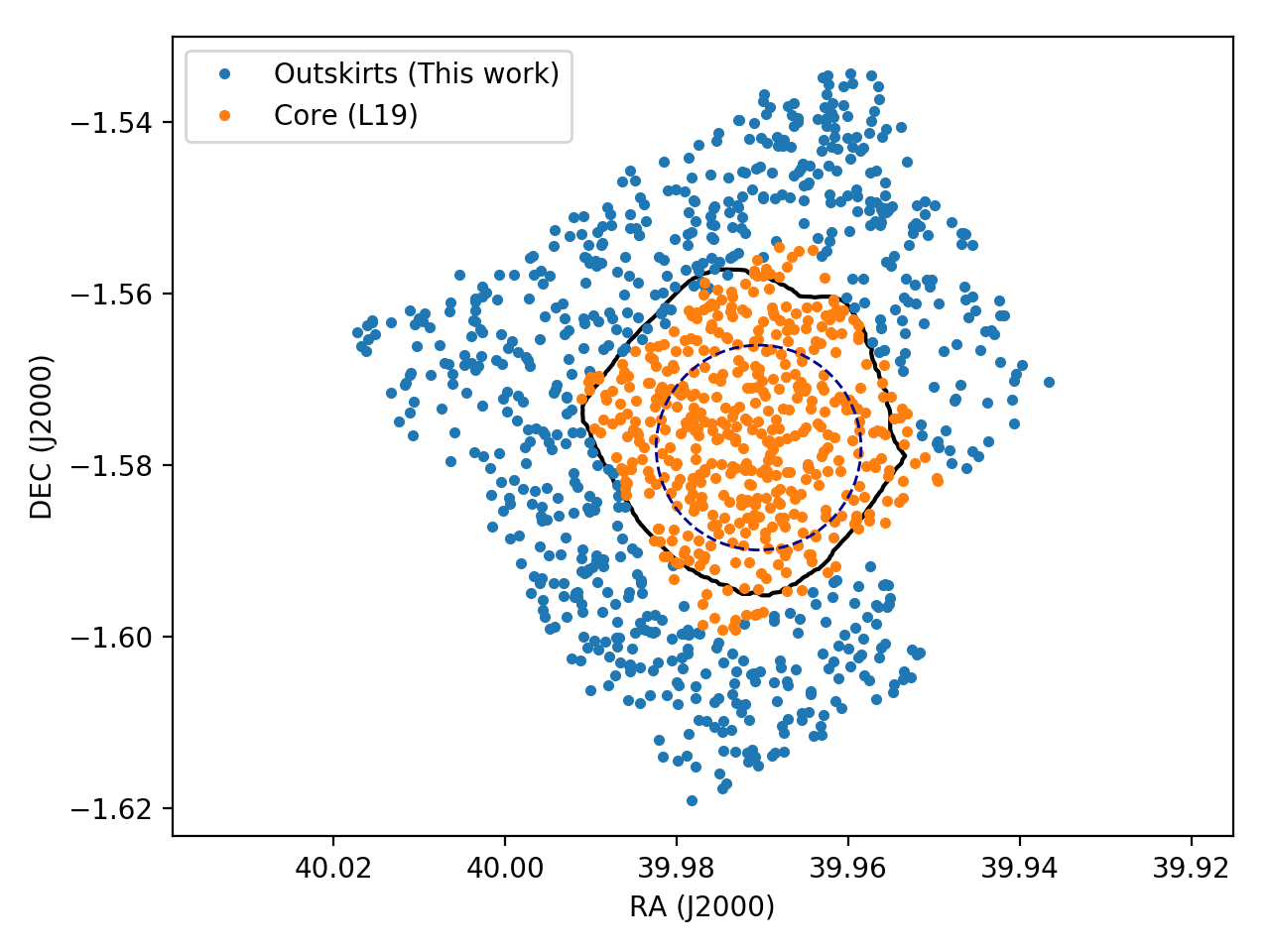}
    \caption{Left: Redshift distribution of the combined MUSE catalogue used in this work. Objects originally identified in the core region \citep[][; L19]{lag19} are shown as an orange histogram, while new objects found in the outskirts are shown in blue. Two dashed lines highlight the systemic cluster redshift ($z = 0.375$) and the median redshift of an overdensity of galaxies at $z\sim1$ ($z = 1.049$, Section \ref{subsec:backGroup}). Right: Spatial distributions (projected on the sky) of all redshifts, separated into the core and outskirts regions, coloured as in the left panel. The thick black line, which roughly traces the boundary between the deep exposures and the new data highlights the extent of the multiple-image region predicted by the L19 lens model. This line marks the maximum possible deflection experienced by multiply-imaged galaxies and provides a lensing-based definition for the core/outskirts boundary in this work. At the cluster redshift ($z = 0.375$) it extends $\sim$350 kpc in physical radius. For reference, the associated model Einstein radius (43\arcsec\ for source galaxies at $z = 10$) appears as a dashed oval.}
    \label{fig:redshifts}
\end{figure*}

We present the overall redshift distribution in Fig.\,\ref{fig:redshifts}, showcasing the cluster core and outskirts regions separately. While there are many ways to define ``core'' and ``outskirts'' in galaxy clusters, \citep[see, e.g.,][]{rei13,wal19} in this work we take a lensing-based approach: namely, we consider the core to be the (projected) extent of the multiple-image region in the lens model, with the outskirts then anything beyond this (2D) border that falls within the cluster redshift range. For our analysis, we use the multiple-image region as defined by L19, which extends over a cluster-centric radius of $\sim$350 kpc. Physically, the multiple-image boundary encloses the densest (i.e., super-critical) regions of the cluster, while also selecting the largest, most luminous cluster member galaxies, including the two Brightest Cluster Galaxies (BCGs). Defining the core/outskirts boundary in this fashion therefore provides a convenient discriminator in terms of both mass environment and stellar properties.

\begin{figure*}
    \centering
    \includegraphics[width=0.49\textwidth]{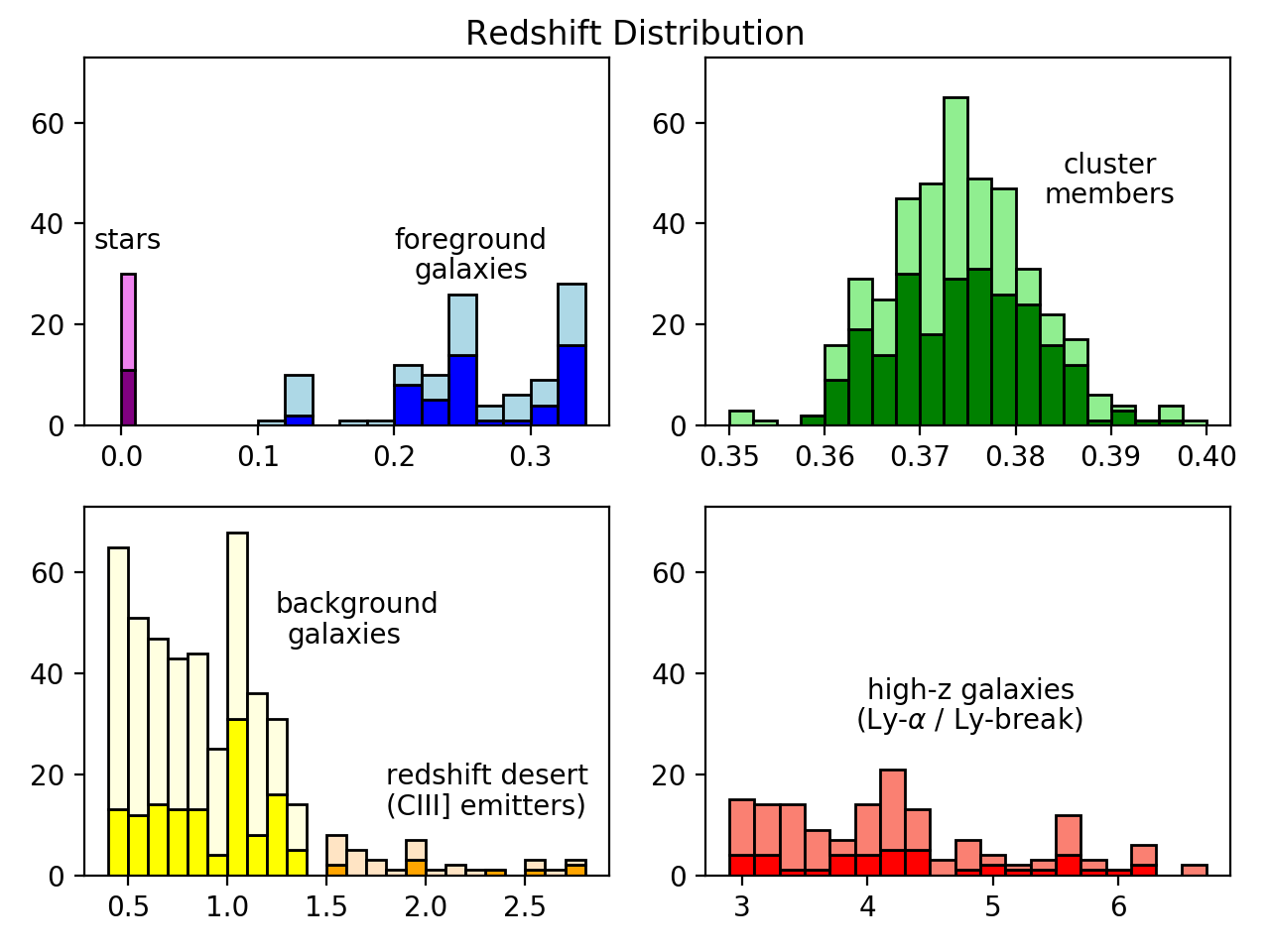}
    \includegraphics[width=0.49\textwidth]{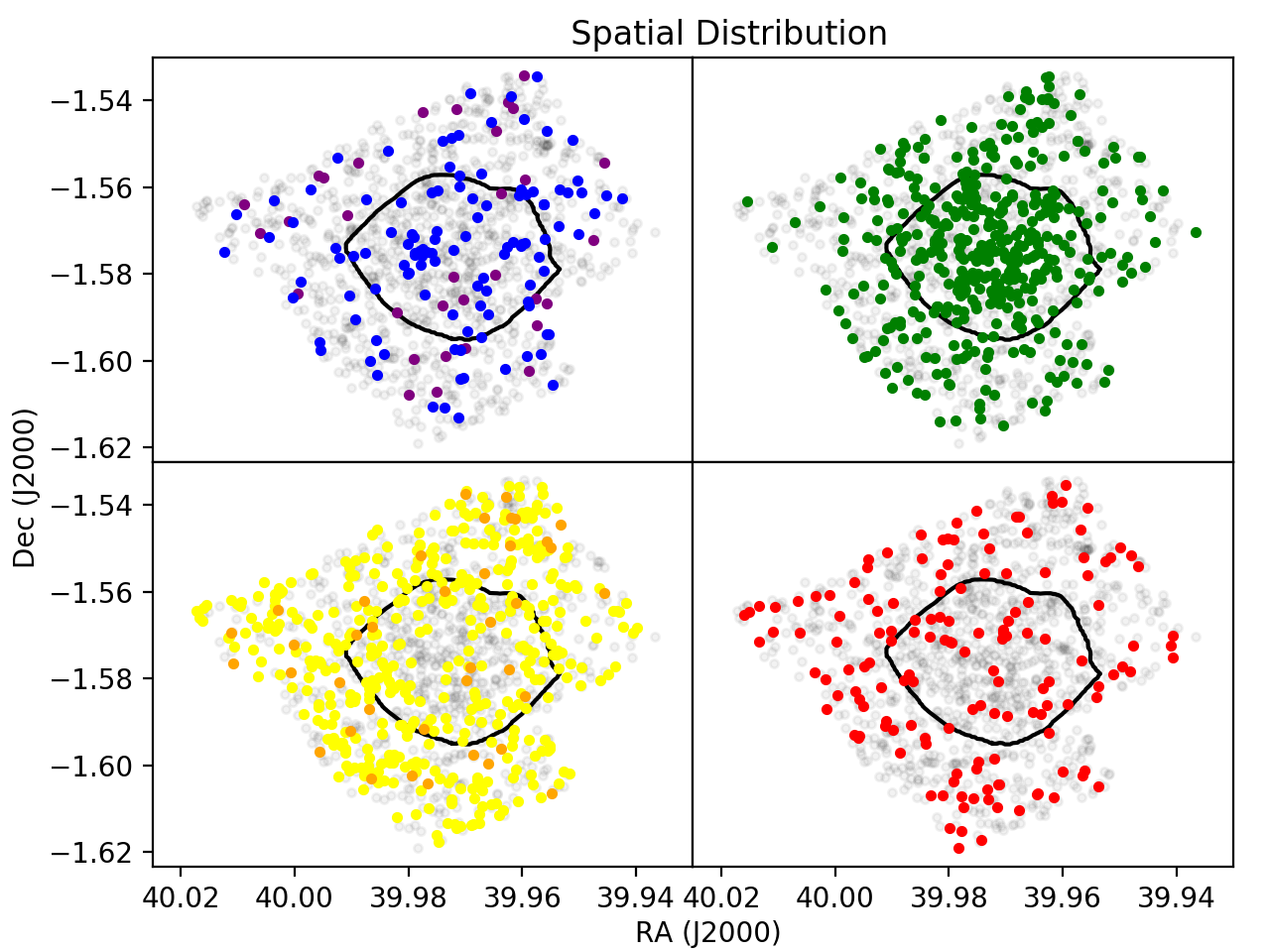}
    \caption{Left: A breakdown of the total redshift distribution based on line of sight. Light-coloured histogram bins represent objects in the outskirts region, while dark-coloured bins are those in the core. Histograms are colour-coded as follows: purple -- stars, blue -- foreground galaxies, green -- cluster members, yellow -- near-background galaxies; orange -- galaxies in the MUSE ``redshift desert'', red -- high-redshift galaxies (typically Lyman-$\alpha$ emitters or Lyman break galaxies). Right: Spatial distribution of the catalogue, with colours matching those in the left-hand panel. Cluster members, as expected, dominate the core region (constituting 45\% of the total population), but become less common in the outskirts (only 28\% of the population).
    }
    \label{fig:z_breakdown}
\end{figure*}

A further breakdown of the combined redshift distribution, highlighting different regions of the line of sight, is presented in Fig.\,\ref{fig:z_breakdown}. Following L17 and L19, objects are broadly classified as being foreground ($z < 0.35$), cluster ($0.35 \leq z \leq 0.4$), or background ($z > 0.4$), with the foreground and background categories further subdivided as follows:\ we find 30 stars (purple bins; $z = 0$), 116 foreground galaxies (blue; $0 < z < 0.35$), 416 cluster members (green; $0.35 \leq z \leq 0.4$), and 623 background galaxies, separated into 434 near-background galaxies, identified largely by [\oii] emission (yellow; $0.4 < z < 1.5$), 39 objects in the MUSE ``redshift desert'' -- a redshift range where typical strong emission lines fall outside of the observable wavelength coverage, instead relying on (usually weak) \ciii] emission/absorption for identification (orange; $1.5 \leq z < 2.9$), 
and 150 distant-background galaxies, identified as Ly$\alpha$ emitters or Ly-break galaxies (red; $z \geq 2.9$). Considering only the outskirts (lighter-coloured bins in Fig.\,\ref{fig:z_breakdown}) we find 19 stars, 61 foreground galaxies, 180 cluster members, and 252, 28, and 109 background galaxy types, respectively. To avoid double counting and maintain an accurate redshift census, we take care to keep only a single entry of multiply-imaged galaxies in these figures, though we note that these objects only constitute a small fraction of the distribution ($<$3\%), and are limited to the core region. We do, however, include and clearly identify all components of each multiply-imaged system in the master catalogue. For reference, we present a small subset of this catalogue (along with a brief description of its contents) in Appendix \ref{sec:appendix}. 

\section{Cluster and Line-of-Sight Structure}
\label{sec:analysis}
With an expanded spectroscopic footprint, we are now able to investigate the structure of A370 and its surrounding environment more thoroughly than either the initial MUSE-based exploration of L17, or the wider mosaic presented in L19. In particular, by moving beyond the dense core, we can probe cluster components that have not yet settled into the central potential, providing insight into mass assembly. At the same time, the lines of sight in these regions are less dominated by cluster members, providing a clearer look at the objects and environments surrounding the cluster itself. Therefore, in this section, we highlight key objects and structures identified in the spectroscopic catalogue, paying special attention to how these systems affect the overall (line of sight) mass distribution in the field. 

\begin{figure*}
    \centering
    \includegraphics[width=0.49\textwidth]{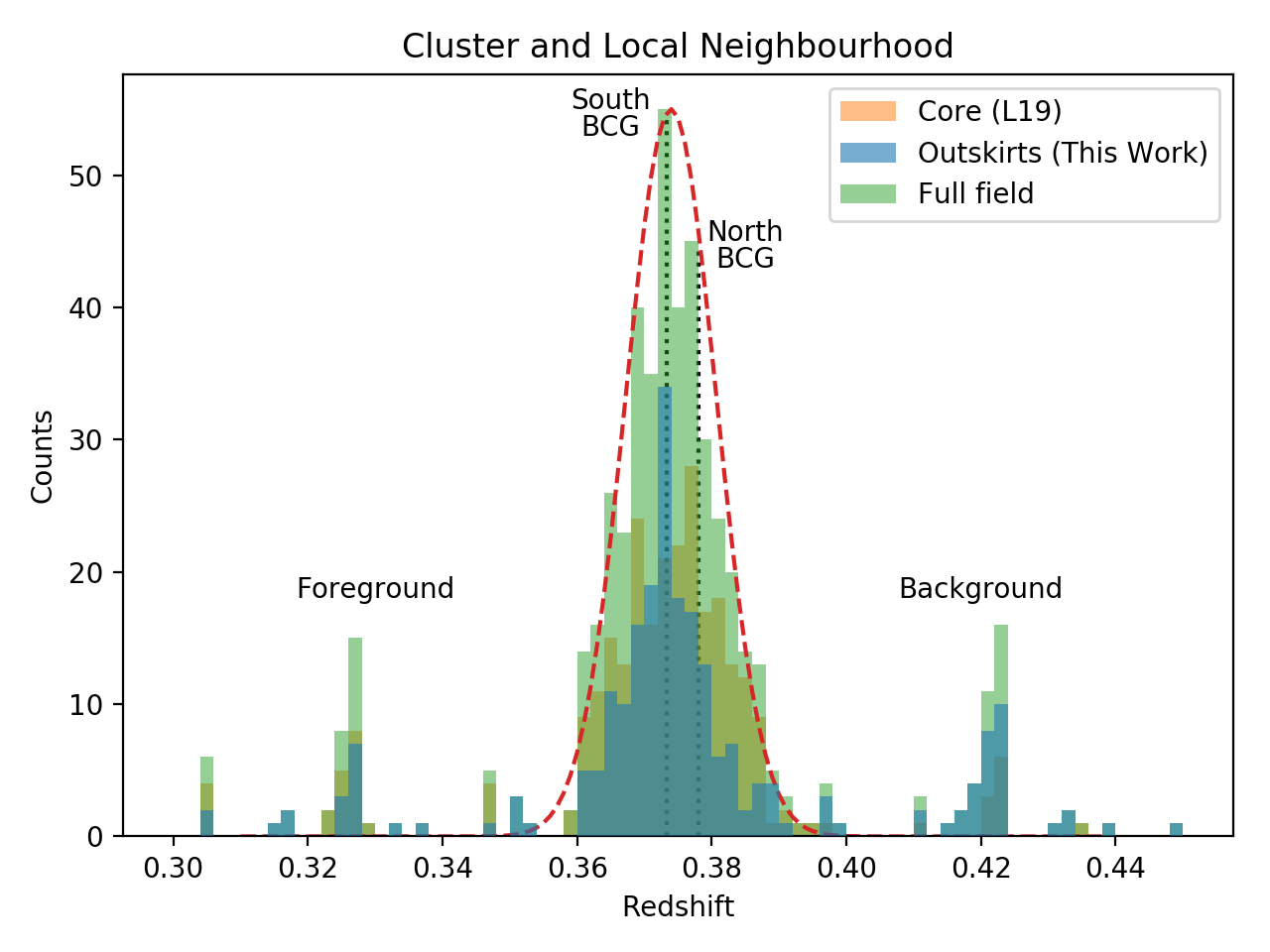}
    \includegraphics[width=0.49\textwidth]{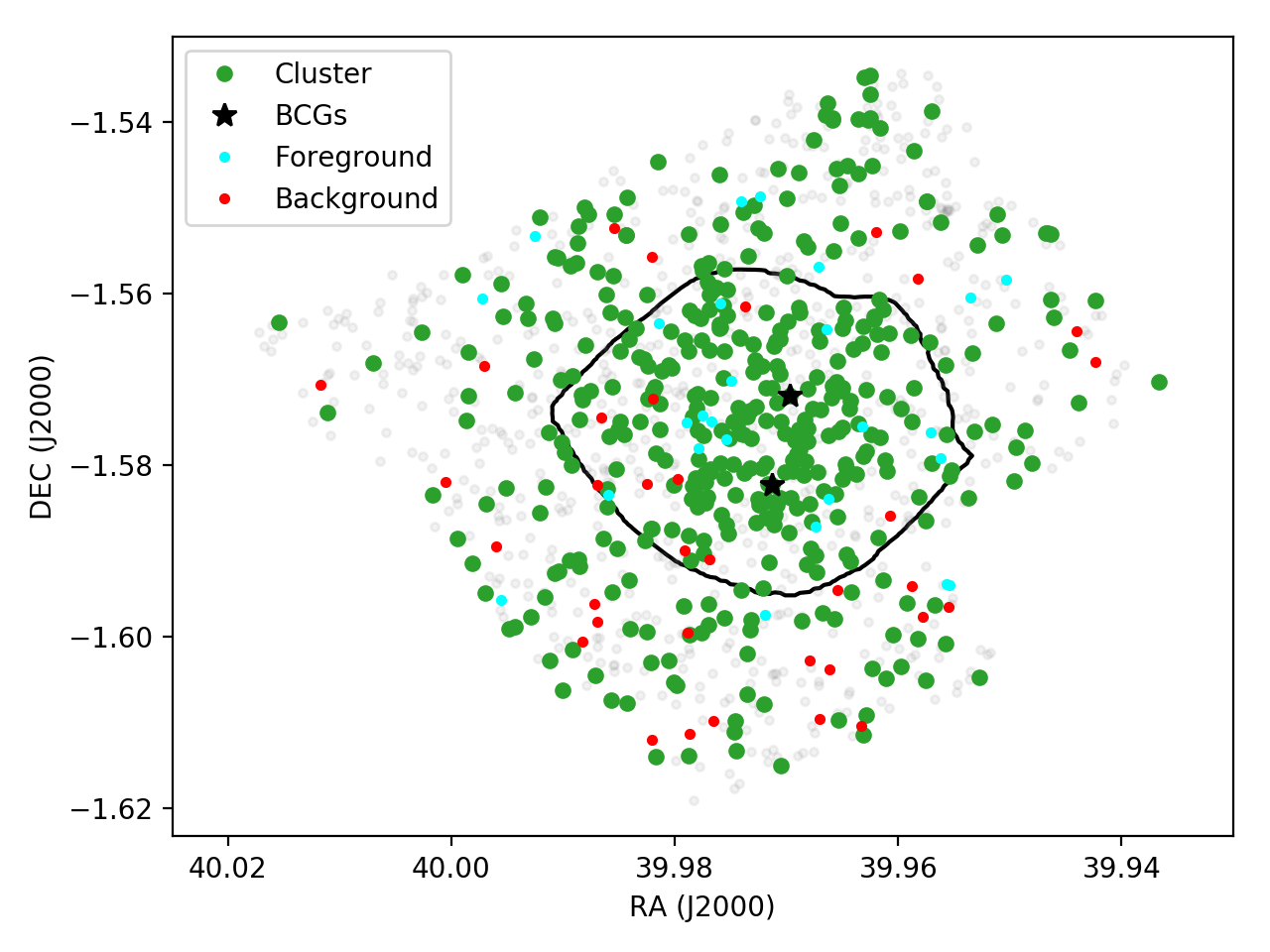}
    \includegraphics[width=0.49\textwidth]{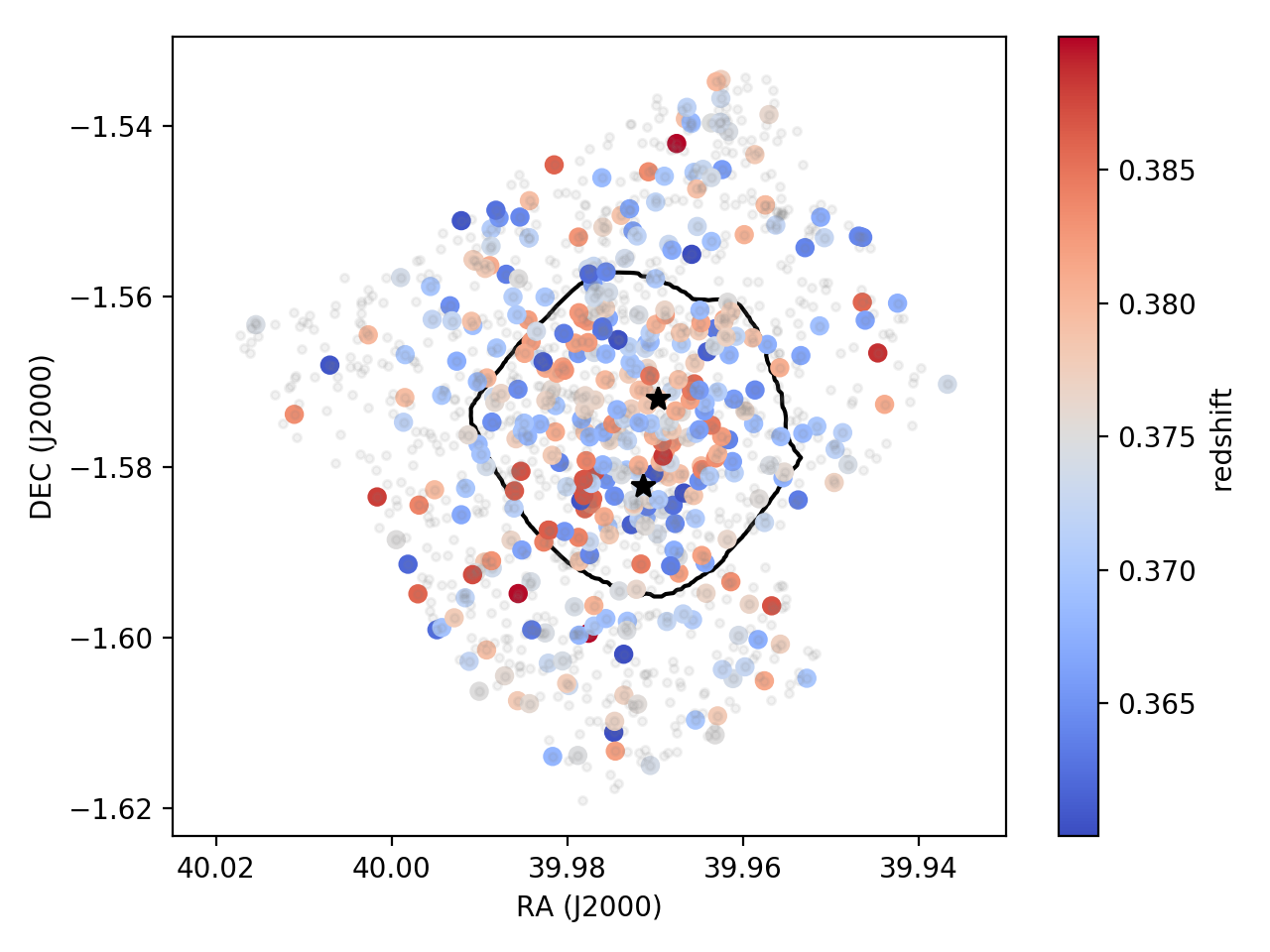}
    \includegraphics[width=0.49\textwidth]{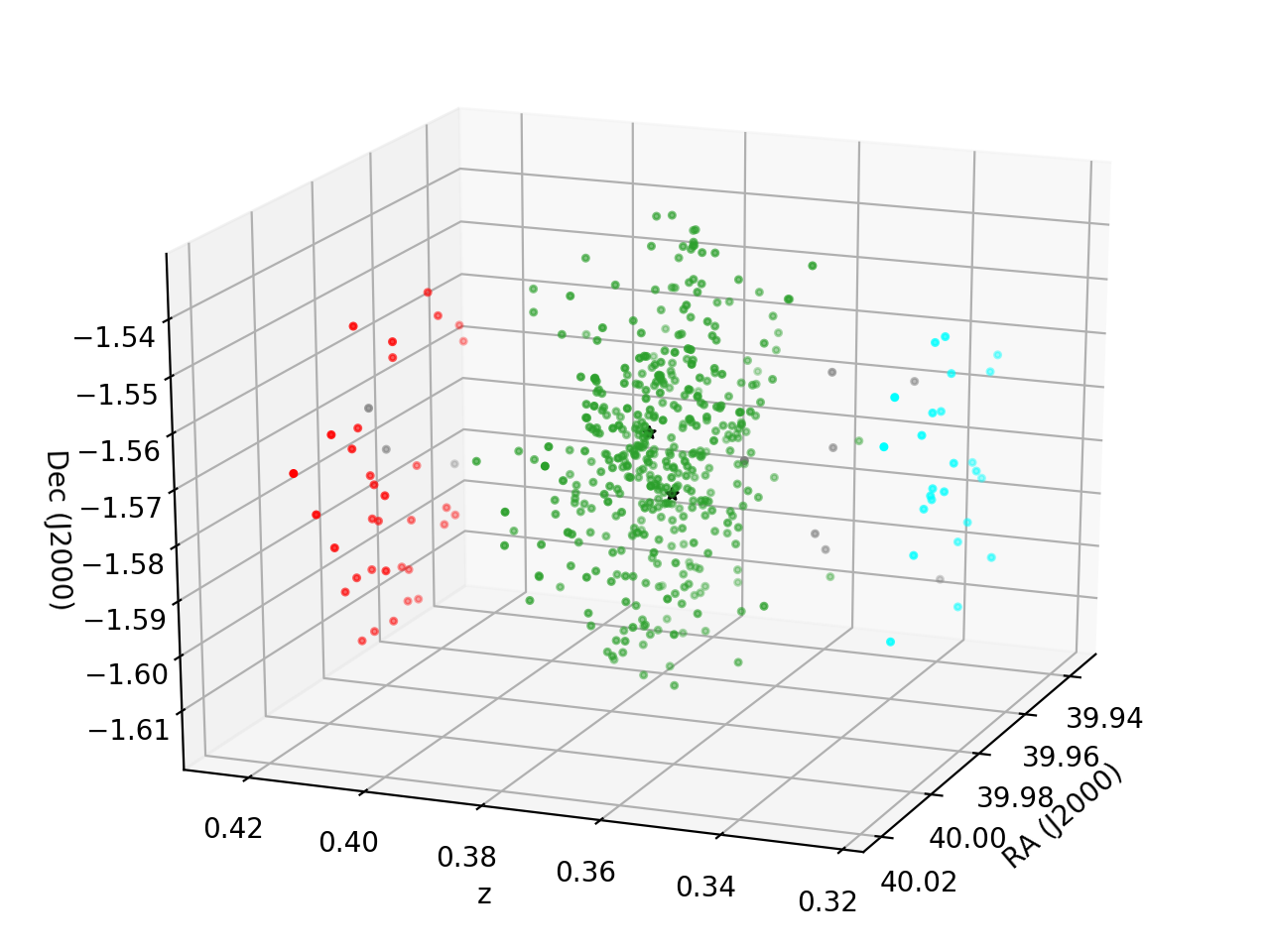}
    \caption{Top left: Redshift distribution of cluster members and close (line-of-sight) neighbors. As in Fig.\,\ref{fig:redshifts} galaxies in the core region appear in orange, while outskirts galaxies appear in blue. The combined distribution is shown in green and is well-fit by a Gaussian (red dashed line; $\sigma_{\rm fit}$ = 1467 km s$^{-1}$). The redshifts of the northern and southern BCGs are highlighted by dashed lines. In addition to the cluster, two small overdensities can be seen at $z = 0.325 $ and $z = 0.42$. Top right: Spatial distribution of confirmed cluster members (green circles) relative to all objects detected in the MUSE footprint (grey dots). The two black stars mark the positions of the BCGs. Galaxies that belong to the narrow line-of-sight overdensities, labelled ``foreground'' and ``background'' in the left-hand panel, appear as small cyan and red dots, respectively. Bottom left: Combined spectral and spatial distribution. Despite undergoing a merger, we do not see distinct populations of cluster members tied to each BCG; this is expected, since the two subcomponents have already passed through each other at least twice, and are likely approacing their next closest pass-through \citep{mol20}. Bottom right: Deprojection of the cluster and its close neighbors into 3D space. We see a clear gap between cluster members and nearby surrounding objects.}
    \label{fig:cluster}
\end{figure*}

\subsection{Cluster members}
\label{subsec:cluster}
Cluster members make up a significant fraction of objects in the A370 field, which is clearly evident in the combined redshift sample. Fig.\,\ref{fig:redshifts} shows a considerable peak of galaxies around the established systemic cluster redshift of $z = 0.375$ in both the core ($r \leq 350$ kpc) and outskirts ($r > 350$ kpc) regions. While the large collection of galaxies serves as a robust statistical sample in its own right (recall, we identify 416 unique objects in the fiducial cluster range of $0.35 \leq z \leq 0.4$), the population as a whole also provides important information about the cluster itself. Combining redshift data with positional information from BUFFALO generates a high-resolution, 3D picture of cluster structure that can trace the overall shape of the host halo and reveal links to the larger cosmic web. At the same time, cluster member dynamics help to probe the central potential well, providing an estimate of cluster mass, while changes between the physical properties of core- and outskirts-based cluster members can track evolutionary effects sensitive to environment.  

\subsubsection{Distribution}
\label{subsubsec:distrib}
We present spectral and spatial distributions of the cluster and its local neighborhood in Fig.\,\ref{fig:cluster}. Selecting all spectroscopically-identified cluster members, we find that the redshift distribution is well-fit by a Gaussian curve centred at $\overline{z} = 0.374$, in excellent agreement with the systemic value. Transforming the redshifts into velocity space, we measure a combined velocity dispersion of ($1465 \pm 67$) km s$^{-1}$, also in good agreement with previous estimates that extend over similar distances, but which use significantly fewer galaxies: e.g., $1340^{+ 230}_{-150}$ km s$^{-1}$ \citep[][1 Mpc, 29 galaxies]{mel88}; $1216 \pm 128$ km s$^{-1}$ \citep[][1.3 Mpc, 75 galaxies]{cri16}. Spatially, the galaxies appear over the entire MUSE footprint.  While some variations in density can be seen, there are no significant galaxy concentrations that would signal the presence of a major substructure -- although there is a $\sim 20\%$ overdensity of objects oriented in a north-south direction through the cluster centre, relative to other position angles. Intriguingly, this matches the orientation and position of a possible filament identified in \citet{gho21}, though at present any relationship between the two is purely speculative.

Focusing on the core region alone, we measure a velocity dispersion that is slightly larger than (but statistically consistent with) the combined sample ($\sigma_{\rm core}$ = 1520$\pm$93 km s$^{-1}$.)  The redshift distribution in this region also appears bimodal, with two maxima at $z = 0.369$ and $z = 0.377$.  Although A370 is a known merging system with two distinct BCGs, the individual BCG redshifts ($z = 0.3733$ and $z = 0.3780$) do not correspond to the distribution peaks. Since recent measurements of x-ray gas surface density in A370 find that the merging components have passed through each other at least twice already and are nearing their third conjunction \citep{mol20}, cluster members originally bound to one or the other of the two BCG halos should now be well-mixed in velocity space rather than tied to distinct sub-populations. As an alternative to a physical explanation, it is also possible that the bimodal appearance is simply a case of statistical noise: after running 100 bootstrap resamplings of the core redshift distribution, we find that bimodality occurs in less than 10\% of the derived histograms, with the remaining samples appearing either flat or centrally peaked.

Moving to the cluster outskirts, the velocity dispersion is slightly lower ($\sigma_{\rm outskirts}$ = 1385$\pm$98 km s$^{-1}$), but still consistent with the combined sample. Unlike in the core, the redshift distribution is singly peaked, with a centre that is closely aligned with the systemic value. Conversely, the spatial distribution is qualitatively less uniform than the core, with the previously-mentioned galaxy overdensities in the north and south, and void-like regions in the northwest and east. The density of cluster members also decreases at large cluster-centric radii; while cluster members greatly outnumber all other galaxy populations in the core (accounting of 45\% of the spectroscopic sample), a combination of decreased number density and increased total volume sees them drop slightly below the background galaxy population in the outskirts (only 28\% of the population).

In addition, Fig.\,\ref{fig:cluster} also reveals two small overdensities along the line of sight, with one just in front ($z = 0.33$) and one just behind ($z = 0.42$) the main cluster. While these concentrations do not currently reside within the cluster halo (based on redshift they are offset $\sim$ 11,000 km/s from the A370 Hubble-flow velocity) they may be infalling, and are likely enhancing the lensing cross-section of the cluster \citep[e.g.,][]{bay14}. It is also possible that they could mark the presence of extended structure such as a large-scale filament along the line of sight, however, more study will be needed to investigate this possibility.

\subsubsection{Role as a mass probe}
\label{subsubsec:mass}
Although the velocity dispersion measurements in the core and outskirts regions are not significantly different, the measured value in the core is nonetheless slightly broader. This is expected, as the galaxies in the core region are on average closer to the (3D) cluster centre than those in the outskirts.  As a result, these galaxies tend to live in deeper parts of the potential well and have more diverse peculiar velocities, leading to a wider redshift histogram.  Variations in histogram width, a measure of the line-of-sight velocity dispersion (LOSVD), therefore indicate changes in potential-well depth.  In principle, by tracking the variations as a function of cluster radius we can map the slope of the potential and estimate the cluster mass profile independently from the lens model. However, using redshifts alone for this purpose can be problematic, because they do not fully account for galaxy position. Since each galaxy is observed along a slightly different line of sight, failure to account for this can bias the LOSVD and subsequently the measured signal. 

To improve the cluster-member-derived mass estimate by better incorporating galaxy position, we turn to a more robust approach: the ``caustic method'' first outlined in \citet{dia97}. Full details of the procedure can be found there and in follow-up work \citep[e.g.,][]{dia99,ser11,gif13a,gif13b}, but briefly, the technique relates the (unknown) 3D velocity $v_{\rm 3D}$ of galaxies to the observed line-of-sight velocity $v_{\rm los}$ through an anisotropy parameter ($\beta$) that accounts for motion in the plane of the sky. By assuming the maximum $v_{\rm 3D}$ at a given radius $r$ is a proxy for the escape velocity $v_{\rm esc}(r)$ (on the basis that galaxies with $v > v_{\rm esc}$ are unlikely to be found within cluster halos), the method further relates $v_{\rm los}$ to the cluster potential $\Phi$. The relationship is given by
\begin{equation}
     v_{\rm esc}^2(r_p) = \left\langle v_{\rm 3D}^2 \right\rangle(r_p) = \langle v_{\rm los}^2 \rangle(r_p) ~ \frac{3 - 2\beta(r_p)}{1 - \beta(r_p)} = -2\Phi(r_p)
     \label{eqn:caustPot}
\end{equation}
where $r_p$ is the 2D-projected radius on the sky, and $\langle v \rangle (r_p)$ is the rms of velocities for galaxies at $r_p$. Like lensing, the caustic technique is expected to be insensitive to the dynamic state of the cluster, thus mass values determined by this method should be unbiased whether the cluster is in a relaxed or merging state \citep[e.g.,][]{gel13}. While recent work suggests that this may not strictly be true -- \citet{mon22} find that caustic simulated cluster mergers approaching pericentric passage (as is the case with A370) have larger caustic mass estimates compared to their true values -- at the total mass level of A370 ($\sim$ 10$^{15} M_\odot$) this systematic effect is expected to be $<$5\%, which is expected to be at the current level of measurement uncertainty.

While in principle, caustic method velocities can be measured in any reference frame, converting them to the cluster rest frame and recasting them as peculiar velocities ($v_p$) relative to the bulk motion is the most practical: by doing so, $\langle v_{p, \rm los} \rangle (r_p)$ is equivalent to the amplitude of the LOSVD, $\mathcal{A}^2(r_p)$. Following the notation of \citet{dia97}, we can simplify equation \ref{eqn:caustPot} to:
\begin{equation}
    \mathcal{A}^2(r_p) ~ g(\beta(r_p))  = -2\Phi(r_p)
    \label{eqn:caustDG}
\end{equation}
where
\begin{equation}
    g(\beta(r_p)) \equiv \frac{3 - 2\beta(r_p)}{1 - \beta(r_p)} 
    \label{eqn:anisotropy}
\end{equation}
is the projection-corrected radial anisotropy term. 

With Equation \ref{eqn:caustDG} relating the observable LOSVD to $\Phi$, we can then use the Poisson equation, $\nabla^2\Phi = 4\pi G\rho(r)$, to convert this into a mass estimate. However, rather than using the Poisson equation directly, \citet{dia97} instead apply a slightly modified differential, $dm = 4\pi\rho(r)r^2 dr$, which is less sensitive to sampling noise in the $\mathcal{A}^2(r_p)$ profile. Using Equation \ref{eqn:caustDG} this differential becomes
\begin{equation}
    dm = -2\pi \mathcal{A}^2(r_p) ~ g(\beta(r_p)) ~ \frac{\rho(r_p)r_p^2}{\Phi(r_p)} d_{r_p}
\end{equation}
and integrating the relationship we arrive at 

\begin{equation}
    GM(< R) = \int_0^R -2G \pi ~ \mathcal{A}^2(r_p) ~ g(\beta(r_p)) ~ \frac{\rho(r_p)r_p^2}{\Phi(r_p)} d_{r_p}
\end{equation}
or
\begin{equation}
    GM(< R) = \int_0^R \mathcal{F}_\beta(r_p) ~ \mathcal{A}^2(r_p) ~ d_{r_p}
\end{equation}
with
\begin{equation}
    \mathcal{F}_\beta(r_p) \equiv -2G \pi ~ g(\beta(r_p)) ~ \frac{\rho(r_p)r_p^2}{\Phi(r_p)} 
\end{equation}
Based on detailed numerical simulations, several independent efforts have demonstrated that the value of $\mathcal{F}_\beta(r_p)$ is nearly constant with radius in clusters, with $0.5 < \mathcal{F}_\beta < 0.7$ \citep[e.g.,][]{ser11,gif13b,gel13}, making the mass estimate largely a function of a single variable, $\mathcal{A}^2(r_p)$. 
However, \citet{ser11} note that this assumption begins to break down at low radii ($r < 0.3\times r_{200}$; the region probed in this work), causing the caustic technique to overestimate the enclosed mass. To avoid this, \citet{gif13a} re-formulate the caustic method, accounting for the bias by assuming a specific mass profile, namely a Navarro, Frenk, and White (NFW; \citealt{nfw97}) distribution. In doing so, $\mathcal{F}_\beta(r_p)$ can be rewritten in terms of NFW parameters:
\begin{equation}
    \mathcal{F}_\beta(r_p) = \frac{(r_p/r_s)^2}{(1+r_p/r_s)^2~\ln{(1+r_p/r_s)}}
    \label{eqn:massProf}
\end{equation}
where $r_s$ is the NFW scale radius. These parameters can then be fit to the observed cluster geometry, generating more robust mass and shape estimates that can be directly compared to previous results. 

The combination of MUSE spectroscopy and \emph{HST} positions provide a full 3D account of the location of individual objects at high quality, enabling us to apply the caustic method to our data set.  We note that the technique was originally designed to work with spectral data extending well beyond the cluster virial radius and the characteristic radius $r_{200}$, both of which can span physical distances of several Mpc.  Although our spectroscopic coverage of A370 does not extend as far as that, spanning only $\sim$ 900 kpc in physical space compared to an estimated virial radius of $\sim$2.3 Mpc \citep[e.g.,][]{app14,lee20}, it has the advantage of sampling the cluster at a much higher density. As a result, while other studies rely on adaptive group-finding algorithms to estimate the cluster centre and reject line-of-sight outliers, these parameters are directly observable in our data.  In particular, Fig.\,\ref{fig:cluster} shows a clear gap between the cluster and its near neighbors in redshift space (recall $\overline{z} = 0.374$), while the projected spatial distribution of cluster members is centred roughly halfway between the two BCGs ($\overline{\alpha}$ = 39.9704672, $\overline{\delta}$ = -1.5779282).  Taken together, we use the coordinate set ($\overline{\alpha},\overline{\delta}, \overline{z}$) as the cluster centre point (which we note is in excellent agreement with the centre of the L17 and L19 lensing mass distributions) to create the ($r_p-v_p$) plane. 

\begin{figure}
    \centering
    \includegraphics[width=0.49\textwidth]{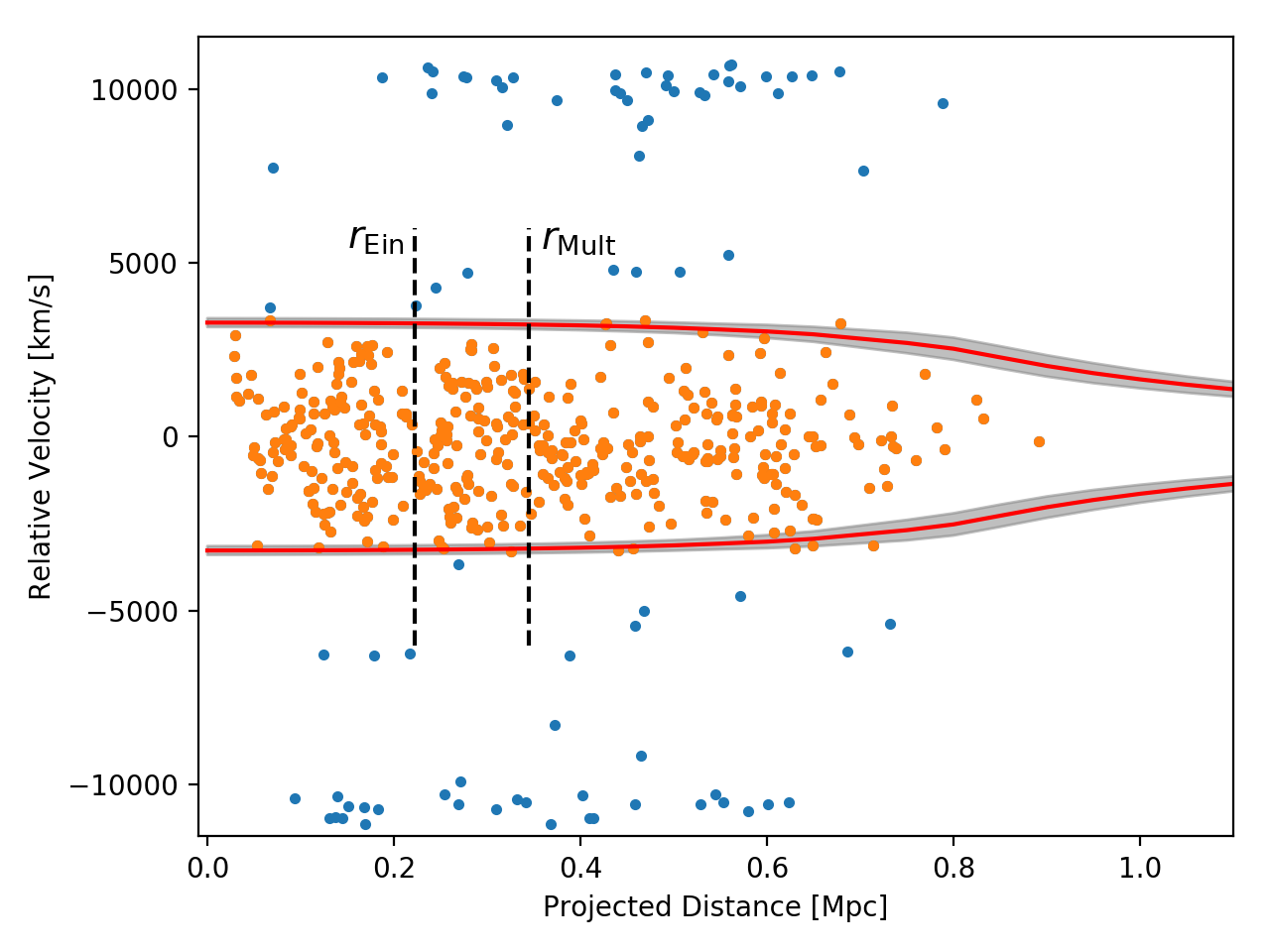}
    \includegraphics[width=0.49\textwidth]{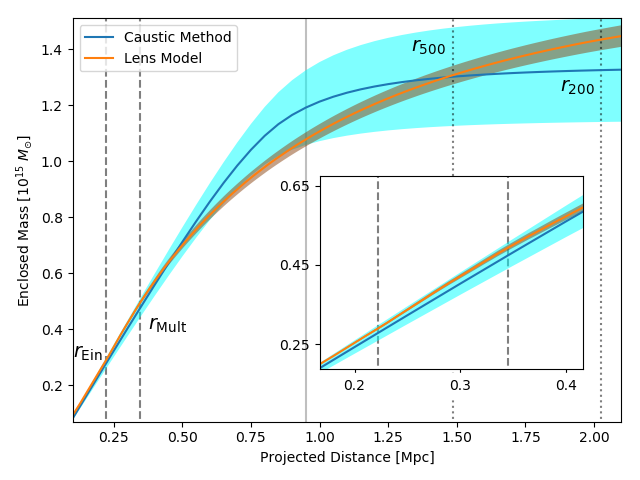}
    \caption{Top: Redshift space diagram for A370 showing line-of-sight velocity as a function of projected radius. Axes are defined relative to the adopted cluster centre ($\alpha$ = 39.9704672, $\delta$ = -1.5779282, $z$ = 0.374) in order to obtain Caustic Method \citep{dia97} estimates of cluster mass. Galaxies within the cluster halo appear as orange points, while external galaxies and interlopers (those with peculiar velocities greater than the cluster escape velocity) are shown in blue. The best-fit velocity caustics appear as red lines, with 1-$\sigma$ uncertainty regions in grey.  For reference, the average cluster Einstein radius for high-redshift ($z>3$) galaxies ($r_{\rm Ein}$) and the radius encompassing the multiple-image region over all redshifts ($r_{\rm Mult}$) appear as dashed black lines. Bottom: Radial mass profiles for A370, determined from both lensing and dynamical methods.  In spite of each technique using different apertures, the two curves broadly agree with one another. $r_{\rm Ein}$ and $r_{\rm Mult}$ are again shown as dashed lines. Two additional radii derived from the caustic fit, $r_{500}$ and $r_{200}$, appear as dotted lines. The solid grey line at $\sim$ 900 kpc marks the edge of the MUSE footprint, highlighting the extent of the caustic fit data.} 
    \label{fig:caustics}
\end{figure}

The caustic method phase diagram, including all confirmed cluster members, can be seen in the upper panel of Fig.\,\ref{fig:caustics}.
From the figure, it is clear our centre is well-chosen, with cluster members distributed symmetrically around $v_p = 0$ and showing the characteristic tapered, trumpet-shaped distribution predicted by theory; a poorly-chosen cluster centre would instead have an asymmetric velocity distribution with a warped trumpet-shape. We tested this explicitly by trialling alternative centroids, including points centred on each of the BCGs. While these alternative points did not change the shape of the distribution considerably, slight asymmetries were still present, giving us more confidence in our initial choice. After populating the plane, we calculate the caustic boundary (red lines in Fig.\, \ref{fig:caustics}) using an algorithm first presented in \citet{gif13b}. This procedure is largely based on the \citet{dia99} method, though it makes some modifications, including a slight simplification of the kernel used to smooth the data, which is less computationally intensive and better matched to our already well-sampled data. For reference, the dividing line between the A370 core region and the outskirts ($r_{\rm Mult}$) and the average cluster Einstein radius ($r_{\rm Ein}$) for high-redshift galaxies ($z > 3$) appear as dashed vertical lines. As expected, the caustic does indeed have a larger amplitude in the core region relative to the outskirts. 

Following Equation \ref{eqn:massProf}, we parameterize the caustic fit with an NFW profile to better compare the results to previous mass estimates. By doing so we derive characteristic values for the cluster, including $r_{200}$, the more compact $r_{500}$, and their associated enclosed masses ($M_{200}$ and $M_{500}$ respectively), finding values of $r_{500}$ = 1.49$\pm$0.06 Mpc,  $M_{500}$ = (1.31$\pm$0.15)$\times 10^{15}$ $M_\odot$, $r_{200}$ = 2.03$\pm$0.08 Mpc, and $M_{200}$ = (1.34$\pm$0.18)$\times 10^{15}$ $M_\odot$. Our measurement of $M_{500}$ agrees well with previous work, falling within 1$\sigma$ of the published values in \citet{mad08},  (1.33$\pm$0.34)$\times 10^{15}$, and \citet{ume11}, (1.32$\pm$0.16) $\times 10^{15}$. Our $M_{200}$ value is nominally lower than the measurement in \citet{ume11}: (2.21$\pm$0.27) $\times 10^{15}$, but still consistent within 2$\sigma$ (\citealt{mad08} does not present an $M_{200}$ value). 
In addition, we plot the enclosed mass as a function of radius in the lower panel of Fig.\,\ref{fig:caustics}, and for comparison we also show the integrated mass profile derived from the L19 lens model. Although the NFW profile has a different functional form than the lensing-based model, which is created from the combined sum of several pseudo-isothermal elliptical mass distributions (PIEMD; \citealt{eli07}), the two profiles trace each other well, with an almost exact agreement at $r_{500}$ and statistically consistent measurements at other radii.  We caution, however, that $r_{500}$ is currently beyond the model-constrained regions of both profiles ($<$ 350 kpc for lensing, $<$ 900 kpc for the caustic method); a larger, more-extended data set would be needed to consider the the result truly robust. Nevertheless, by comparing general trends between the profiles we can still extract additional information about large-scale mass distribution. This is because the caustic profile is constructed from spherical mass shells, while the lensing profile is instead constructed in cylindrical apertures; discrepancies between the two therefore constrain the mass distribution along the line of sight. In the case of A370, we see that, within the well-constrained core region (Fig.\,\ref{fig:caustics}, inset), the best-fit lensing mass profile (orange line) is slightly above the caustic mass profile (blue line), suggesting that there may be an extended mass distribution perpendicular to the plane of the sky. We note however, that both profiles are still fully consistent with their 1-$\sigma$ errors (brown and cyan shaded regions) at these distances. Therefore, we do not draw strong conclusions from the result at this time, but this interesting possibility could be studied further in future work.

\subsubsection{Physical properties}
\label{subsubsec:properties}
By measuring specific characteristics of cluster members, such as broad-band colours and spectral line strengths, we can derive information about their physical properties, including star-formation history and gas content. Combining this information with 3D position, we can map the distribution of these quantities and note how they change, providing a window into galaxy evolution as a function of local environment throughout the cluster. Galaxy colour is particularly useful for this purpose, since it not only acts as a proxy for stellar activity, but can also reveal new information about cluster substructure. Specifically, a smoothed map of cluster member light can reveal the location of potential substructure candidates, many of which lie at considerable distances from the cluster core (see, e.g., Fig.\,11 in L19). However, these estimates are often made with sparsely-populated galaxy samples using colour-cuts based on the cluster red-sequence (as defined in the central core). This may bias the final result, especially since cluster members are expected to be bluer at larger radii \citep[e.g.,][]{dre80,gil08}, due to increased star formation and a higher gas fraction. Accounting for this behaviour can therefore refine the cluster member selection, leading to a more accurate light map at large distances.

\begin{figure*}
    \centering
    \includegraphics[width=0.49\textwidth]{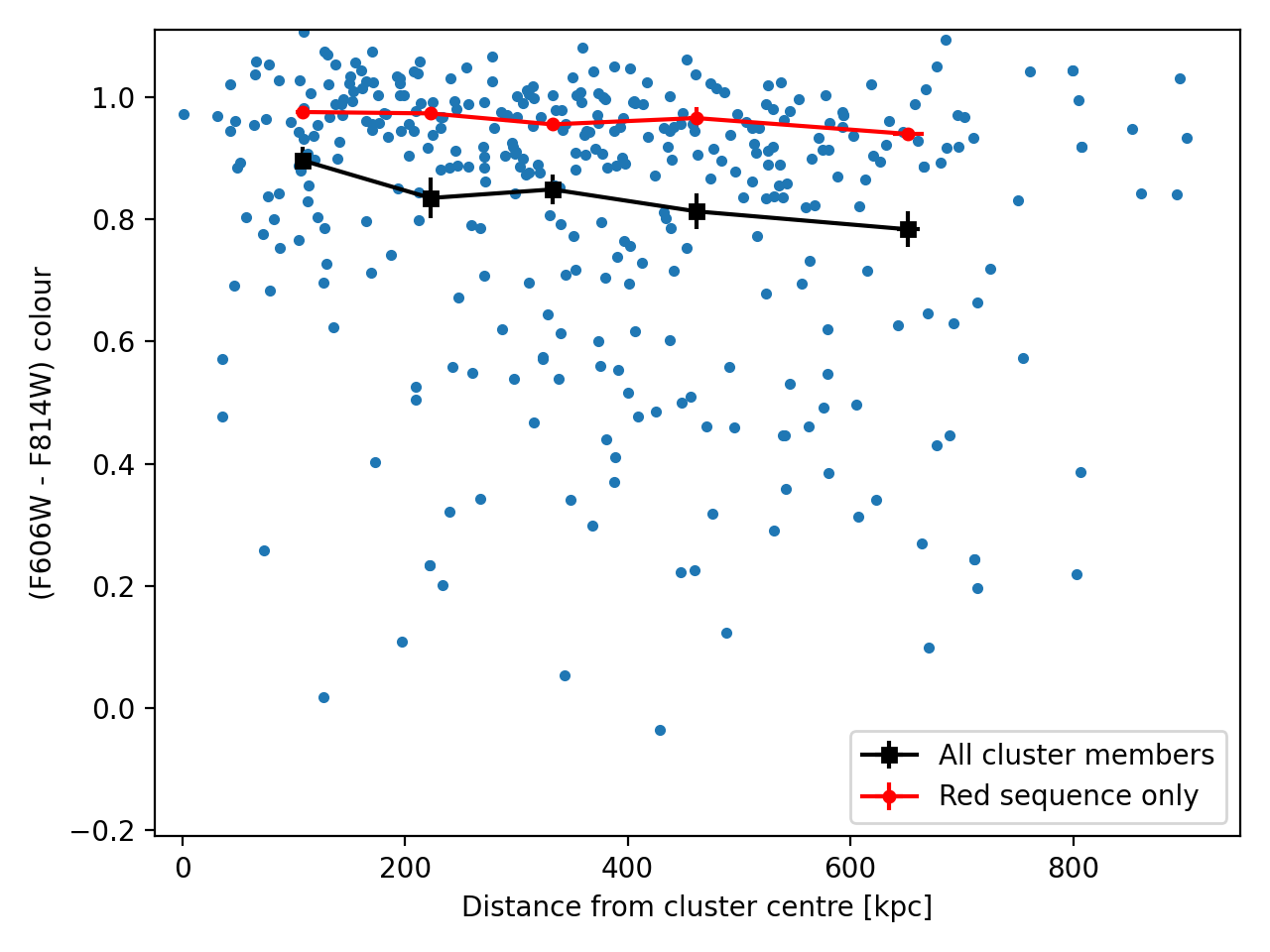}
    \includegraphics[width=0.49\textwidth]{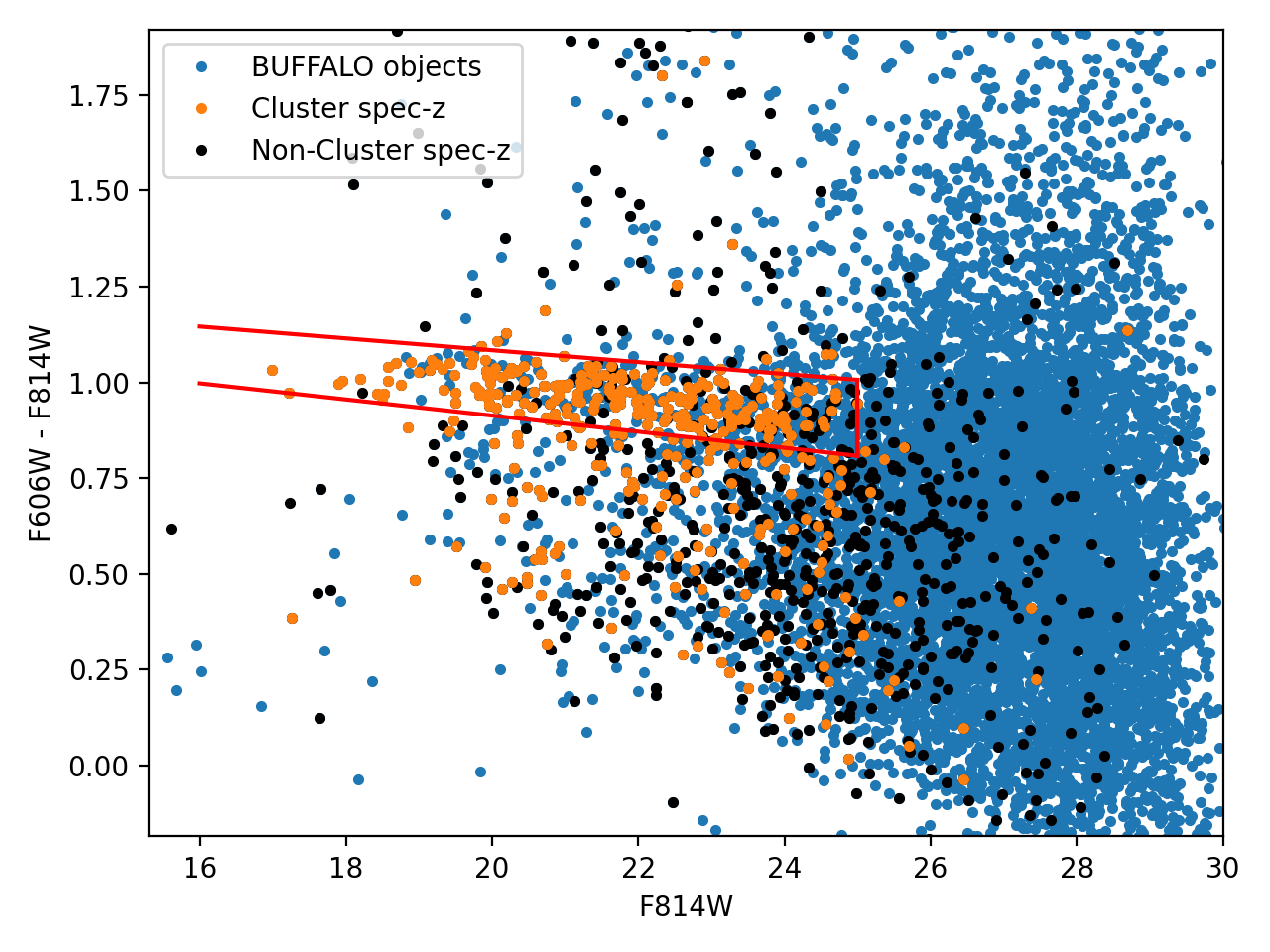}
    \caption{Left: Optical (F606W - F814W) colours of spectroscopically-confirmed cluster members as a function of projected cluster-centric radius. Individual galaxies are shown as blue dots, while binned averages of all cluster members (black squares) and red-sequence objects alone (red diamonds) are overlaid. Both samples trend to bluer colours at larger radii, but the red-sequence-only subset has a shallower slope. Right: Colour-magnitude diagram of the A370 BUFFALO field, constructed from the \citet[][]{pag21} photometric catalogue (blue points). The red sequence region is highlighted by red lines, and contains a majority of spectroscopically-confirmed cluster members (orange points). As galaxies become fainter, the red sequence itself shifts towards bluer colours. }
    \label{fig:radColor}
\end{figure*}

To characterize colour in the A370 field, we collect photometric information for all spectroscopically confirmed cluster members. We ensure the most accurate colour information by matching our galaxies to the catalogue of calibrated, isophotal magnitudes presented in \citet{pag21}, which are also derived from BUFFALO data. For this exercise we explicitly limit our focus to optical bands (F606W and F814W) both because the optical (ACS) BUFFALO frames cover the entire spectroscopic footprint and because these two bands bracket the D4000\AA\ break, providing the largest contrast in colour changes and hence the most inclusive red sequence. Plotting (F606W-F814W) colour as a function of cluster radius (Fig.\,\ref{fig:radColor}, left), it is apparent that the bulk of cluster member colours do shift to bluer values with increasing cluster-centric distance, though the large scatter in colour space makes the effect difficult to quantify. To reduce the scatter, we combine the data into a series of radial bins extending from the cluster centre, placing a roughly equal number of galaxies into each bin to maintain an equivalent sample size at each point. We then measure the average galaxy colour of each bin (black line), which shows a steady blue-ward trend with increasing radius. 

However, by simply averaging the colour of all cluster members in a given bin we are, in all probability, overestimating the effect: along with an intrinsic shift to bluer colours, the fraction of galaxies in the blue-cloud likely increases at larger radii, skewing the trend.  To investigate this possibility, we search for the cluster red sequence by constructing a colour-magnitude diagram from all objects in the \citet{pag21} catalogue (Fig.\,\ref{fig:radColor}, right). Thanks to the large size of the catalogue, we can identify the red sequence using photometric data alone (blue points), but to refine the overall shape we over-plot the spectroscopic cluster sample (orange points). As expected, a majority of secure cluster members lie in the red-sequence locus, though we find a non-negligible fraction of galaxies in either the blue cloud or the region redder than the red sequence. As a further check, we also plot all objects in the spectroscopic catalogue that do \emph{not} fall in the cluster redshift range (black points), which we find largely avoid the red sequence region. Using all available data, we define magnitude and colour cuts to select red sequence galaxies (red lines), and using only the spectroscopic cluster members that fall in this defined region, we estimate the average red sequence galaxy colour in the same radial bins as before. Compared to the measurement of all galaxies, the average red-sequence-only colour is redder by 0.1-0.2 magnitudes and has a shallower fitted slope ($\Delta_{\rm colour, red}$ = -0.067 mag/Mpc; $\Delta_{\rm colour, all}$ -0.207 mag/Mpc). This implies that the increasing blue fraction is a much stronger driver of the radial colour change than the intrinsic galaxy colour shift, at least over the radii we probe here. This agrees well with \citet{con19}, who find that the colour slope of the (F625W-F814W) red sequence of CLASH clusters \citep{pos12} at z$\sim$0.35 is consistent with zero to at least $\sim$1 Mpc.

\begin{figure*}
    \centering
    \includegraphics[width=0.49\textwidth]{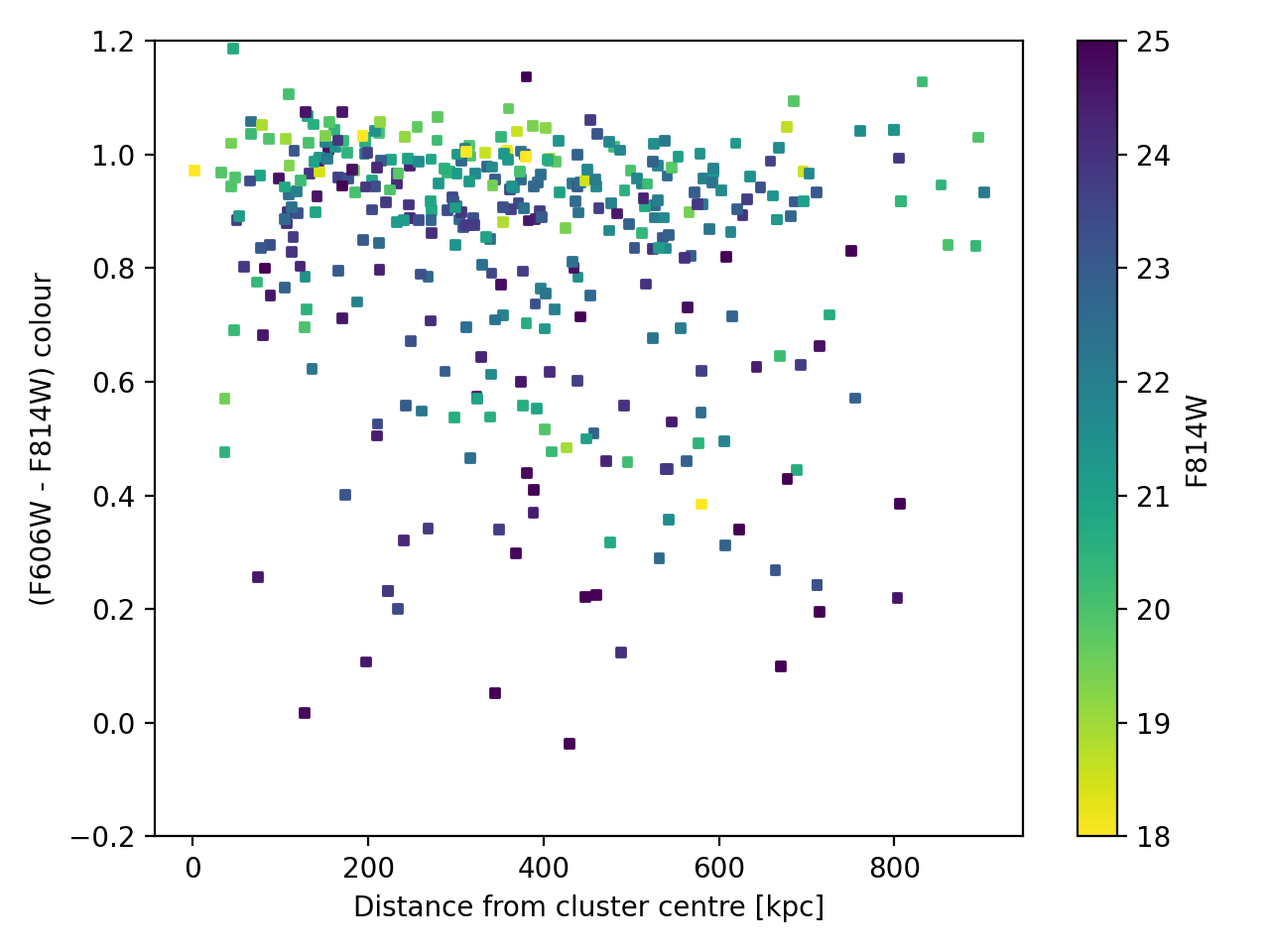}
    \includegraphics[width=0.49\textwidth]{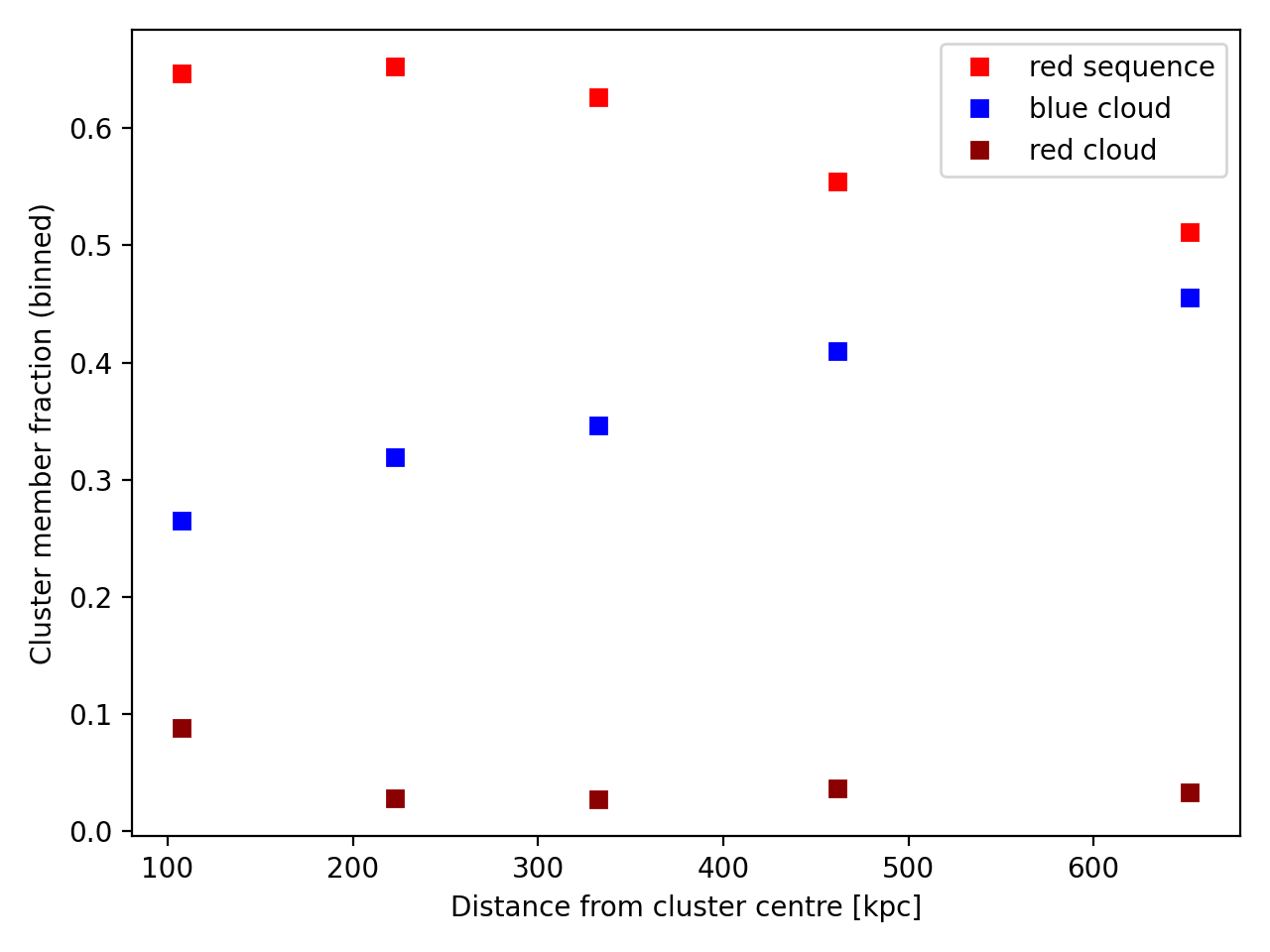}
    \caption{Left: Colour-radius plot from Fig.\,\ref{fig:radColor}, but with galaxies shaded according to their F814W magnitudes. Brighter galaxies tend to be redder (in agreement with the colour magnitude diagram) and lie closer to the cluster centre, suggesting the observed radial trend of cluster member colours is at least partially due to galaxy luminosity. Right: Relative fraction of cluster members in the red sequence and blue cloud as a function of radius; red cloud galaxies (objects redder than the red sequence) also exist, but are rare throughout the cluster. While red sequence galaxies dominate the core, there is a steady and significant increase in blue cloud fraction at larger radii. An increasing blue fraction can also contribute to the observed colour trend.}
    \label{fig:magFrac}
\end{figure*}

\citet{con19} additionally find that red sequence deviations within 1 Mpc are largely tied to total magnitude, with fainter galaxies appearing bluer than brighter ones. Testing this in A370, we recreate the colour-radius plot of Fig.\,\ref{fig:radColor}, but this time we scale each galaxy by its F814W magnitude (Fig.\,\ref{fig:magFrac}, left). From the plot, we see that the red sequence galaxies do generally appear brighter than those in the blue cloud, and (focusing specifically in the red sequence region) cluster members in the outskirts are qualitatively fainter than those in the core. Fitting a trend line to the red sequence galaxies shows this to be quantitatively true as well, with $\overline{m}_{\rm F814W} = 21.69 \pm 0.26$ at $\overline{r} = 100$ kpc, and $\overline{m}_{\rm F814W} = 21.85 \pm 0.18$ at $\overline{r} = 650$ kpc, where $\overline{m}_{\rm F814W}$ and $\overline{r}$ are the average galaxy magnitude and cluster-centric radius of a given bin.  Considering the uncertainties, however, the slope is fully consistent with zero, which further strengthens the idea that blue cloud fraction, rather than red sequence evolution, is responsible for any colour changes found over the data footprint.  To underscore this point, we explicitly plot the relative fraction of red sequence, blue cloud, and redder-than-red-sequence (or red cloud) galaxies as a function of radius (using the same bins as in all previous exercises) in the right-hand panel of Fig.\,\ref{fig:magFrac}. The plot clearly shows a steadily increasing blue cloud population with radius, and while red sequence galaxies form the largest subgroup in all bins, the red-sequence/blue-cloud ratio approaches 1/1 by the edge of the MUSE footprint (with red cloud galaxies remaining very uncommon at all radii). Identifying additional blue-cloud cluster members (that would likely be missed by traditional spectroscopic surveys) has an impact on the caustic method mass estimate as well, especially given that many of these galaxies appear at larger physical radii. Re-running the caustic technique using only the red-sequence galaxies, we find mass values that are consistently lower than the full cluster sample.  In some cases this difference can be considerate:\ the red-sequence $M_{500}$ measurement is 25\% lower than the full-sample, though we remind the reader that the radius enclosing $M_{500}$ is well outside of the MUSE footprint. Inside the regions constrained by the the caustic data ($<$900 kpc) the difference is a more moderate  (12\%, less than the 1-$\sigma$ uncertainty envelope), but the result still highlights the importance of a comprehensive cluster-member census.

In addition to colour information, we also use the spectroscopic data to explore trends in stellar activity in cluster members. As a first step, we separate the spectroscopic cluster member sample into broad ``star-forming'' and ``non-star-forming'' categories, using a simplistic assumption that star-forming galaxies have nebular emission lines ([\oii], [\oiii], etc.), while non-star-forming galaxies do not. This assumption slightly overestimates the star-forming fraction by including emission from active galactic nuclei (AGN), though we note that AGN fraction is extremely low in clusters -- between 0.2\% and 2\% of member galaxies at $z\sim 0.4$ \citep[e.g.,][]{buf17,mis20}, suggesting a negligible contamination rate. After identifying star-forming galaxies, we examine the remaining non-star-forming galaxies and further subdivide them into ``passive'' and ``post-starburst'' categories using stellar template matching. Specifically, we fit each galaxy's spectrum with the MILES library \citep{san06,fal11} and identify the spectral type of the best matching template. Galaxies that are best-fit with younger stars (i.e., A-type templates) are classified as post-starburst, with all others being classified as passive. Unsurprisingly, the vast majority of cluster members fall into the passive category (315 galaxies), followed by star-forming (88 galaxies), and finally post-starburst (13 galaxies).

\begin{figure*}
    \centering
    \includegraphics[width=0.49\textwidth]{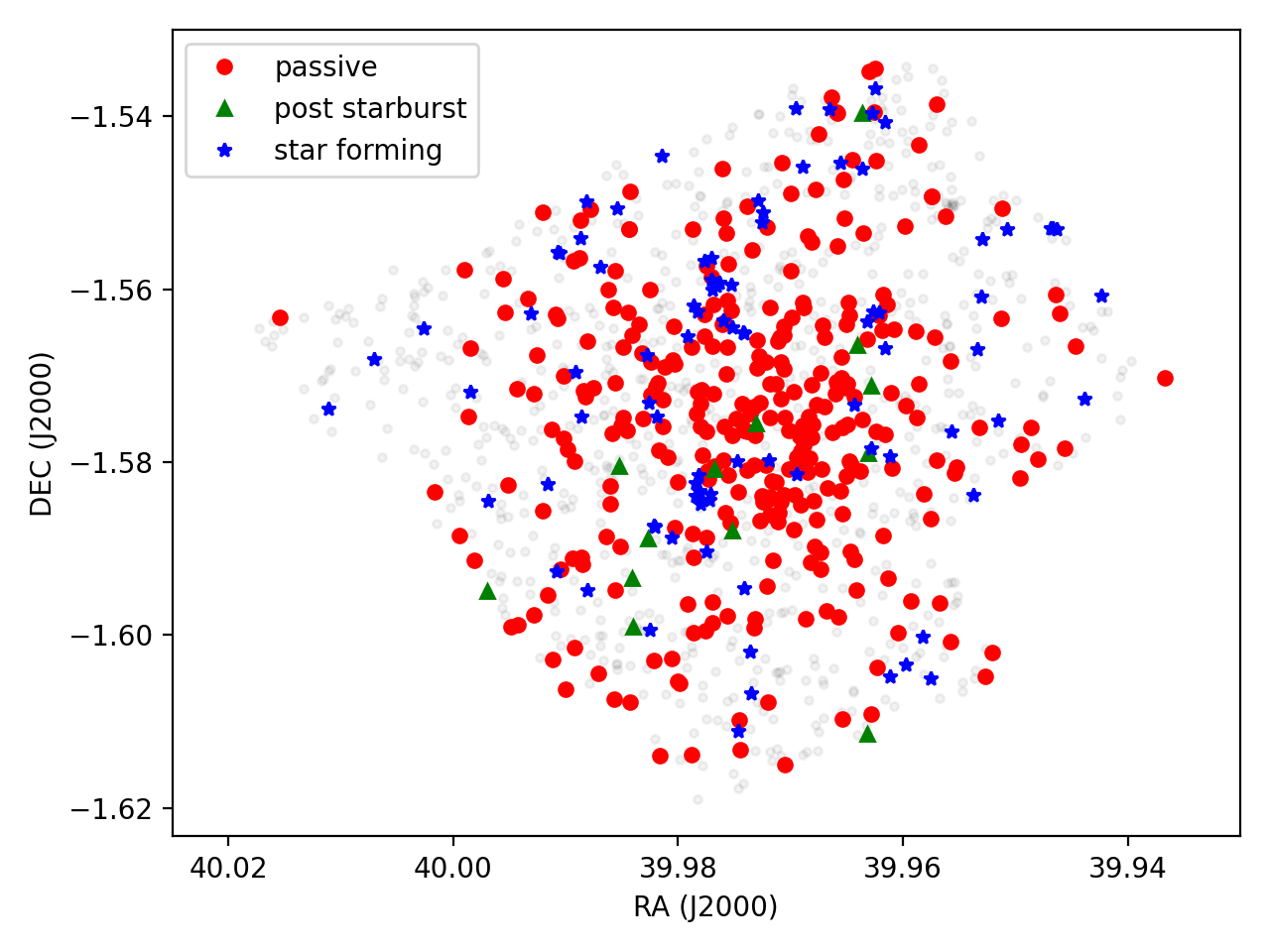}
    \includegraphics[width=0.49\textwidth]{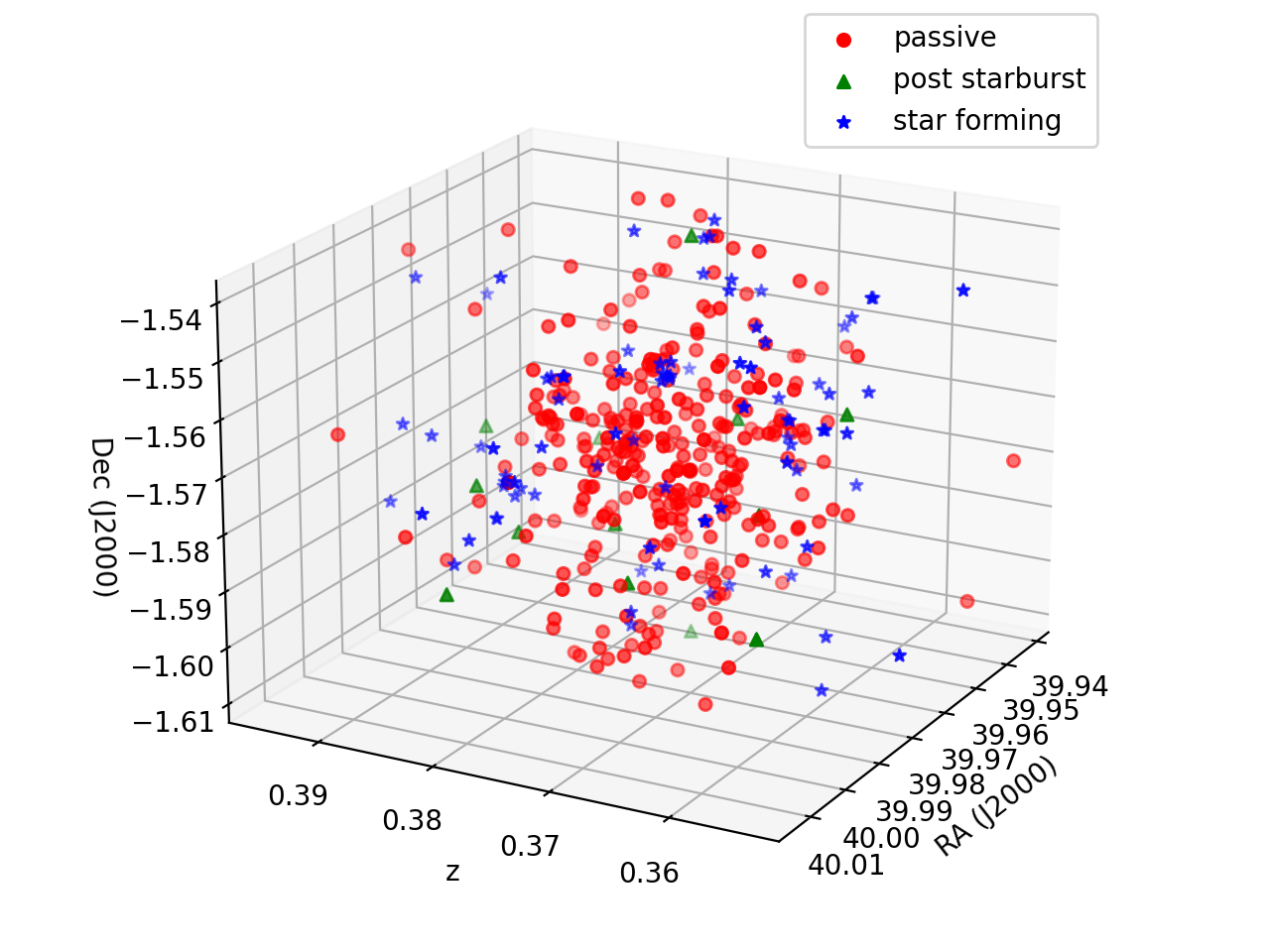}
    \includegraphics[width=0.49\textwidth]{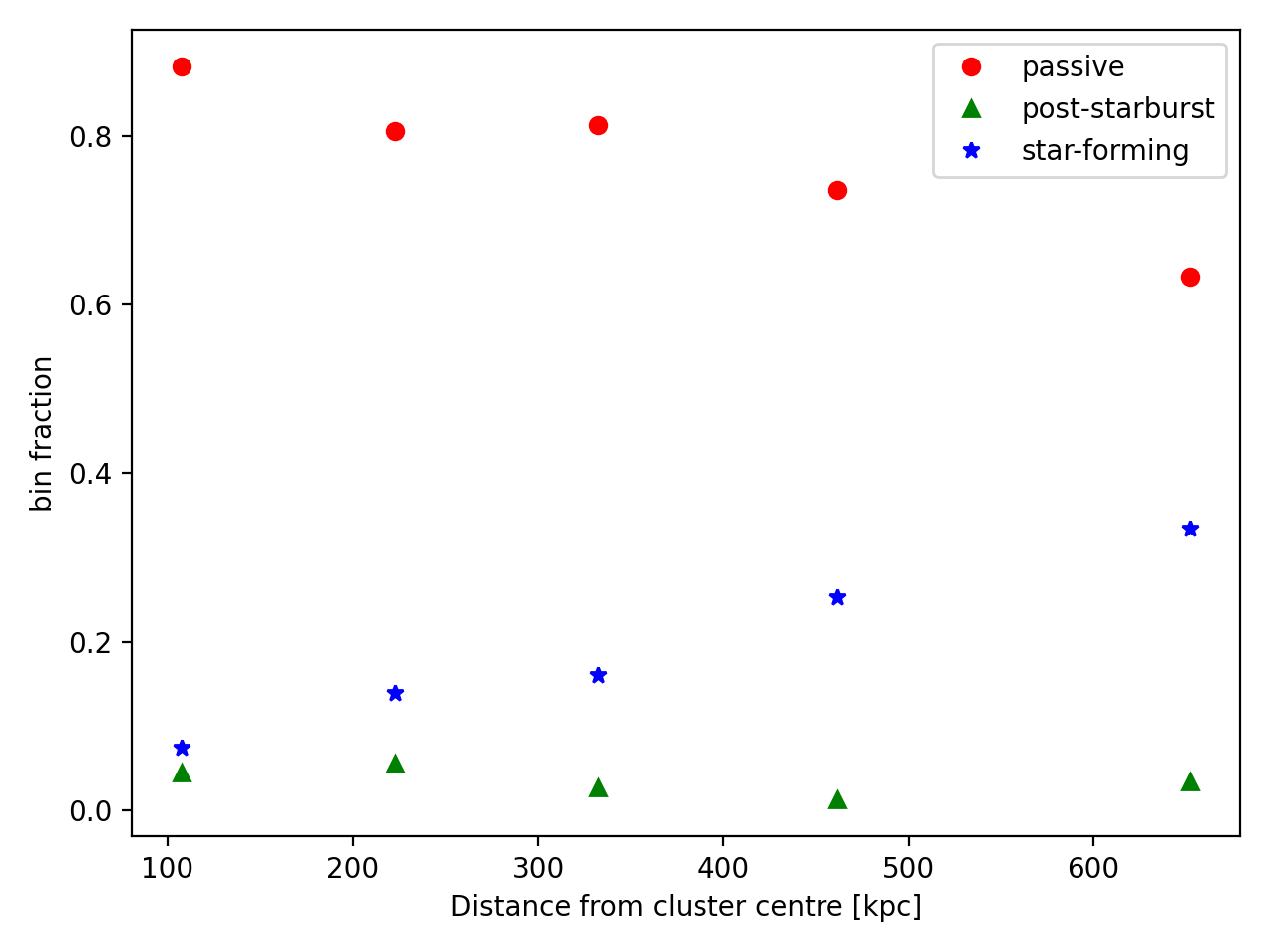}
    \includegraphics[width=0.49\textwidth]{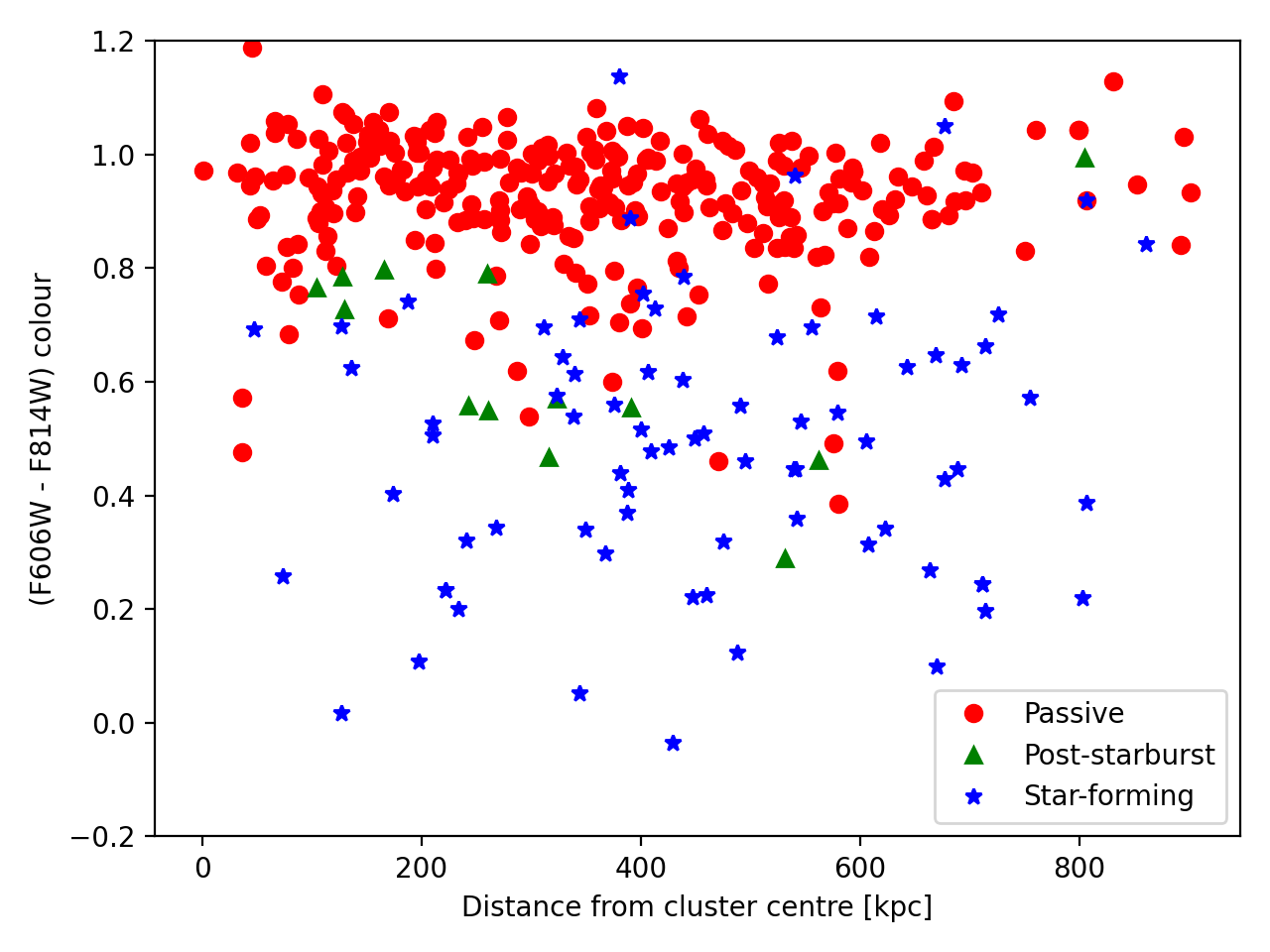}
    \caption{Stellar activity of spectroscopic cluster members, divided into passive, post-starburst, and star-forming categories (section \ref{subsubsec:properties}. Top panels: distribution of galaxies in the plane of the sky (left panel) and deprojected into redshift/velocity space (right panel). In both cases passive galaxies are strongly concentrated close to the cluster centre, while post-starburst and star-forming galaxies are more diffuse. Bottom left: Relative fraction of each galaxy type as a function of radius. Much like the version of this plot showing broadband colour trends (Fig.\,\ref{fig:magFrac}) we see that the fraction of passive (red) galaxies decreases with radius, while star-forming (blue) galaxies become more prominent. Bottom right: Colour-radius plot of Fig.\,\ref{fig:radColor}, but now galaxies are coded by stellar activity. Passive and star-forming galaxies are well-separated in colour space but have largely flat colour gradients with radius. Conversely, post-starburst galaxies show significant colour evolution, with galaxies near the core up to 0.4 magnitudes redder than those in the outskirts.}
    \label{fig:starform}
\end{figure*}

Once cluster members are classified, we investigate their behaviour throughout the data footprint, presenting the results in Fig.\,\ref{fig:starform}. In the plane of the sky (top left panel) we see that passive galaxies dominate the central regions of the cluster, with the first significant population of star-forming objects only appearing at the boundary between the core and outskirts regions. Within the outskirts, star-forming galaxies -- and to a lesser extent post-starburst galaxies -- become more common, though they are still sub-dominant to the passive population.  This distribution holds in 3D space as well (top right panel), with star-forming galaxies primarily located beyond 350 kpc. Spatially, there does not appear to be any significant substructure within any of the three galaxy populations, though we do identify some minor clustering of star-forming galaxies near the positions of known ``jellyfish'' galaxies \citep{ebe14} experiencing significant ram pressure stripping by the intra-cluster medium, suggesting that local turbulence may be enhancing star-formation activities in these regions. Though outside the scope of this paper, we note that these overdensities appear as narrow trails behind the jellyfish galaxies in 3D space; future work could therefore use this information to probe the complex kinematics of ram pressure stripping events within the cluster. 

Looking more explicitly at radially-averaged distributions (bottom left panel), we see that the passive galaxy fraction decreases as a function of cluster-centric radius, while star-forming galaxies increase. This is similar to the general trend of galaxy colour, which, given the close ties between colour and star-formation is not surprising. The relative increase of star-forming galaxies to passive galaxies is nominally greater than that of the blue cloud to the red sequence, with star-forming galaxies increasing by 26\% over the data footprint compared to a 19\% increase of the blue cloud. Given the much larger starting fraction of passive galaxies though, (88\% in the central-most bin vs. a 65\% red sequence population) the population difference is still on the whole larger. Care must be taken with this result, however, since the outskirts data are significantly shallower than the core. Because MUSE is more sensitive to emission line features than absorption features, a shallower exposure time in the outskirts may mean that we are biased against the detection of passive galaxies. Deeper spectroscopic data in the outskirts regions would allow us to investigate this possibility more thoroughly.

Finally, we again reproduce the color-radius plot of Fig.\,\ref{fig:radColor}, but this time we label each point according to its stellar category (bottom right panel). The strong correlation between stellar activity and colour is evident in the plot, with most passive galaxies located within the red sequence and most star-forming galaxies found in the blue cloud. Overall, we see little change in the average colour of both passive and star-forming galaxies as a function of radius, providing further evidence that observed bulk changes in colour are due to an increased fraction of blue galaxies rather than a change in the stellar populations of those galaxies. However, the same cannot be said for the post-starburst galaxies, which do show a strong colour gradient: galaxies located closer to the centre are $\sim$ 0.4 magnitudes redder than those at the edge of the footprint, with a clear linear trend between these regions. Given the relatively small number of identified post-starburst systems, it is difficult to know whether this result is real or simply a statistical coincidence. Nonetheless, it is in intriguing possibility, and considering the transitional nature of post starburst galaxies \citep[e.g.,][]{qui04,fre18}, the gradient could constrain the timescale over which galaxies shift from the blue cloud to the red sequence, which -- once a galaxy enters the post-starburst phase -- is expected to happen within its first radial pass-through of the cluster \citep{cen14a,cen14b}. However, additional study is needed to investigate this possibility properly.

\subsection{Galaxy-galaxy lens populations}
\label{subsec:ggl}
While multiple-image constraints map the enclosed mass within the entire strong-lensing region, cluster members acting as galaxy-galaxy lenses (GGLs) also supply localized mass estimates. Specifically, GGLs yield complementary information to more broadly-separated multiple-image systems by probing mass at both smaller physical scales -- closer to the baryon-dominated regions at the centres of galaxies -- and larger radial distances, extending beyond the boundaries of the cluster Einstein radius. GGLs provide additional data points for measuring the radial slope of the cluster mass density profile \citep{tu08}, while perturbations in observed GGL image patterns (relative to their field-galaxy counterparts) can reveal substructure within the cluster halo \citep{lim10}. At the same time, population counts of cluster GGLs can be used to highlight tensions in the ``standard'' cosmological model (\citealt{men20}; though see caveats in \citealt{bah21} and \citealt{rob21}). Through these applications interest in cluster-based GGLs has considerably increased: \citet{des18} recently presented a large sample of GGL candidates identified in CLASH clusters, and a similar search targeting the BUFFALO fields is actively ongoing.

\subsubsection{Census of GGLs in the A370 field}
\label{subsubsec:census}
In the BUFFALO GGL search, members of the BUFFALO Lens Modeling Working Group first visually inspect colour images of each cluster field and identify potential lens candidates. After compiling a master candidate list, individual Working Group members independently rank each candidate from 1 (a very likely lens) to 4 (probably not a lens). The average of each object's rank serves as its final score, with lower-scoring objects considered more likely to be true GGLs.  For the purpose of this paper, we take objects with a final score $<$ 1.5 to be our probable candidates. In the A370 field, there are five objects (two previously known and three newly-identified) that meet this criterion, which we again stress is based only on visual inspection.

\begin{figure*}
    \centering
    \includegraphics[width=0.3\textwidth]{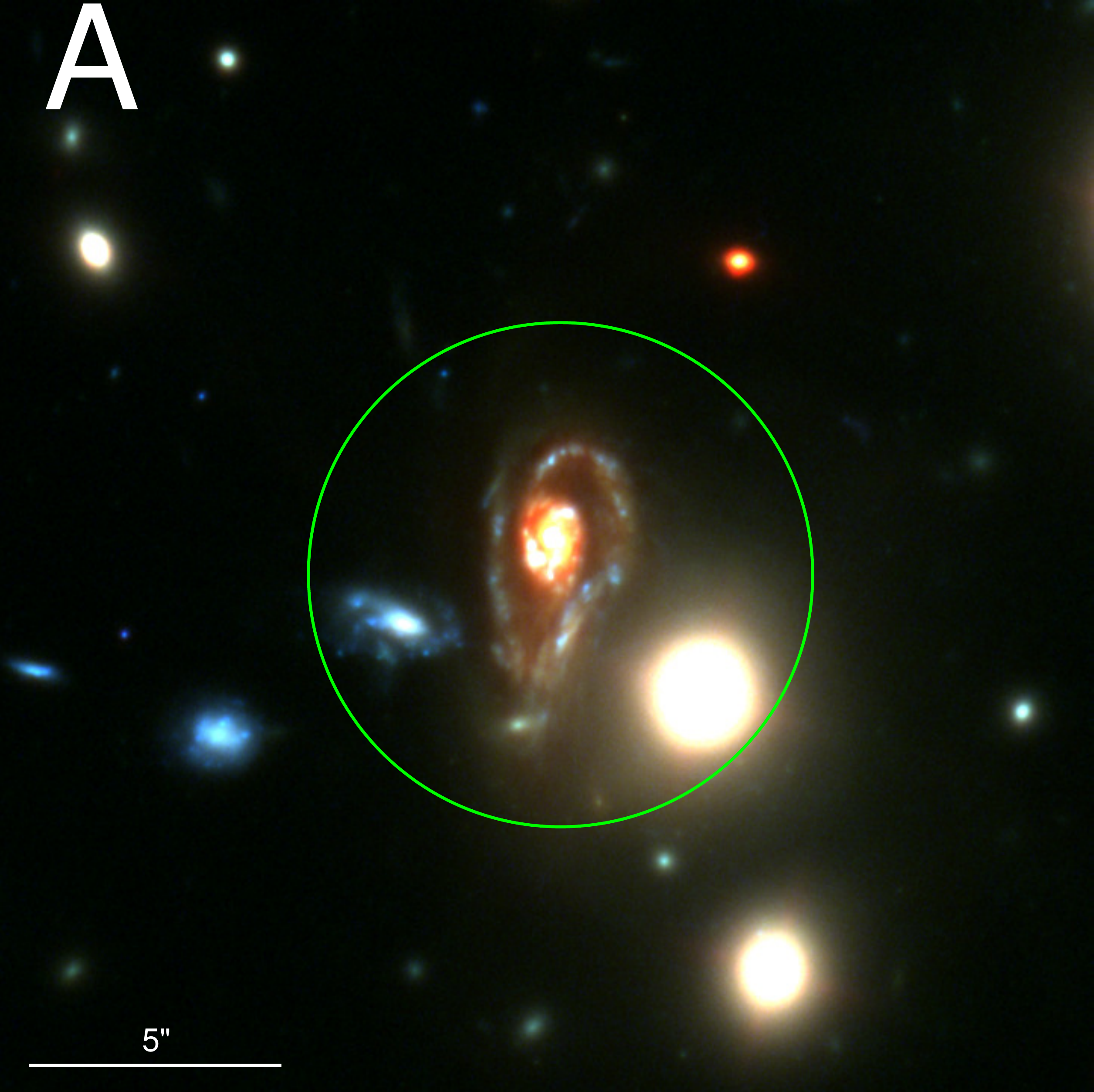}
    \includegraphics[width=0.3\textwidth]{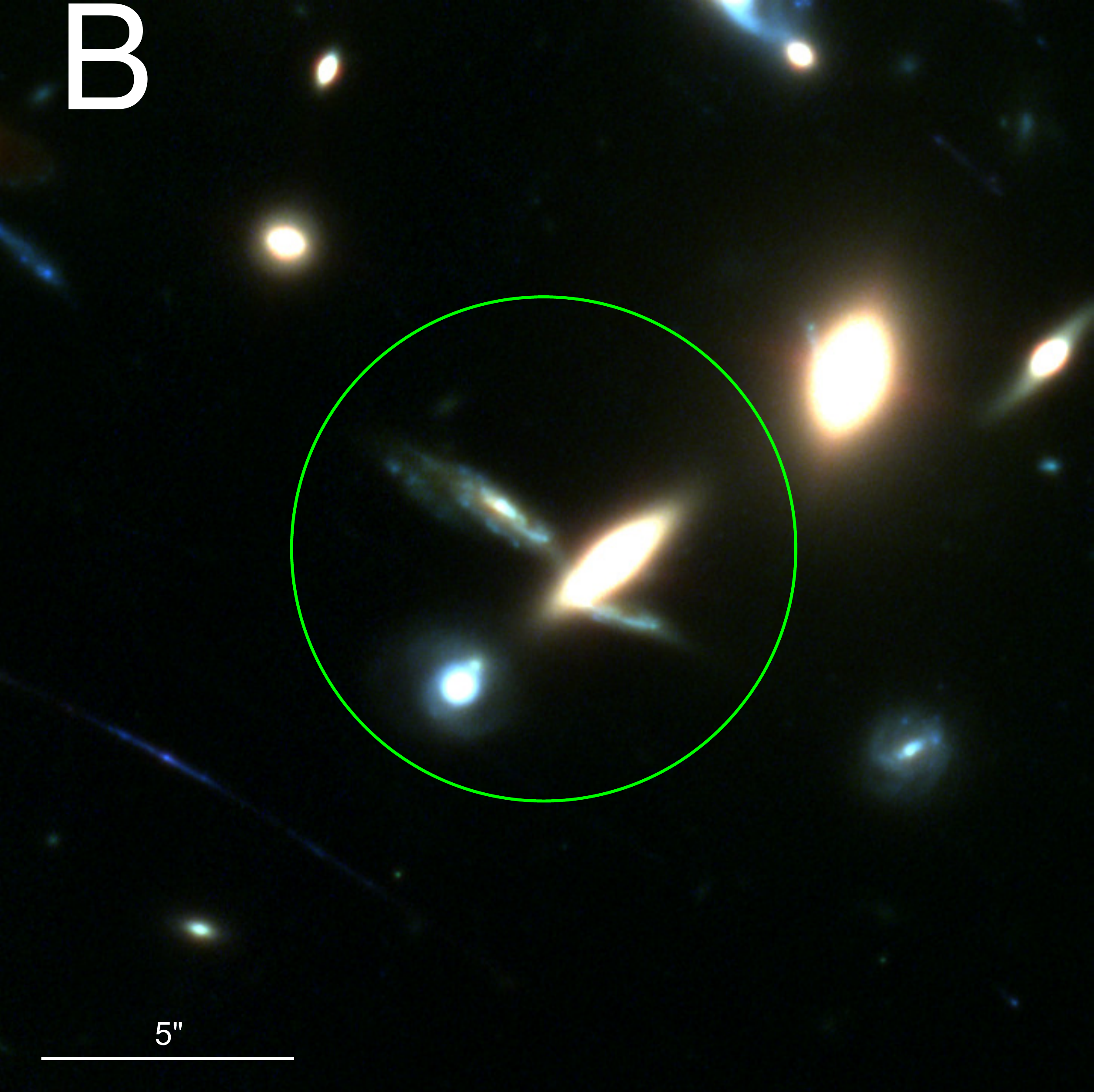}
    \includegraphics[width=0.3\textwidth]{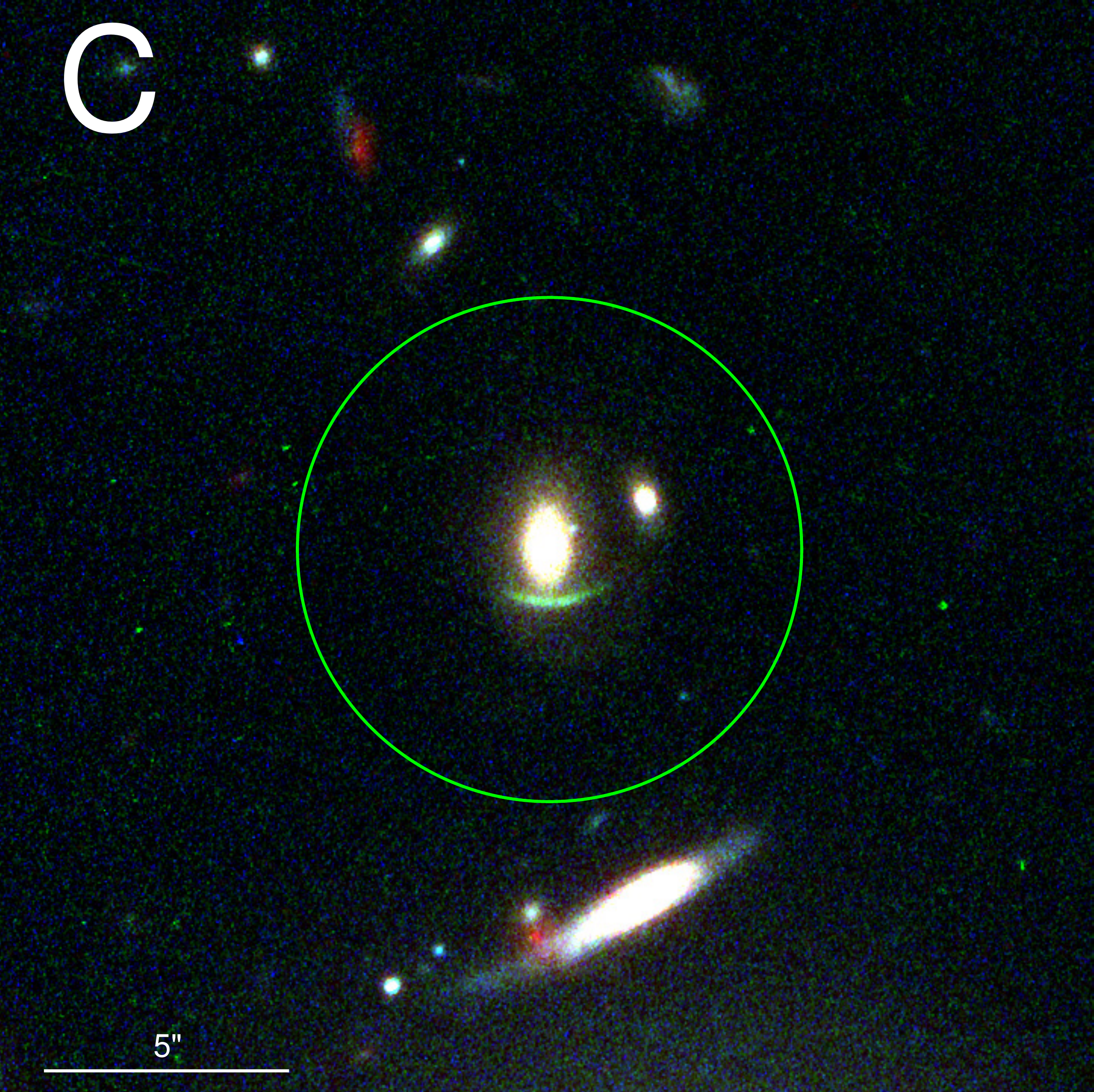}\\
    \centering
    \includegraphics[width=0.3\textwidth]{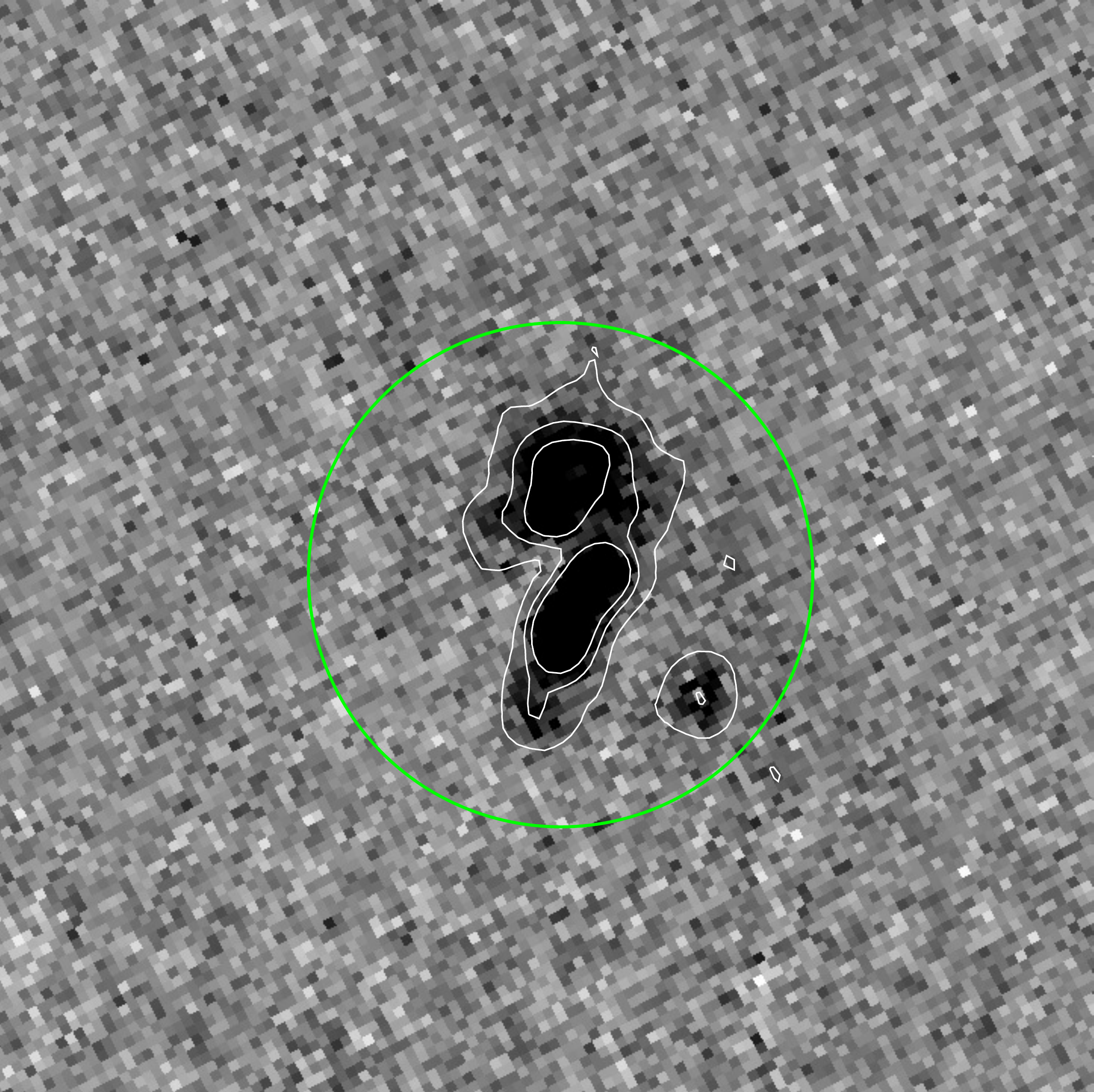}
    \includegraphics[width=0.3\textwidth]{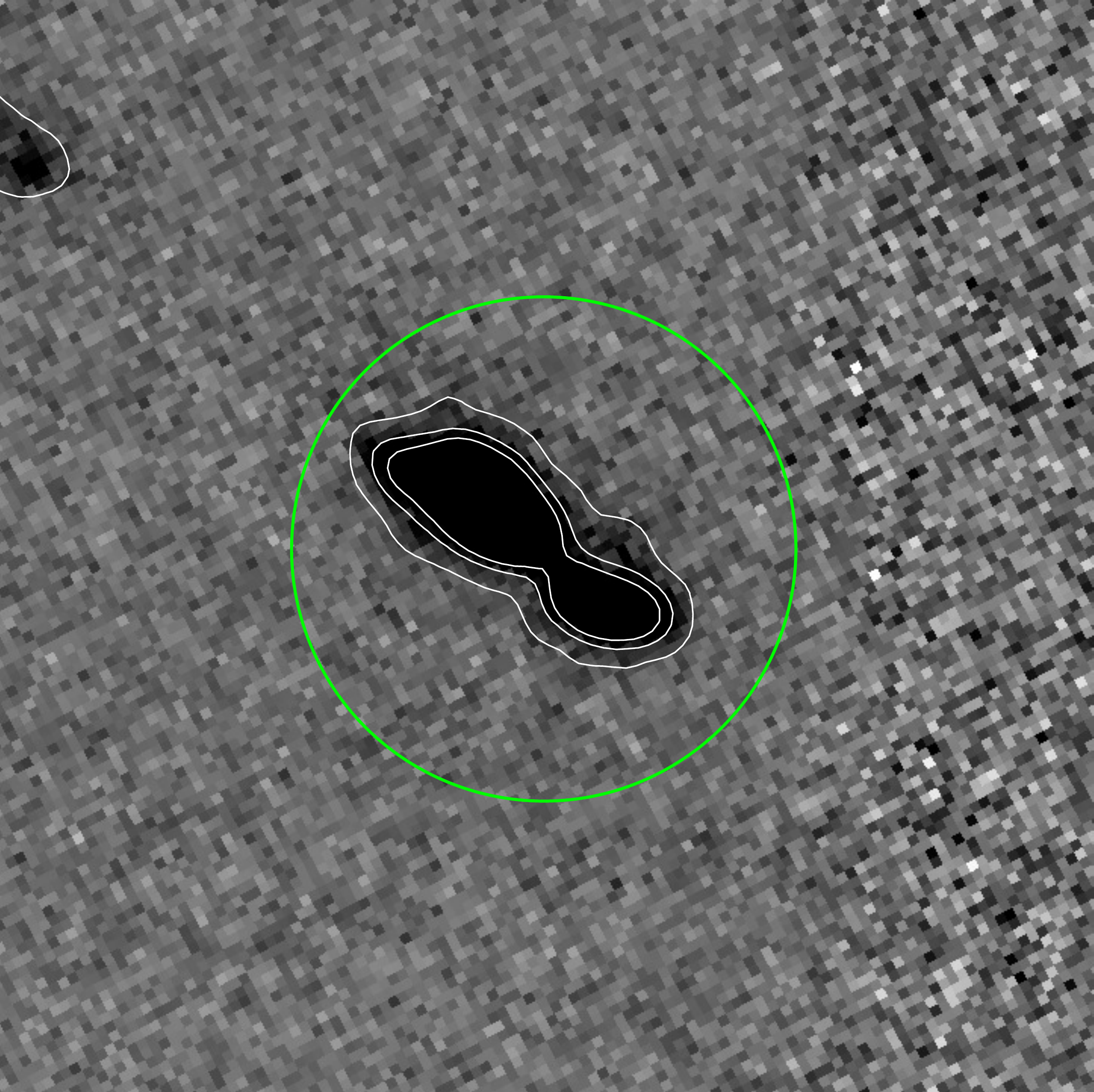}
    \includegraphics[width=0.3\textwidth]{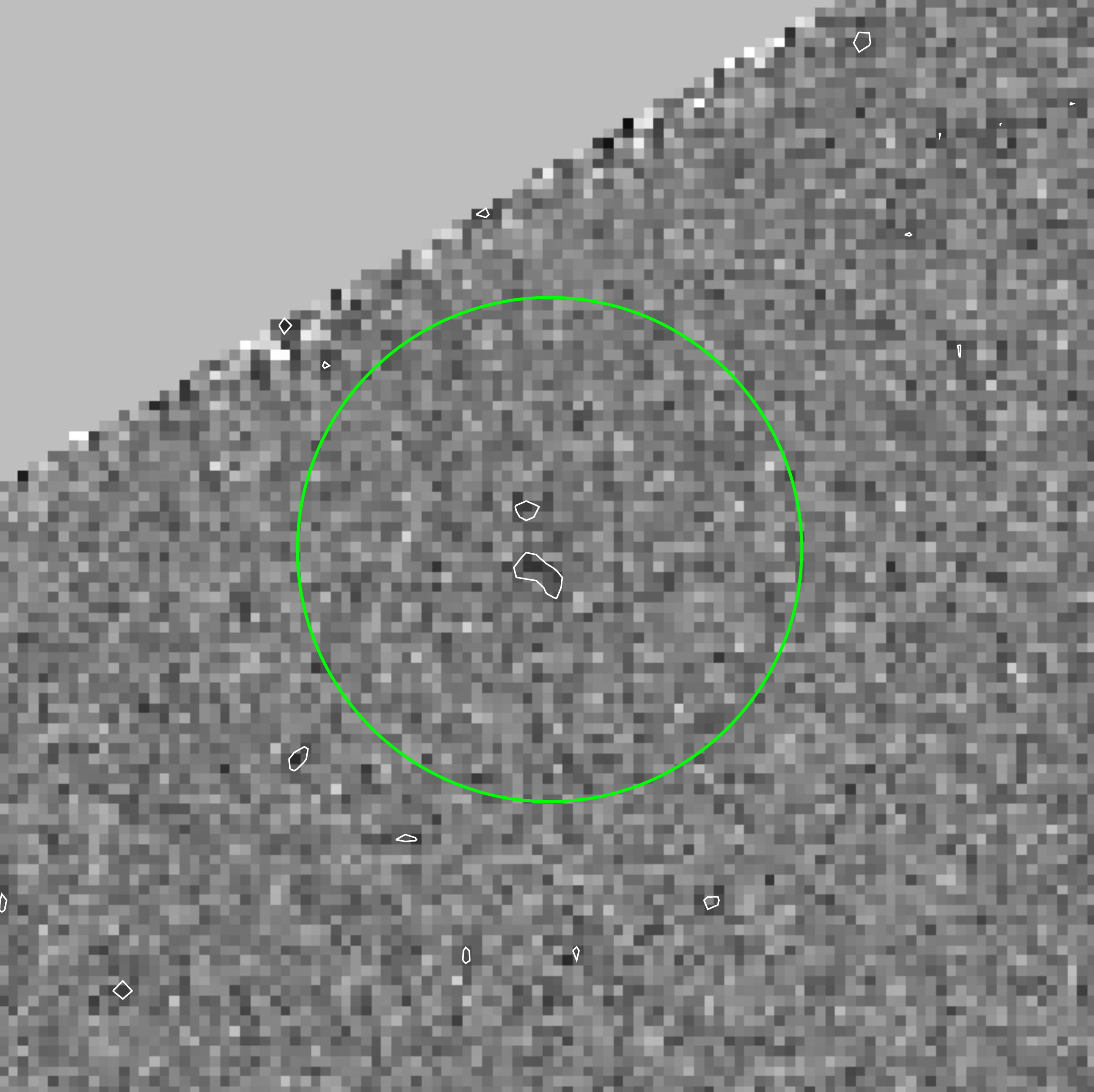}\\
    \medskip
    \centering
    \includegraphics[width=0.3\textwidth]{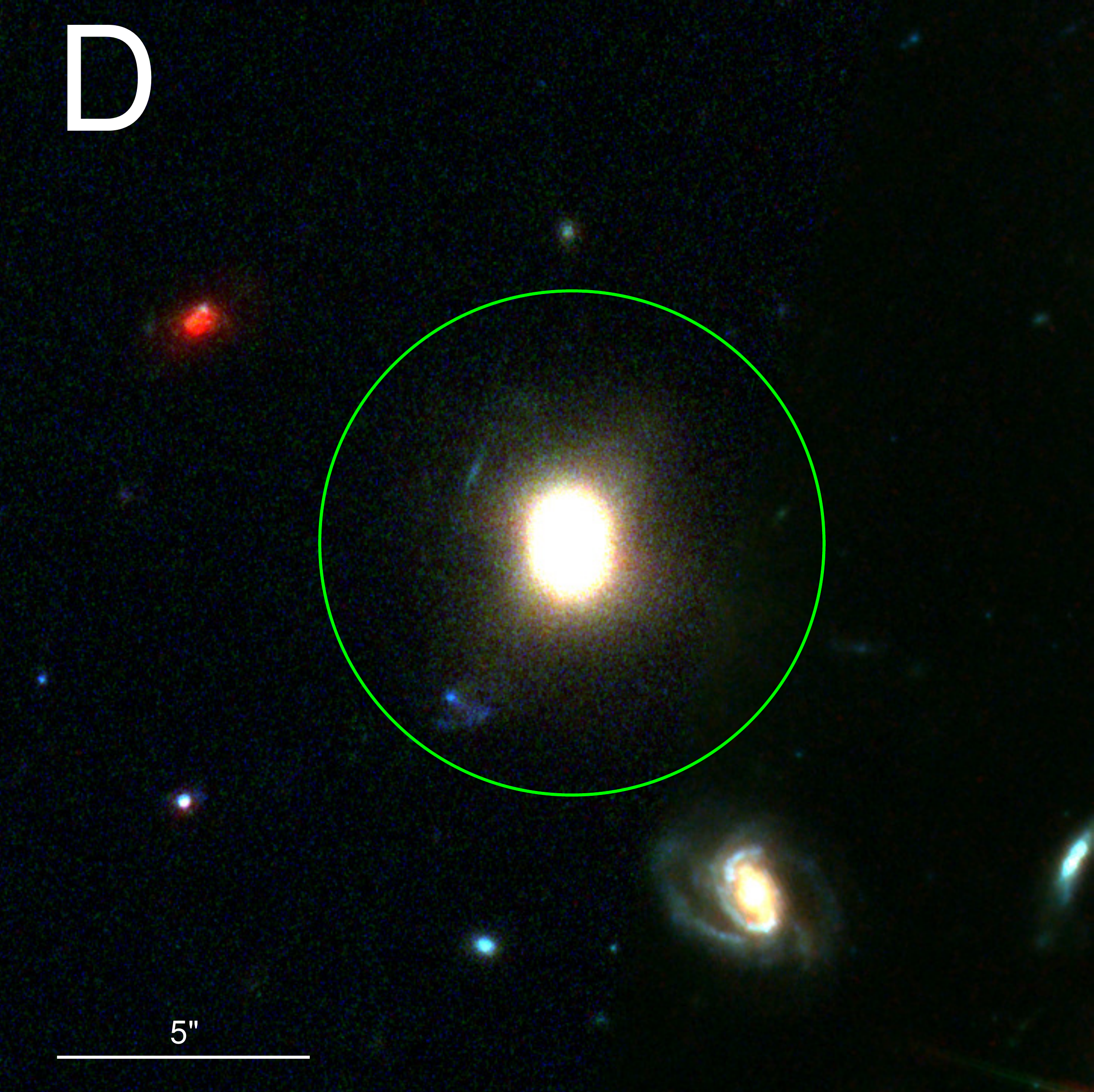}
    \includegraphics[width=0.3\textwidth]{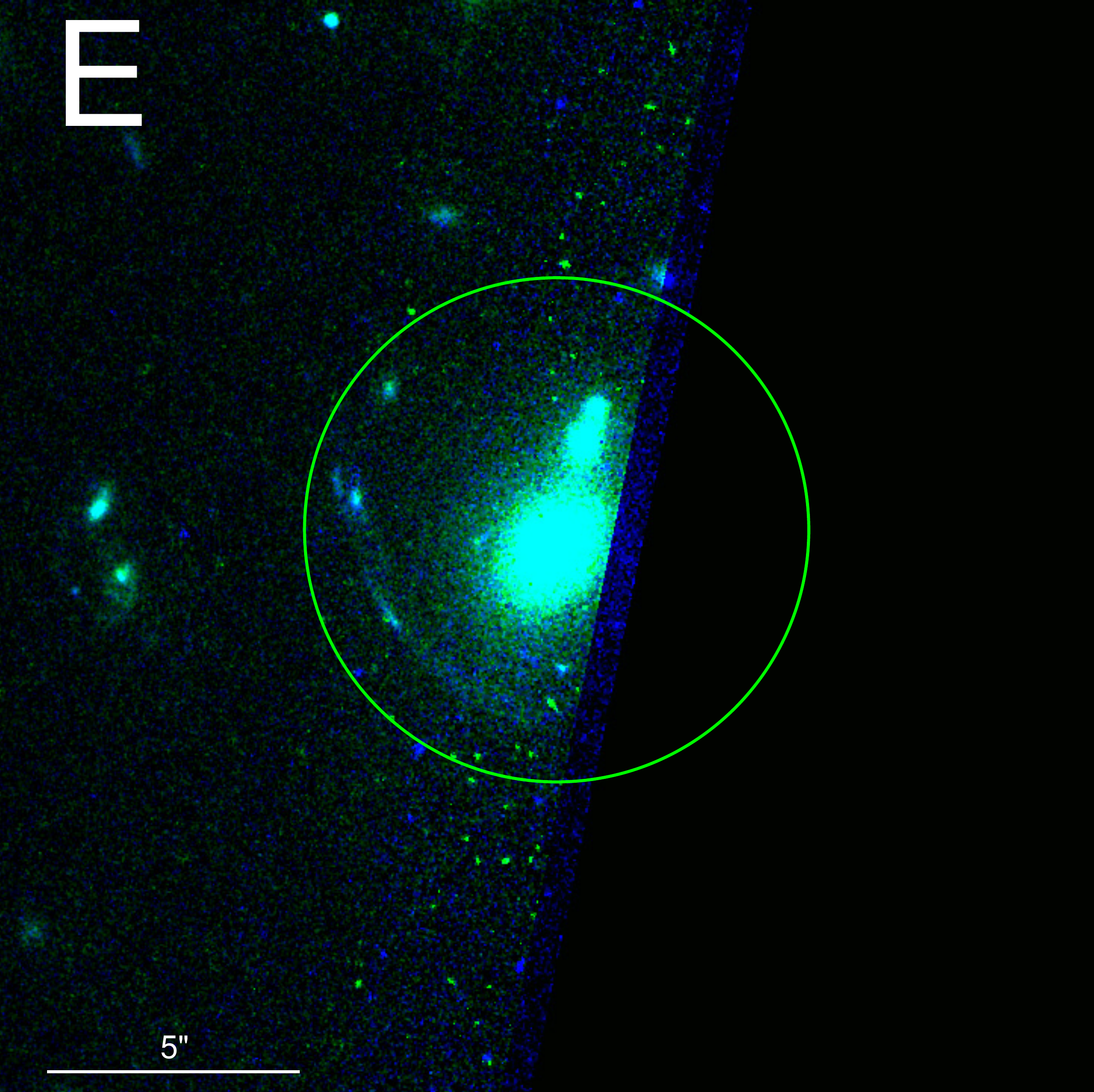}
    \includegraphics[width=0.3\textwidth]{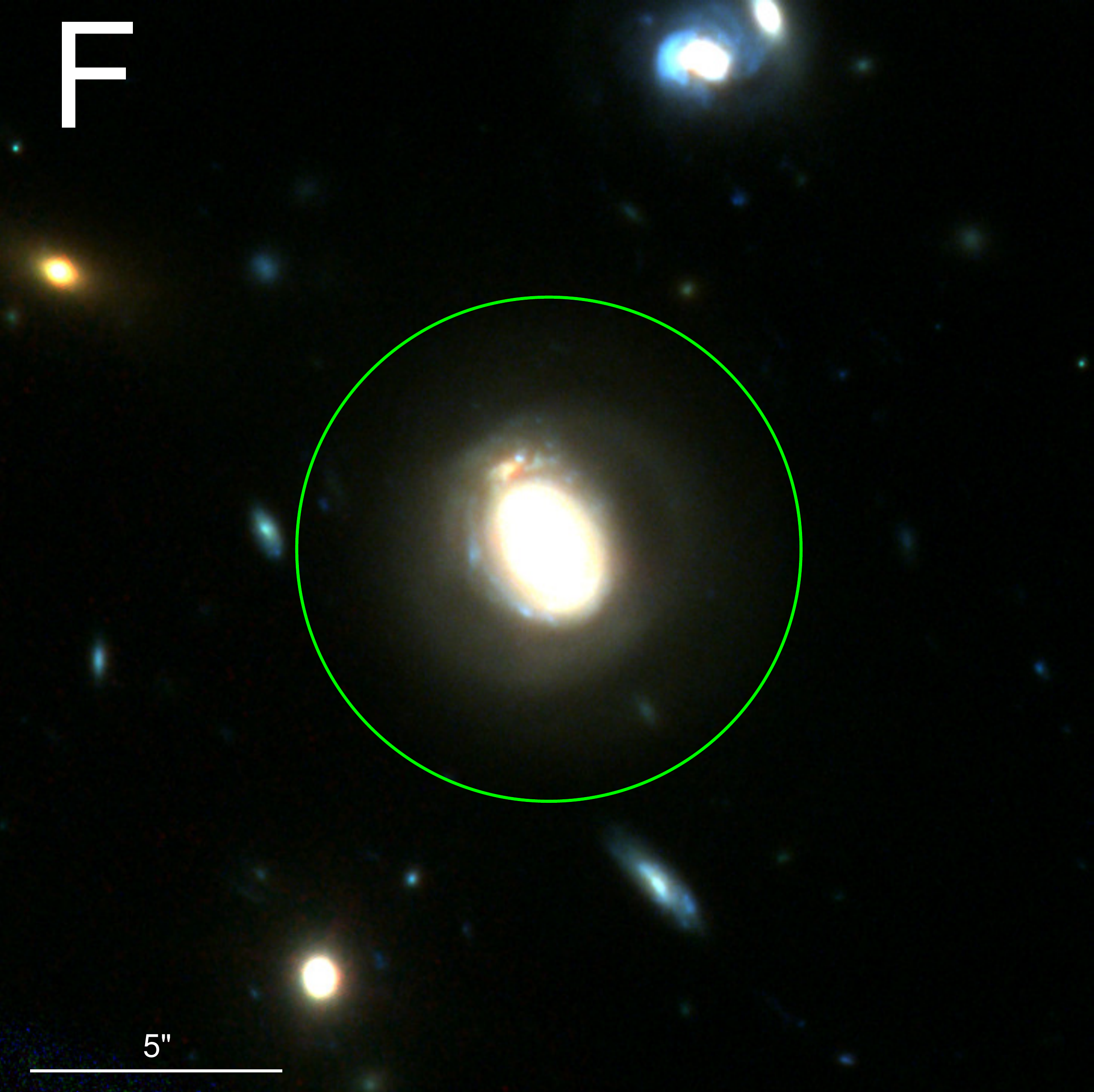}\\
    \centering
    \includegraphics[width=0.3\textwidth]{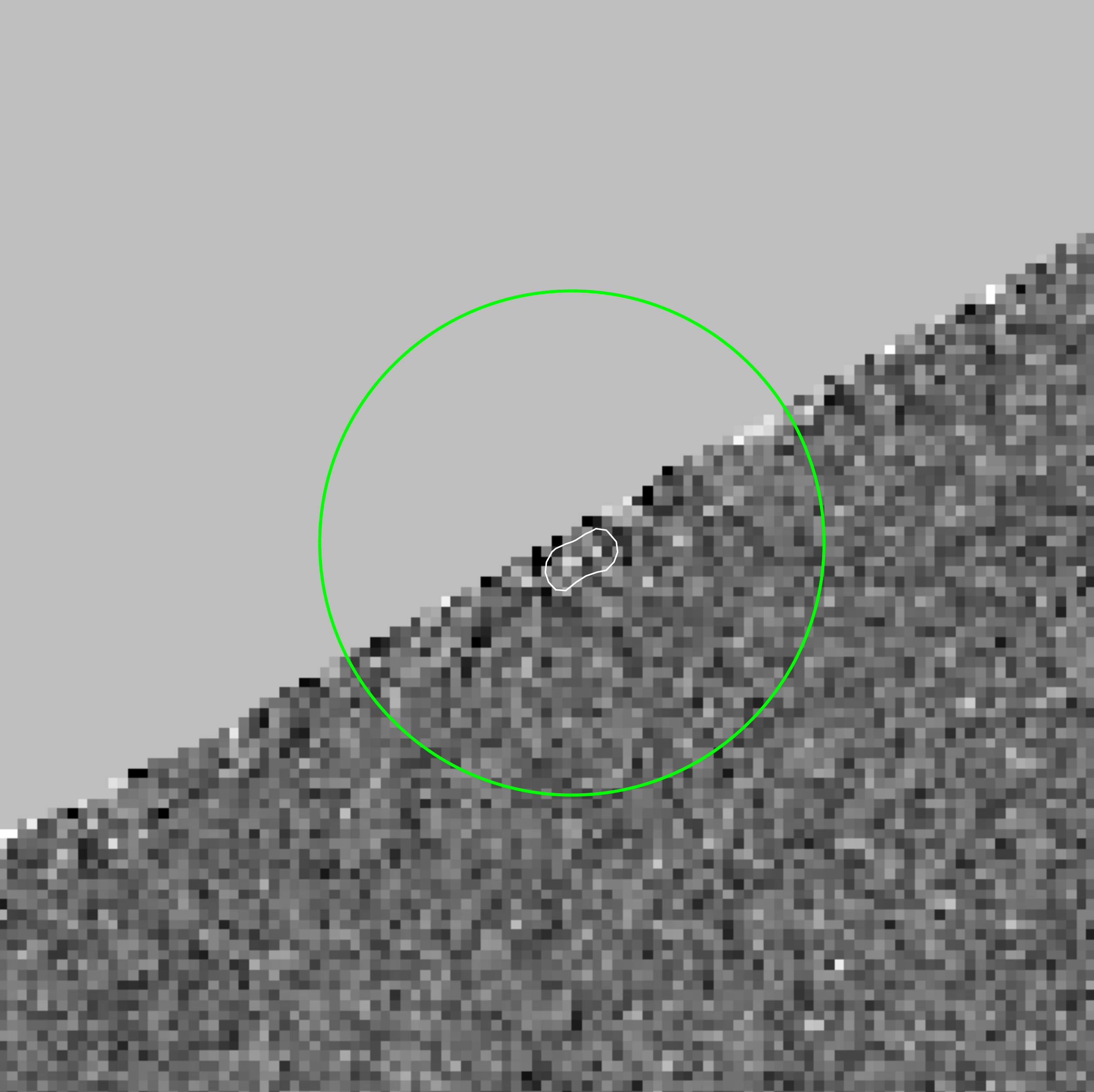}
    \includegraphics[width=0.3\textwidth]{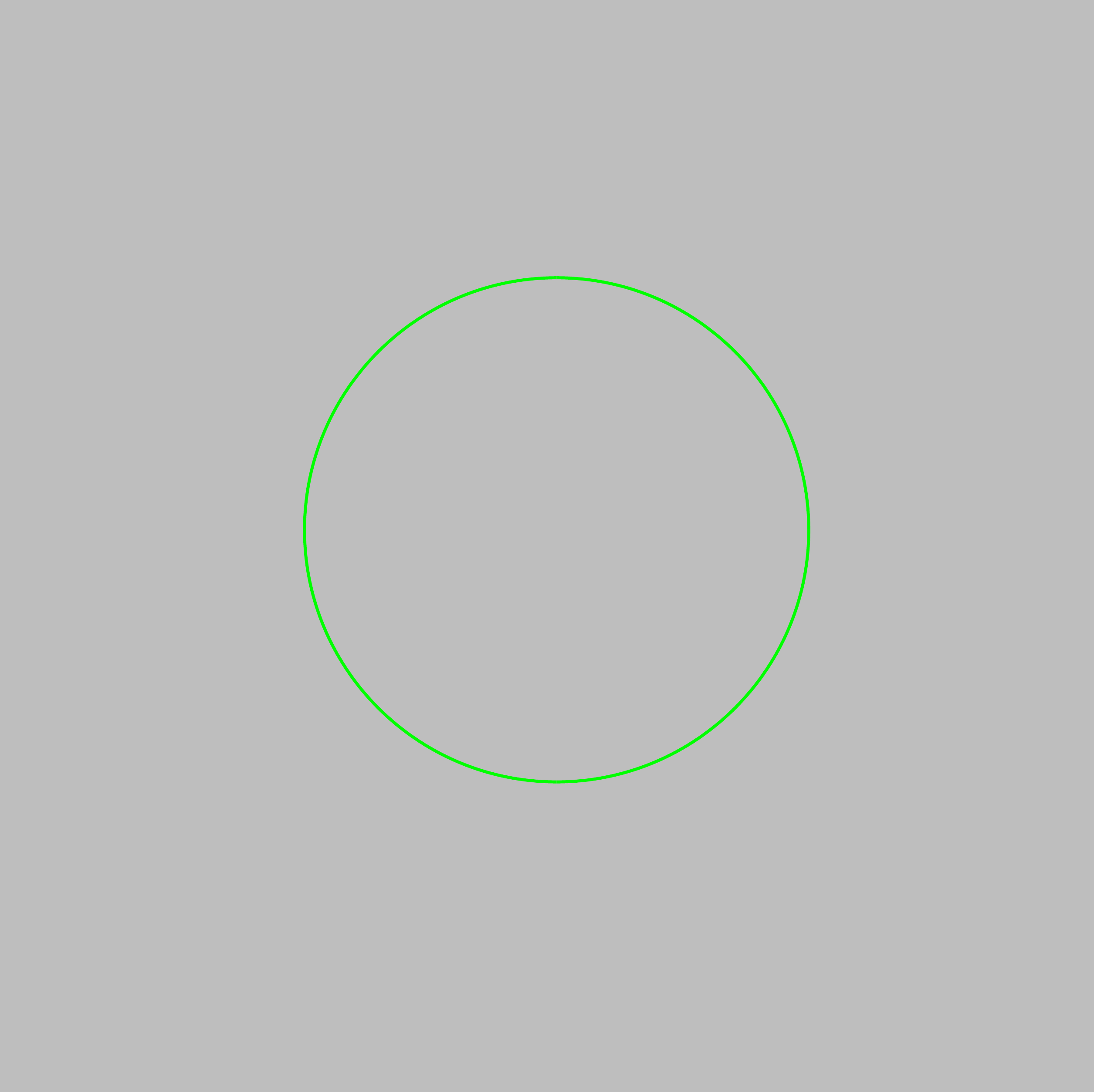}
    \includegraphics[width=0.3\textwidth]{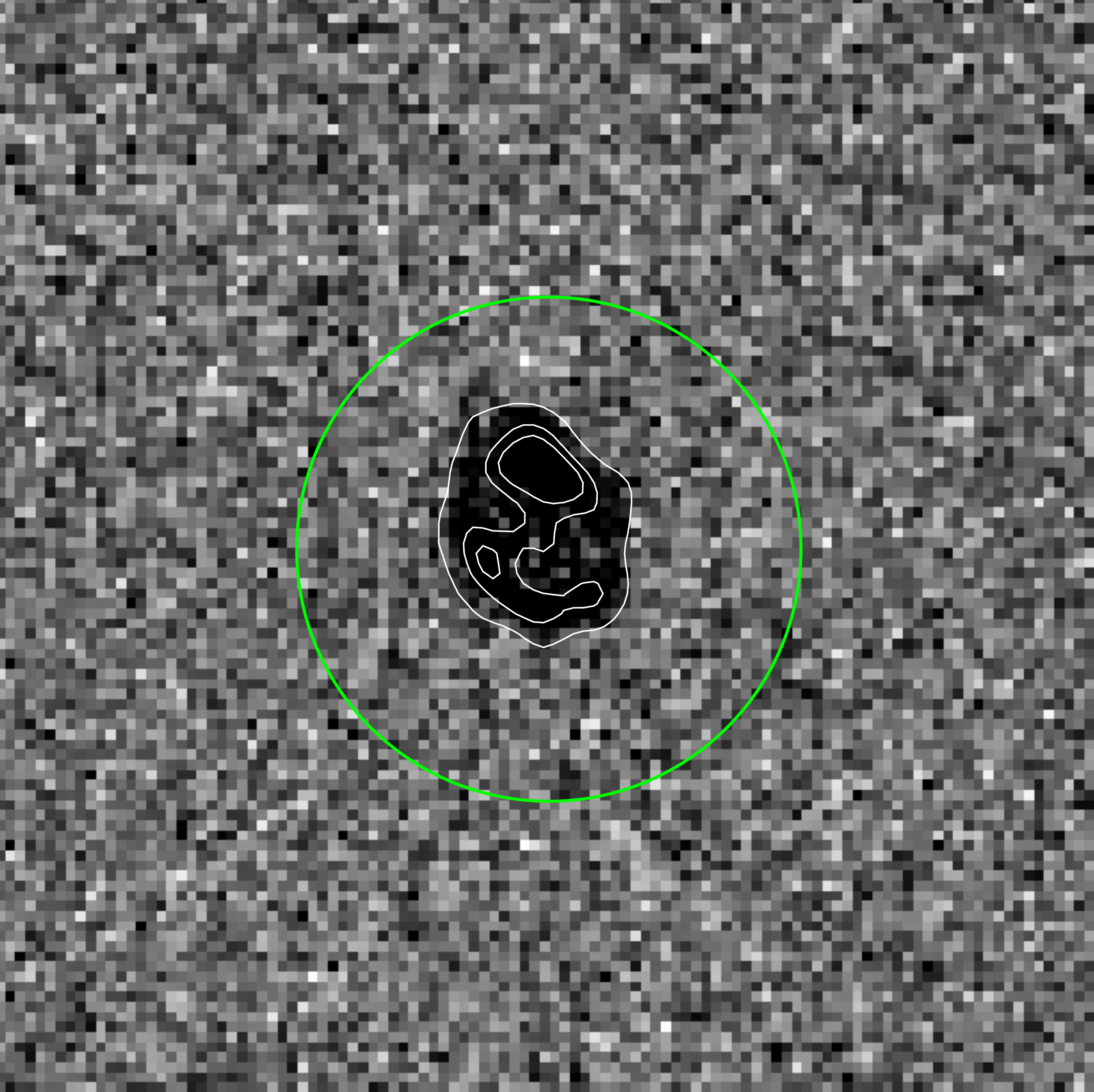}\\    
    
    \caption{Gallery of galaxy-galaxy lenses (GGLs) identified in the Abell 370 field. Colour images show (F606W/F814W/F160W) cutouts from BUFFALO data, while greyscale images show narrow-band emission of the background galaxy extracted from MUSE. To highlight the relative sizes of each GGL, a 5\arcsec circle surrounds each target. Contours on the narrow-band images show 1- 3- and 5$\sigma$ emission levels. Source galaxies in GGLs A, B, and F show strong [\oii] emission, while a tentative \ciii] line can be seen in GGL-C and an unknown line not associated with the lens galaxy appears in GGL-D. We note that GGL-F was identified as a GGL \emph{only} by looking at the MUSE data -- it was missed in a dedicated imaging-based lens search.}
    \label{fig:gglens}
\end{figure*}

However, like other lensing-based techniques, GGL mass estimates require accurate distance measurements of both the foreground and background galaxies.  While many searches make effective use of multi-band imaging to estimate photometric redshifts, spectroscopic redshifts still provide significantly tighter results, leading to more precise mass values. Additionally, narrow-band spectroscopic data (in particular IFU spectroscopy) can reveal ``hidden'' GGL events of emission-line regions that broad-band imaging alone can miss (see, e.g., Lens C in Fig.\,1 of \citealt{men20}), providing even more constraints. Therefore, we take advantage of the A370 MUSE footprint to both follow-up known candidates and look for new objects the imaging-based method may have missed. Results of the combined search are shown in Fig.\,\ref{fig:gglens}.

While we inspect the entire MUSE field, we note that few GGLs are expected in the core: because of the high densities in this region, critical lines of individual cluster members often merge with the larger cluster halo, contributing to broad multiple-image constraints rather than creating true GGL events \citep[e.g.,][]{men20}. Nevertheless, there are two previously-identified GGLs contained within the cluster multiple-image region.  The first (GGL-A) is a singly-imaged but highly distorted ring-like galaxy at $z = 1.062$, first identified by \citet{sou99}. The second (GGL-B) is a late-type galaxy at $z = 1.032$, where a spiral arm is multiply-imaged by a cluster member. Due to its multiplicity, this system is used as a mass constraint in several lens models (e.g., System 42 in L19). While technically, the giant arc system in the south of the core (System 2; e.g., \citealt{ric10}) also shows signs of galaxy-galaxy lensing, we do not consider it as a true GGL because (due to its extreme distortion) it is affected by several cluster members simultaneously. 

Moving beyond the core, two of the three additional BUFFALO candidates (GGL-C and GGL-D) fall -- at least partially -- within the MUSE footprint.  The final candidate (GGL-E) lies completely outside of both the MUSE and BUFFALO WFC3 regions, making it difficult to measure even an accurate photo-z. Nevertheless, the morphology and orientation of the system strongly suggest the system is a GGL, so for completeness we include it in the final count. Examining the objects' spectroscopic data, we easily measure redshifts for the lens galaxies of GGLs C and D ($z_l$ = 0.3798 and 0.3607, respectively; both cluster members), but the source redshifts are less clear. In GGL-C, we tentatively identify faint \ciii] emission at $\lambda$ = 6425\AA, though this detection is just at the noise limit. If confirmed, this would place the source redshift at $z = 2.371$. Conversely GGL-D shows a significant emission line -- not associated with the lens galaxy -- at $\lambda$ = 6135\AA, which appears as a compact source embedded within the lens galaxy continuum. Although the line itself is noisy (at its peak its SNR is only slightly below 2-$\sigma$), the feature spans $\sim$ 20 \AA, making it far wider than a typical noise fluctuation. At the same time, the peak of the flux remains compact and fixed rather than moving in space, giving us confidence that the detection is a real feature. However, the line shape is ambiguous (\ciii], \civ, and Lyman-$\alpha$ are all possible solutions), preventing a definitive redshift estimate. Unfortunately, the likely counterimage of the feature (the small blue arc northeast of the lens galaxy) falls just outside the MUSE footprint. Given the lensing configuration, this image should be both brighter and less contaminated by lens-galaxy light, which would make redshift identification easier. Considering these limitations, additional data could help our analysis in key ways: a wider MUSE mosaic would extend coverage over the missed GGL components, while deeper data would increase the SNR of tentative detections, improving our ability to identify line shapes. We discuss the benefits of a potential follow-up program (called BUFFALO-WINGS) that would cover the entire BUFFALO footprint with 2-hour MUSE pointings, in Section \ref{sec:conclusions}.  

\subsubsection{Detection and Lens Model of GGL-F}
\begin{figure*}
\centering
\includegraphics[width=0.3\textwidth]{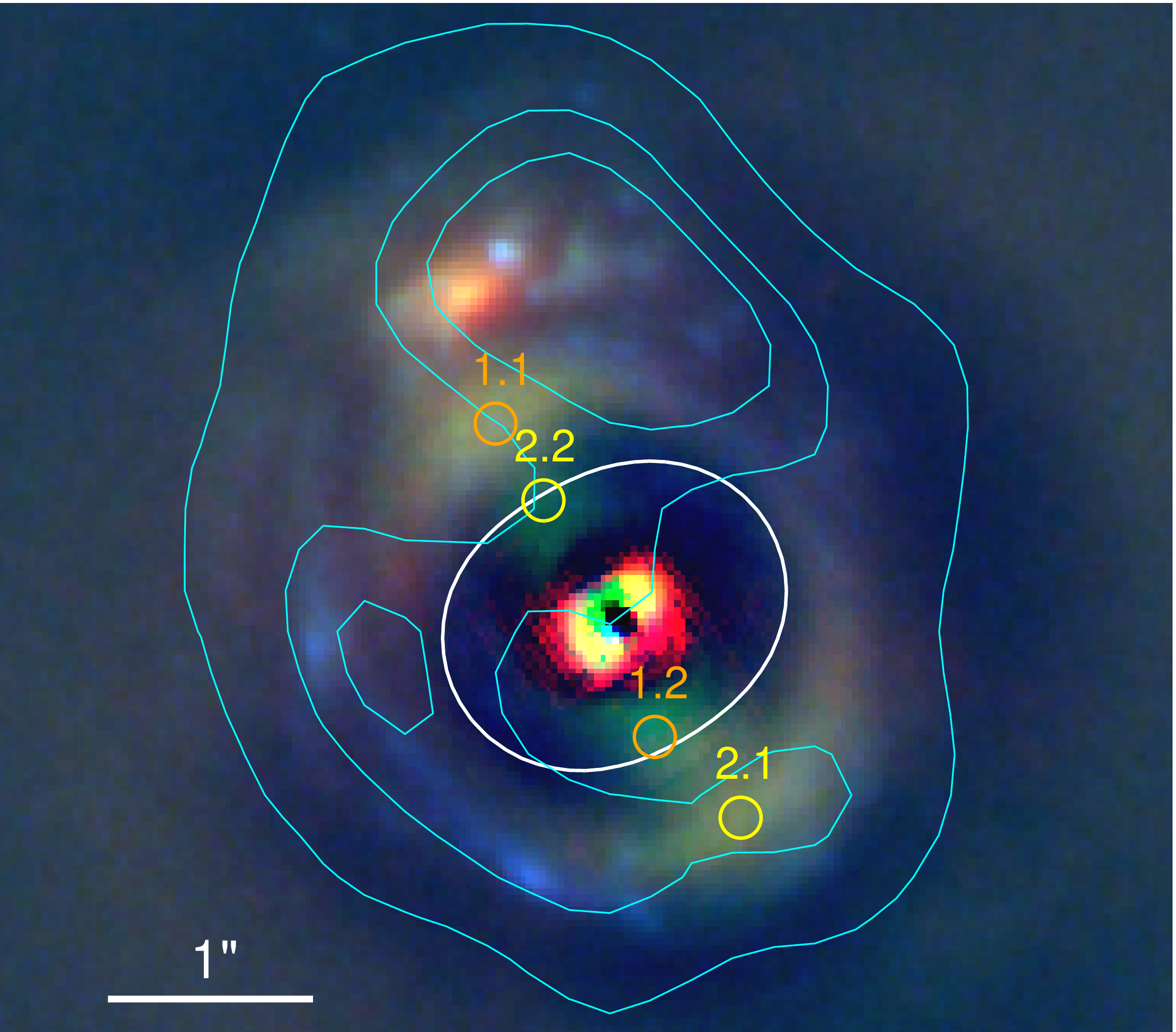}
\hspace{0.15in}
\includegraphics[width=0.65\textwidth]{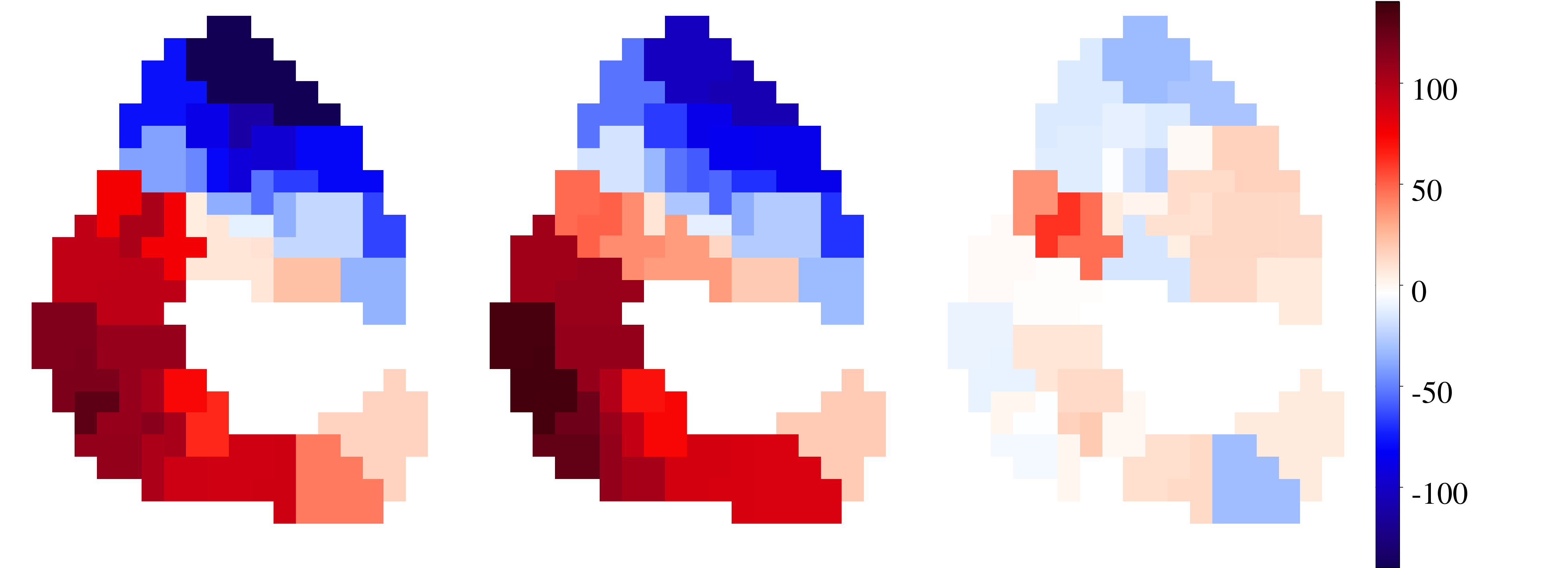}
\caption{(Left) F435W / F814W / F160W RGB color image of the system GGL-F, where the central lens galaxy has been subtracted using a Bulge+Disk fit in each band. The white line gives the critical line of the lens model at the source redshift $z=1.0655$. Colored circles mark the location of the two multiple systems used as constraints for the lensing model. (Right) Modelling of the MUSE [\oii] kinematics, in reading order best fit line velocity, model velocity and residuals (km\,s$^{-1}$).}
\label{fig:gglf}
\end{figure*}

In addition to the three optically-identified candidates, we also highlight an additional spectroscopically-identified GGL, which we label GGL-F. Due to the bright nature of the cluster member, this system was not identified in the optical search, though after modeling and subtracting the lens light (Fig.~\ref{fig:gglf}, left panel), evidence of lensed structure becomes more apparent in the continuum. However, even stronger evidence can be seen in the MUSE data, where narrow-band emission from the background galaxy (in this case the [\oii] line) forms an almost complete Einstein ring, clearly indicating a lensed system. Using the full MUSE data cube, we map the velocity of the [\oii] line emission around the galaxy in the image plane, which shows a velocity gradient across the ring (Fig.~\ref{fig:gglf}, second panel).

We identify two pairs of arcs in the galaxy-subtracted image that serve as constraints on the GGL-F lens model. (We also note for clarity that unless otherwise specified, the term ``lens model'' in this section explicitly refers to the localized model surrounding GGL-F and not an overall cluster-scale lens model.) Both arc systems have identical colours and appear in an expected lens-like configuration in the image plane; at the resolution of the MUSE data, counterimage pairs also have matching velocity values, strengthening the idea that they are multiple images of the same object. The lens model itself consists of a single PIEMD galaxy halo representing the cluster member, described by seven parameters:\ the mass centroid ($\alpha$ and $\delta$), position angle ($\theta$), ellipticity ($\varepsilon$), central velocity dispersion ($\sigma_0$), and two characteristic radii, the core radius ($r_{\rm core}$), and cut radius ($r_{\rm cut}$).  These radii set where the mass profile diverges from a purely isothermal slope, with $\rho(r < r_{\rm core}) \sim$ constant and $\rho(r > r_{\rm cut})$ falling off as $\sim r^{-4}$. To increase the flexibility of the model, we also include an additional systematic term (``external shear'') that characterizes other (possibly unknown) mass components in the local neighborhood and along the line of sight. The external shear adds two parameters to the model, the magnitude ($\Gamma$) and position angle ($\Gamma_{\theta}$) of the shear field.  Furthermore, to explicitly account for the significant mass contribution of the cluster environment, we embed all of these parameters within the best-fit (``Copper'') model presented in L19. Specifically, we include all of the halos in that model as additional mass components, but for computational ease (and because they will be practically unaffected by an individual galaxy-scale halo) we leave them fixed to the final L19 values. 

We optimize the lens model using the publicly available \texttt{LENSTOOL} software. During model optimization we fix the centroid of the halo to the observed position of the cluster member, and following previous modelling efforts (e.g., L17 and L19) we also fix the core radius to be 0.15 kpc. The remaining six parameters ($\theta$, $\varepsilon$, $\sigma_0$, $r_{\rm cut}$, $\Gamma$, and $\Gamma_{\theta}$) are allowed to freely vary. The resulting fit is excellent, with almost negligible rms residuals between the predicted and observed positions of multiple-image constraints ($<$ 0\farcs03). We present the final best-fit model parameters in Table \ref{tab:modParams}. With this model we then make use of the public software {\tt Frapy} \citep{Patricio2019} to adjust the velocity map with a parametric model of a rotating disk accounting for the lensing distortion (Fig.~\ref{fig:gglf}, third panel). The resulting low residuals (Fig.~\ref{fig:gglf}, fourth panel) unambiguously confirm the lensing system, and showcase the strength of using MUSE as a GGL detector.

\begin{figure}
    \includegraphics[width=0.5\textwidth]{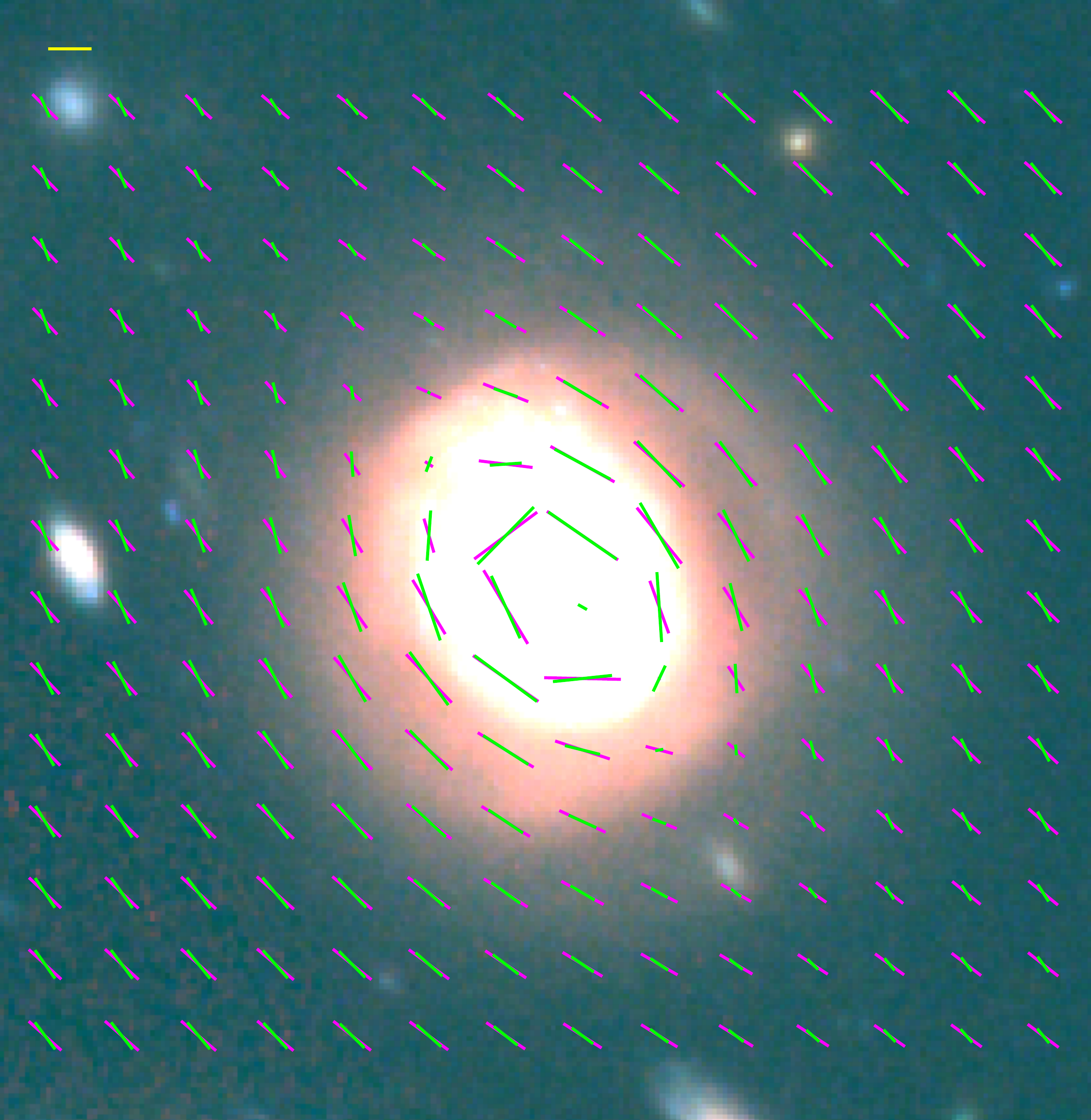}
    \caption{Local shear field comparison between the L19 lens model (magenta) and the new model including GGL-F (green). The yellow stick at the top left represents an induced ellipticity of 0\farcs5. The updated model favours a lower shear magnitude in the neighbourhood of the GGL, suggesting that the original L19 model overestimates the total mass in this region.}
    \label{fig:shearComp}
\end{figure}

The external shear term of the best-fit model is significant, with a value that is more than 5$\sigma$ above a null result. The magnitude of the shear is large ($\Gamma = 0.2$), with an orientation ($\Gamma_{\theta} = 65.2$) pointing towards the cluster core, nearly parallel to the line connecting GGL-F to the cluster centre. Since the ``regular'' (i.e., non-external) shear field generated between two massive galaxy halos is oriented perpendicular to their connecting line, the external shear parameter in the GGL-F model effectively reduces the global shear in the region. We can see this effect directly in Fig.\,\ref{fig:shearComp}, where the elliptical distortions induced by the GGL-F model (green lines) are clearly smaller than those of the original L19 model (magenta lines), especially along the major axis of the cluster member. Physically, this suggests that the total mass in the vicinity of GGL-F (the quantity responsible for inducing the shear) is lower than what was predicted in L19, likely because the $r_{\rm cut}$ parameters of the cluster-scale halos are too large. In some respects this is not surprising:\ since the strong-lensing constraints used in the L19 model do not extend as far as GGL-F, L19 were forced to set an arbitrarily large (i.e., unconstrained) cut-radius of 800 kpc for each cluster-scale component. 

With the additional information of GGL-F, however, it is now possible to explore limits on these parameters, though we note that with only a single GGL, it is difficult to overcome degeneracies in the model. Including additional constraint points (such as GGLs C, D, and E) would overcome these degeneracies, but we are not yet in a position to add them, due to practical limitations on the current data. An expanded MUSE footprint (see Section \ref{sec:conclusions}) could improve this situation considerably, allowing us to access valuable insight on large-scale cluster mass distributions.

\begingroup
\setlength{\tabcolsep}{12pt} 
\renewcommand{\arraystretch}{1.4} 
\begin{table}
    \caption{Best-fit Model Parameters of GGL-F}
    \centering
    \begin{tabular}{l|r|r}
    \hline
        parameter & value  & units \\[0.5pt]
    \hline
        $\alpha$ & 39.991642 & deg\\
        $\delta$ & -1.5952948 & deg\\
        $\varepsilon$ & 0.12$^{+0.04}_{-0.06}$ & --\\
        $\theta$ & -40.18$^{+26.08}_{-26.21}$ & deg\\
        $\sigma_0$ & 166.75$^{+6.63}_{-7.64}$ & km/s\\
        $r_{\rm core}$ & 0.15 & kpc\\
        $r_{\rm cut}$ & 100.00$^{+79.81}_{-41.13}$ & kpc\\
        $\Gamma$ & 0.20$^{+0.01}_{-0.03}$ & --\\
        $\Gamma_{\theta}$ & 65.21$^{+18.81}_{-7.99}$ & deg\\
    \hline    
    \end{tabular}
    \label{tab:modParams}
\end{table}
\endgroup

\begin{figure*}
    \centering
    \includegraphics[width=0.49\textwidth]{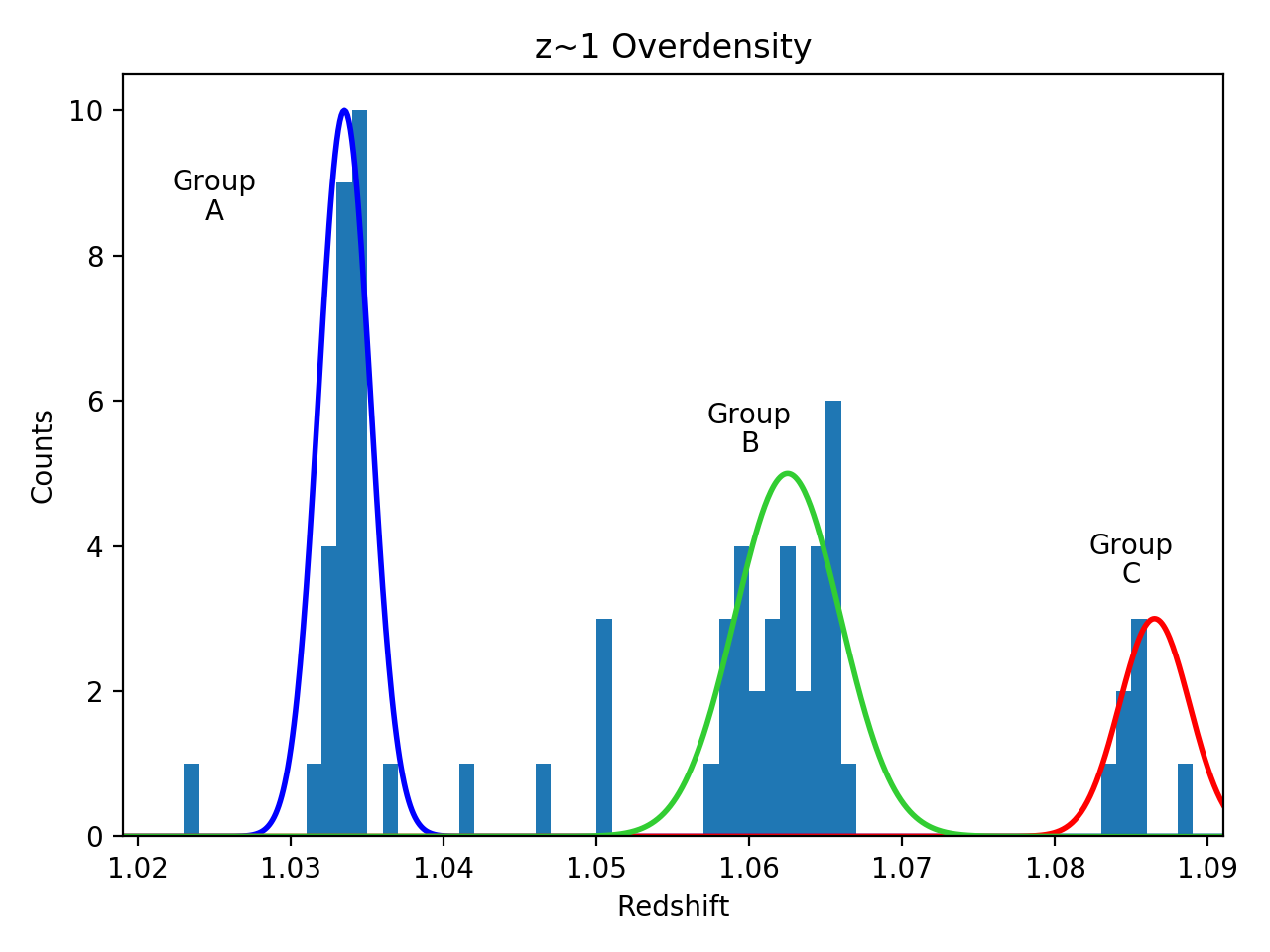}
    \includegraphics[width=0.49\textwidth]{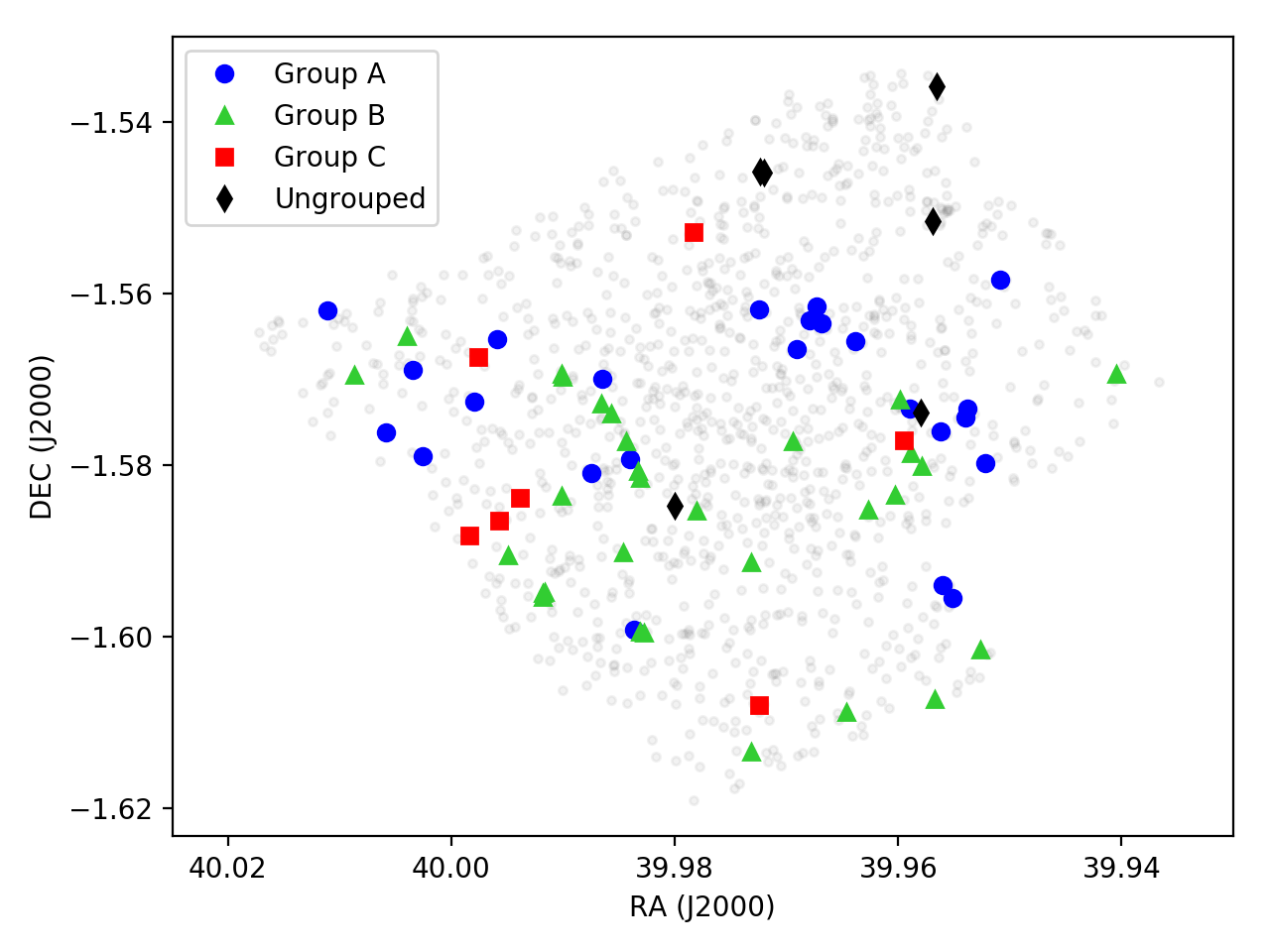}
    \includegraphics[width=0.49\textwidth]{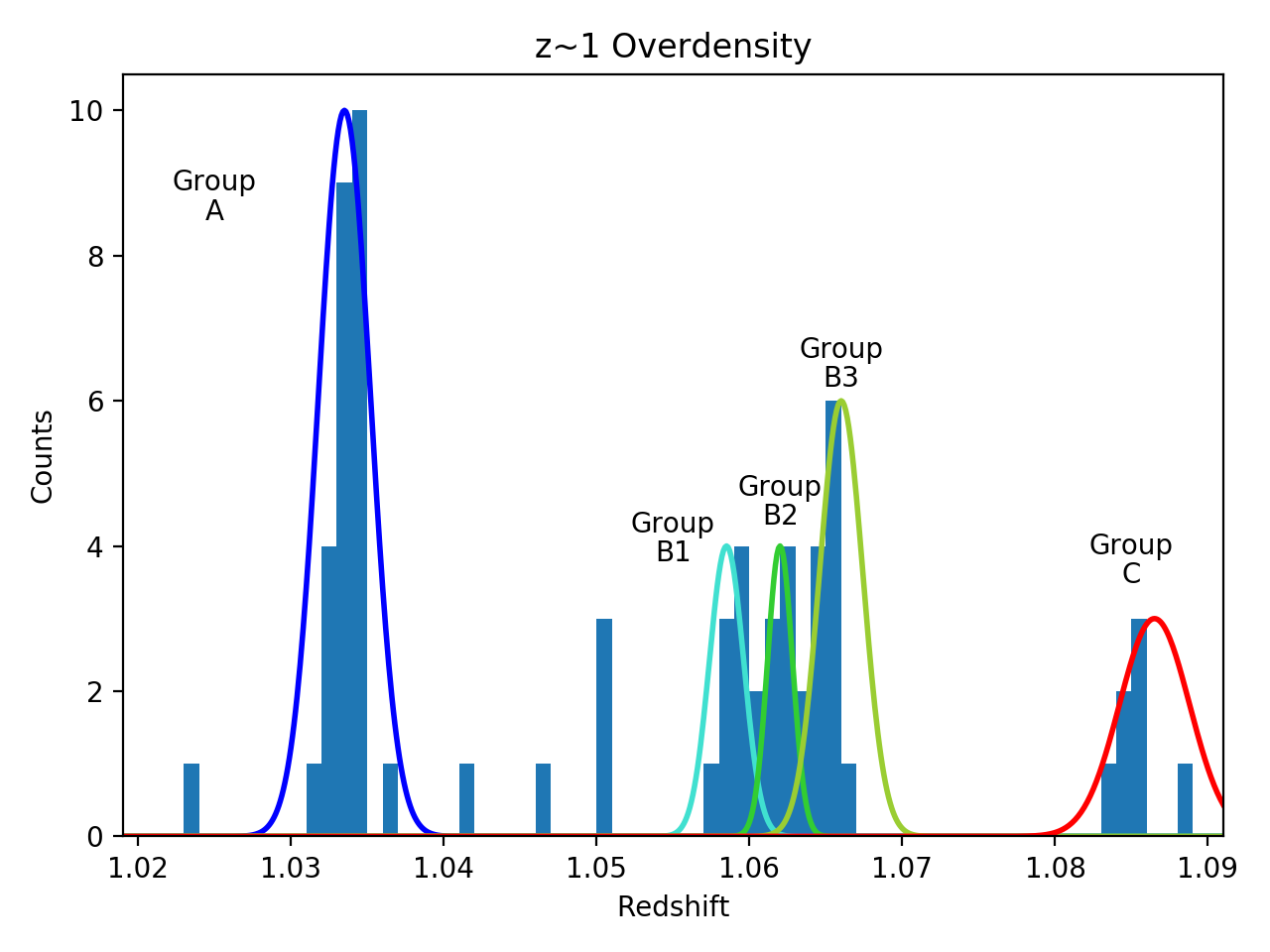}
    \includegraphics[width=0.49\textwidth]{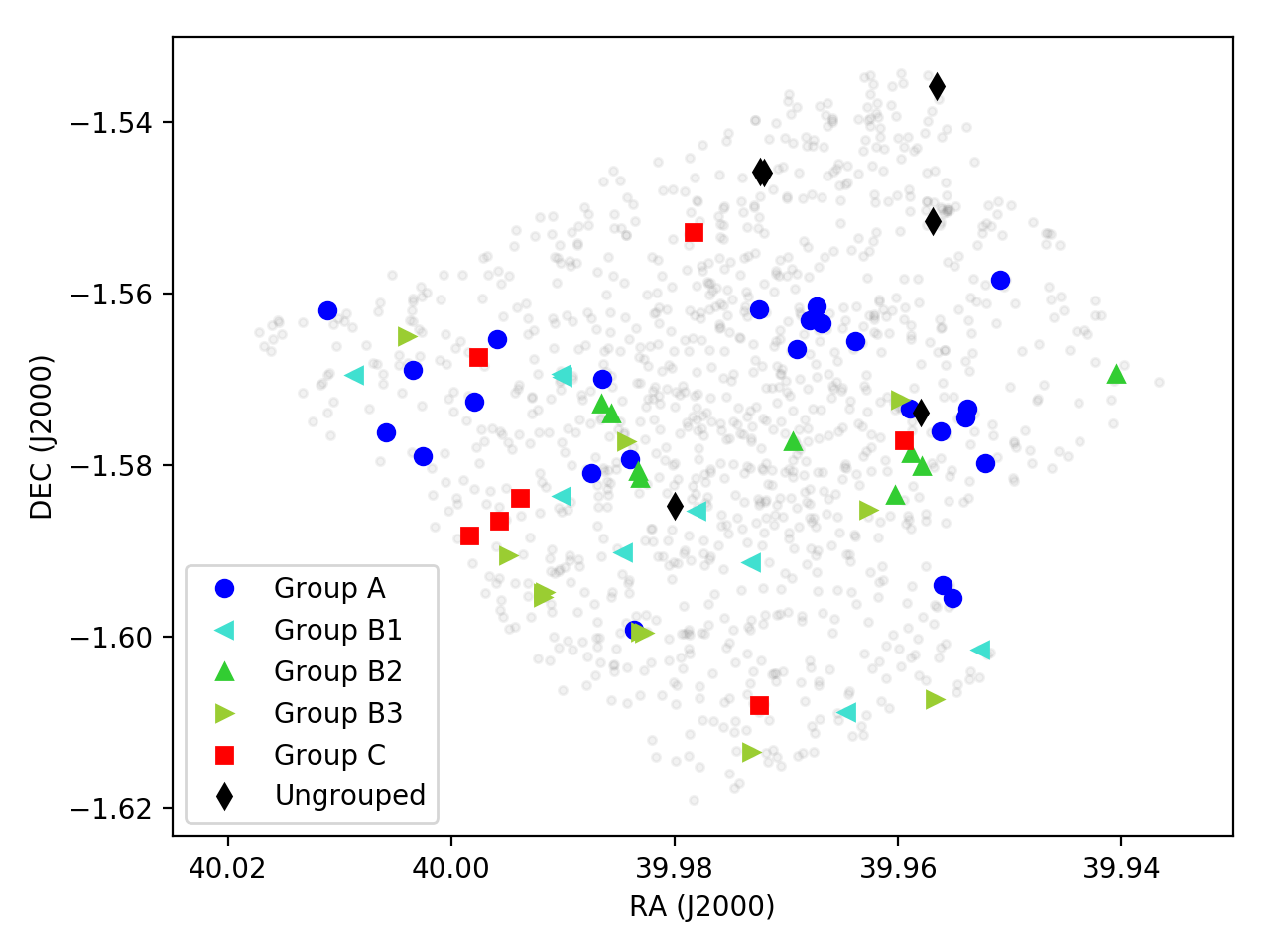}
    \caption{Top left: Redshift distribution of the 68 galaxies in the $z \sim 1$ overdensity. Rather than a single structure, we identify and fit velocity dispersion curves to at least three distinct galaxy populations, dubbed Groups A, B, and C. The best-fit dispersion value of each structure is low (Table \ref{tab:subgroups}), with only Group B approaching group-scale potential level ($\sigma > 350$ km s$^{-1}$). Right: Spatial distribution of galaxies in the redshift bin (coloured shapes), relative to all other spectroscopic objects (faint grey dots). Galaxies are coloured by group membership, corresponding to the solid lines in the left-hand panel. ``Ungrouped'' galaxies (those not assigned to a specific group) are coloured in black. There is little spatial overlap between groups, which suggests that they are distinct systems rather than a coherent structure, such as a galaxy filament. Bottom panels: Same as top panels, but with Group B further subdivided into three separate subgroups.}
    \label{fig:overdensity}
\end{figure*}

\subsection{Galaxy overdensity at redshift $\sim$1}
\label{subsec:backGroup}

Thanks to the considerable increase of confirmed strong-lensing constraints over the past few years, recent lens models have been able to capture the complex nature of the A370 mass distribution better than ever before. A common feature in many of these models is a systematic ``external shear'' term, needed to properly fit constraints lying at the edge of the multiple-image region \citep[e.g.][]{str18,kaw18,ran20}. While many physical phenomena can induce such an effect, one common explanation is the presence of additional (unidentified) mass concentrations either within the cluster plane or along the line of sight. Efforts to identify this mass are ongoing, but one potential candidate is an overdensity of galaxies located behind the cluster (but in front of most strongly-lensed features) at $z \sim 1$. Using the new MUSE mosaic, we have spectroscopically identified 68 galaxies as part of this group. 

First identified by \citet{die18} and later refined in L19, the initial candidate structure consisted of $\sim$30 galaxies (including one multiply-imaged galaxy) between $z=1$ and $z=1.1$, located along the line of sight of the A370 central core. After transforming the observed positions of these galaxies to the unlensed source plane, they appeared as a compact group behind the main cluster, centred at ($\alpha$ = 39.9709852, $\delta$ = -1.5751906; see Fig.\,10 in L19) with a physical diameter of 534 kpc (assuming the median redshift $z = 1.054$) -- prompting L19 to include them as additional mass components in their lens model. 

While including the group members did slightly improve the model fit (rms$_{~\rm no\ group} = 0.78\arcsec$; rms$_{~\rm group} = 0.75\arcsec$), their presence did not statistically affect the magnitude of the shear ($\Gamma \sim 0.1 \pm 0.01$ in both cases), suggesting that the identified structure was not a major contributor to the systematic effect. However, this conclusion was somewhat limited: because confirmed group members had been identified up to the edge of the MUSE field, an alternative possibility -- that additional group members (and hence an extended mass distribution) existed beyond the available data footprint -- could not be discounted. The extended MUSE footprint allows us to test this hypothesis, and the histogram of the data in the outskirts alone (Fig.\,\ref{fig:redshifts}) shows evidence of a slight bump at $z \sim 1$, with 37 galaxies in the bin. When combined with the 31 galaxies in the original core data (Fig.\,\ref{fig:z_breakdown}), it is evident that the overdensity of objects in the $z = 1.0 - 1.1$ bin still persists, although the gap is smaller than before, implying that a larger relative fraction of the group is still localized behind the cluster core. Nonetheless, with the outskirts data more than doubling the number of objects in the bin, the idea of an extended structure is still feasible, and can be further explored.

Investigating more closely, we expand the $z = 1.0 - 1.1$ bin to higher resolution, allowing us to better identify its constituent parts. Doing so reveals a complex redshift structure, with several small overdensities spanning the full redshift range (Fig.\,\ref{fig:overdensity}), as opposed to a single Gaussian-distributed system. In particular, we identify at least three distinct groupings, which we label A ($z \sim 1.035$), B ($z \sim 1.06$), and C ($z \sim 1.085$), and which could increase up to five if group B is actually a combination of separate sub-groups. Physically, each system contains between 10-30 galaxies and spans a narrow range of redshifts. Gathering the individual members of a given group together, we calculate an average redshift for each collection, which we take to be the systemic redshift. Converting these into recessional velocities, we also fit a velocity dispersion to each group, measuring values between $\sim$100 and $\sim$400 km s$^{-1}$, with the sub-systems of group B again measuring only a fraction of the total. A summary of all measurements can be seen in Table \ref{tab:subgroups}.

While non-negligible, the measured velocity dispersions are still relatively low, and when combined with the large redshift separations between systems suggests that groups A, B, and C are not gravitationally bound to each other. Observing the physical locations of each group (shown in the right-hand panel of Fig.\,\ref{fig:overdensity}) seems to strengthen this notion. From the figure, the objects of group A appear predominantly to the north of the cluster core (with distinct sub-clumps in the west and east), while the objects that make up group B are instead seen mainly in the south. The subsets of group B also appear to locally cluster themselves, with objects in group B2 (which includes the multiply-imaged galaxy) situated closest to the cluster core, followed by group B1 and finally group B3, whose members are largely found near the edge of the MUSE footprint. With only seven objects in total, group C does not appear constrained in any one place, though there is  a small clustering of members in the east. We also remind the reader that the apparent ``hole'' seen near the cluster centre is caused by lensing deflection effects; in the source plane, the galaxies would appear closer together. Regardless, when coupled together, this suggests that the main overdensity itself is not a coherent structure, but rather an illusion created by a coincident set of small, physically distinct groups aligned in narrow redshift slices along the line of sight. 

The low velocity dispersions of the $z \sim 1$ structures further imply that they are all low-mass, especially compared to the cluster potential dominating the line of sight.  Because of this, it is not surprising that including these objects in the mass model (even as a combined ``pseudo-cluster'') has only a limited effect on the overall result, and it seems much clearer now that these background galaxies are not a significant source of the systematic shear.

\begin{table}
    \caption{Substructure in the $z \sim 1$ overdensity}
    \centering
    \begin{tabular}{c|c|c|c}
        \hline
        group name  & members  & systemic redshift & $\sigma$\\
               &    &   &  (km/s) \\   
        \hline
          A    & 25 & 1.034 & 116\\
          B    & 30 & 1.062 & 394\\
          \emph{\hspace{3mm} B1}    &  \emph{\hspace{3mm} 9} & \emph{\hspace{3mm} 1.059} &  \emph{\hspace{3mm} 127}\\
          \emph{\hspace{3mm} B2}    &  \emph{\hspace{3mm} 9} & \emph{\hspace{3mm} 1.062} &  \emph{\hspace{3mm} 90}\\
          \emph{\hspace{3mm} B3}    &  \emph{\hspace{3mm} 12} & \emph{\hspace{3mm} 1.065} &  \emph{\hspace{3mm} 125}\\ 
          C    &  7 & 1.085 & 195\\
          Unattached &  6 & -- & --\\
          \hline
    \end{tabular}
    \label{tab:subgroups}
\end{table}

\section{Conclusions and Future Prospects}
\label{sec:conclusions}

In this work, we have examined the structure of the strong-lensing galaxy cluster Abell 370 (A370), using a powerful combination of high-resolution \emph{HST}/BUFFALO imaging and MUSE spectroscopy. The wide-area data mosaics provide a clear look at the core and outskirts regions simultaneously, revealing new insights that previous efforts (focusing mainly on the core alone) could not access. In particular, our MUSE mosaic, a deep (2-8 hour exposure time) 2$\times$2 block of pointings in the core surrounded by 10 shallow (45 min. exposure time) pointings in the outskirts, provides dense spectroscopic coverage over 14 arcmin$^2$ of the field. The footprint extends to $\sim$ 900 km in projected radius from the cluster centre at the systemic redshift ($z$ = 0.375) and produces a robust 3D look at the region. It is, to date, the largest contiguous area of a galaxy cluster ever probed by MUSE. Taking advantage of this rich data set, we have characterized properties of galaxies, both within the cluster and along the line of sight, that constrain aspects of mass distribution and galaxy evolution. Reporting these efforts, we summarize our main results as follows:

\begin{itemize}
    \item We have constructed a new, extended redshift catalogue from the MUSE data, containing 30 stars, 116 foreground galaxies, 416 cluster members, and 623 background galaxies, including 153 at high-redshift ($z > 3$). The 1185 spectroscopically-confirmed objects more than doubles the previous A370 MUSE spec-z catalogue, presented in \citet{lag19}.\smallskip
    
    \item Using the 3D positions of identified cluster members, we determine the coordinates of the cluster centre ($\alpha$ = 39.9704672, $\delta$ = -1.5779282, $z$ = 0.374). Constructing a redshift space diagram with these coordinates, we measure the radial mass profile and enclosed halo mass of A370 using the caustic method, finding $M_{500}$ = (1.31$\pm$0.15)$\times 10^{15}$ $M_\odot$ and $M_{200}$ = (1.34$\pm$0.18)$\times 10^{15}$ $M_\odot$. Our measurement for $M_{500}$ is in good agreement with estimates derived from gravitational lensing and other values in the literature \citep[e.g.,][]{mad08,ume11}, while the $M_{200}$ measurement is in moderate agreement. \smallskip
    
    \item Tracing the physical properties of confirmed cluster members, we see that galaxies in the outskirts have, on average, bluer broadband colours than those in the core. This is due to the steady increase in blue cloud galaxies at larger radii, with up to 50\% of cluster members lying off of the red sequence by the edge of the data footprint.  This reveals that a significant subset of cluster galaxies would be missed by traditional photometric identification.\smallskip
    
    \item Classifying galaxies according to stellar activity, we also find that star-forming cluster members become more common further away from the cluster centre. However, while there is a distinct colour difference between star-forming and passive galaxies, we see little change in their intrinsic colours with radius. Conversely, post-starbust galaxies show a strong colour gradient with radius, possibly constraining the timescale over which galaxies transition to the red sequence. \smallskip
    
    \item Mirroring an ongoing galaxy-galaxy lens (GGL) search in the BUFFALO fields, we perform a follow-up search in the spectroscopic data. Of the five identified GGL candidates in the A370 \emph{HST} imaging, four are at least partially contained within the MUSE footprint. We easily identify lens redshifts for all four candidates (all confirmed cluster members) and source redshifts for two: GGL-A ($z_s$ = 1.062) and GGL-B ($z_s$ = 1.032); see Fig.\,\ref{fig:gglens}). We measure a tentative redshift of $z_s$ = 2.371 for GGL-C, though the line is noisy and falls at the edge of the data field where sensitivity is low. GGL-D shows a prominent emission feature at $\lambda$ = 6138\AA\ that is not consistent with the lens galaxy redshift, though the exact line interpretation is ambiguous. Since the expected counterimage of this system falls outside of the MUSE field, we cannot use spectroscopic data to confirm if the two images are related, making it difficult to constrain the redshift. The fifth GGL (GGL-E) falls completely outside of the MUSE footprint, so we are unable to constrain either of its redshifts at the present time.\smallskip
    
    \item Additionally, we discover a new GGL in the outskirts region (GGL-F) that shows strong [OII] emission ($z_s = 1.0655$), forming a near complete Einstein ring around the lens. Due to the small apparent separation of the lens and source, this system was not identified in the BUFFALO imaging search. We identify two multiply-imaged constraint points in the background galaxy, allowing us to construct a lens model with extremely low (rms < 0\farcs03) residuals.\smallskip
    
    \item Embedding the GGL information into the global lens model, we find that the induced shear field in the space around GGL-F is lower than what was predicted in past work (L19).  We use this result to investigate constraints on the cluster-scale halo components, providing valuable insight into the large-scale mass distributions in A370.\smallskip 
    
    \item Finally, taking advantage of the spectroscopic information along the line of sight, we also identify additional galaxy overdensities in front of and behind the cluster. The most prominent of these is a collection of 68 galaxies at $z\sim1$. While previous work suggested it could be a compact galaxy group, our analysis reveals a complex structure that is more likely a chance alignment of several unbound objects, providing additional clarity to the lensing mass model.
\end{itemize}

While the topics we explore in this paper are broad reaching, they are not by any means the only science this data set can address. Though we have not presented new lens models in this work we are actively involved in such efforts, incorporating elements of the data presented here into updated models that will be released as part of the upcoming BUFFALO public modelling challenge (an analogue to the previous HFF modelling project\footnote{\url{https://archive.stsci.edu/prepds/frontier/lensmodels/}}. We are particularly interested in probing the effects of adding additional cluster members -- especially the blue-cloud galaxies identified at larger radii -- and exploring alternative origins for the external shear term, given the likely low influence of the $z \sim 1$ overdensity. Similar to \citet{gra21} we are also integrating cluster member velocity dispersion measurements into the model, allowing us to reduce our reliance on luminosity-based scaling relations in favour of the more robust Fundamental Plane (Mahler et al., in prep.) Furthermore, using the full MUSE data set as a template we are beginning to calibrate background galaxy distributions more accurately in weak- and combined strong+weak-lensing models, providing important improvements to BUFFALO science outputs.

Besides modelling, the high density of spectroscopy combined with \emph{HST}-resolution imaging makes this data set a valuable resource for studying elements of galaxy evolution, especially for the significant background galaxy sample at medium and high redshifts ($z > 1$). Along similar lines, the spectra of the complete data set (which we are making publicly available\footnote{\url{https://astro.dur.ac.uk/~hbpn39/pilot-wings.html}}) will add to the growing library of SED-fitting standards, and in particular our investigation into the luminosity and spatial distributions of spectroscopically confirmed red and blue cluster members can serve as a template for generating more accurate cluster-member selection functions in similar clusters without comprehensive spectroscopic coverage.

We stress again that the combination of imaging and spectroscopy used in this work is the \emph{only} plausible way to obtain many of these results, and by exploring areas beyond the compact core we strengthen the conclusions we can draw from them. Yet, in spite of the benefit that a larger mosaic provides, our analysis still reveals limitations in the current data set. While the 3D catalogue of cluster members provides a detailed look at cluster dynamics and structure, the sample does not yet extend to characteristic radii such as $r_{500}$ ($\sim$1.5 Mpc), forcing us to rely on model inference to describe extended mass distributions. Similarly, with several GGL candidate source galaxies located just at (or beyond) the edge of the MUSE field, we are unable to reliably measure their spectroscopic redshifts, which are critical for investigating small-scale mass concentrations in the host halo. This suggests that there is need for additional coverage, and highlights the value in continuing our campaign to expand the A370 footprint. Broader spectroscopic coverage would be especially helpful: even though the 14 arcmin$^2$ MUSE footprint presented here is considerably larger than any of its predecessors, it accounts for only 40\% of the primary A370 BUFFALO field -- there is still significant ground to cover. 

Extending the MUSE mosaic to fill the BUFFALO field of view would maximize the effectiveness of our data combination, allowing us to use the analysis techniques outlined in this paper over a $\sim$32 arcmin$^2$ region of the sky, reaching distances $\sim$1.3 Mpc from the projected cluster centre. This would easily encompass all of the existing GGL candidates, and possibly reveal other hidden systems (much like GGL-F) that are not readily apparent in imaging alone. Likewise, by identifying cluster members to the edge of the BUFFALO footprint, we can continue quantifying the physical properties of galaxies out to even larger radii, and also push the caustic method analysis into the predicted range of $r_{500}$, providing for a more accurate estimate of $M_{500}$ and $M_{200}$; this would allow us to directly compare our enclosed mass contours to complementary techniques (such as x-ray gas temperature and weak lensing) that provide mass estimates at similar cluster-centric distances. We note that spectroscopic coverage will be particularly useful for validating the redshifts of all galaxies in the outermost regions of the BUFFALO footprint, since the imaging in these areas lack significant high-resolution IR coverage, limiting photo-z effectiveness.

Wider coverage is not the only way we can enhance our view of A370, however, as deeper MUSE data will also improve the results. While the existing outskirts pointings have been specifically designed to be shallower than the core, there are quantifiable benefits to increasing the current 1-hour exposure time. With the data in hand, we can reliably extract continuum/absorption-line redshift features from galaxies to a magnitude limit of $m_{\rm F814W} \sim$ 22.7. By doubling the exposure time, we will instead be able to push to $m_{\rm F814W} \sim$23.3, based on observed magnitude limits in the $\sim$2-hour data regions of the \citet{ric20} sample. A fainter magnitude cut will broadly benefit the robustness of our statistics:\ consulting the BUFFALO photometric catalogues, we find that a $\sim$23.3 magnitude limit should yield up to $\sim$100 more cluster members (a 20\% increase from the current sample), improving both lens modelling and caustic mass analyses while also increasing the SNR of source galaxies in the GGL sample. At the same time, it will allow us to approach the 0.01 $L_*$ limit used by \citet{nat17} when calibrating substructure fractions in the Frontier Fields (for the redshift of A370, this should be mag $\sim$23.5.) This in turn will allow us to compare a spectroscopically confirmed selection of cluster subhalos to the theoretical predictions made in that work.

Finally we recognize that, because our results are based on a single cluster, they are sensitive to the underlying galaxy population, and may be subject to systematic effects arising from, e.g., differences in formation history. To make this analysis more general then, we will need to study a larger sample of clusters. With deep imaging and some MUSE coverage already, the remaining BUFFALO clusters are ideal candidates for this purpose, and to that end, we are proposing a new survey:\ the BUFFALO \textcolor{red}{W}ide-area \textcolor{red}{IN}tegral-field \textcolor{red}{G}alaxy \textcolor{red}{S}urvey (\textbf{BUFFALO-WINGS}), to help overcome these limitations. Following the arguments outlined in the previous few paragraphs, BUFFALO-WINGS would cover the full \emph{HST} footprint of each BUFFALO cluster, using at least two OBs of MUSE data in each pointing. This would significantly expand the capabilities of this already promising pilot study, and with an estimated $\sim$8000 redshifts, would stand as a powerful legacy product in its own right. While this would require a non-trivial amount of telescope resources, this work demonstrates such an effort would be well spent, providing considerable benefit to the larger astronomical community. At the very least, the high density of WINGS spectra will considerably improve our understanding of cluster member populations, providing a more accurate selection function for identifying cluster members with traditional/MOS instruments such as Hectospec (even though, by definition, this coverage would still be incomplete compared to the IFU). Thus, we can even envision a hybrid/extended WINGS program, where (in places where universal IFU coverage is impractical) the WINGS region of the field can be supplemented by traditional spectroscopy, providing some level of spectroscopic coverage out to the very edge of the cluster itself.

\section*{Acknowledgements}

DJL, ACE and RM are supported by STFC grants ST/T000244/1 and ST/W002612/1.
MJ is supported by the United Kingdom Research and Innovation (UKRI) Future Leaders Fellowship `Using Cosmic Beasts to uncover the Nature of Dark Matter' (grant number MR/S017216/1).
MS acknowledges financial contribution from contract ASI-INAF n.2017-14-H.0 and from contract INAF mainstream project 1.05.01.86.10. RC is supported in part by HST-GO-15117. FEB acknowledges support from ANID-Chile Basal AFB-170002 and FB210003, FONDECYT Regular 1200495 and 1190818, 
and Millennium Science Initiative Program  – ICN12\_009.
GM received funding from the European Union’s Horizon 2020 research and innovation programme under the Marie Skłodowska-Curie grant agreement No MARACAS - DLV-896778.
The work of TC was carried out at the Jet Propulsion Laboratory, California Institute of Technology, under a contract with NASA.


\section*{Data Availability}
The data underlying this article are available in the article itself, its online supplementary material, and the following url: \url{https://astro.dur.ac.uk/~hbpn39/pilot-wings.html}
 



\bibliographystyle{mnras}
\bibliography{references} 




\appendix

\section{Sample Redshift Catalogue}
\label{sec:appendix}

In this appendix we present an abbreviated version of the master catalogue, which is included in the online supplementary material to this manuscript. The catalogue itself contains a combination of spectroscopic, photometric, and spatial information, and for the interested reader we now briefly provide a description of each data column. 

\begin{itemize}
    \item \textbf{ID}: a numerical identifier (typically matched to a SExtractor detection run) for each galaxy. (see Note A) \smallskip
    
    \item \textbf{Source}: An indication of how the object was detected. \emph{Prior} sources are identified in HST images, while \emph{muselet} sources are only found in the MUSE data. (see Section \ref{sec:redshifting})\smallskip
    
    \item \textbf{Field}: The data set containing the galaxy. ``CORE'' objects are located in the core region of the cluster, which is covered by deep MUSE data, while ``Pxx'' objects are found in the shallower outskirts regions. (see Fig. \ref{fig:footprint})\smallskip
    
    \item \textbf{RA, Dec}: Spatial coordinates for each object.\smallskip
    
    \item \textbf{z}: The measured redshift of each object.\smallskip
    
    \item \textbf{z$_{\rm conf}$}: An assessment of the reliability of the redshift measurement, from low ($z_{\rm conf} = 1$) to high ($z_{\rm conf} = 3$). (See Note B)\smallskip
    
    \item \textbf{m$_{\rm FxxxW}$, ~m$_{\rm FxxxW_{err}}$}: The measured flux of the object in a given \emph{HST} band and its uncertainty. For clarity, in this sample table we only display magnitudes in the F606W and F814W bands. However, the full catalog contains entries for 7 bands (F435W, F606W, F814W, F105W, F125W, F140W, and F160W).\smallskip 
    
    \item \textbf{Mult ID}: If the object is part of a multiply-imaged system, this column provides a unique identifier for each image. Numbering is taken from the lensing catalogue presented in \citet{lag19}. \smallskip

\end{itemize}

\textbf{\emph{Note A}}: While individual ID numbers can be repeated between objects in different datasets or detection methods, a given (ID + Source + Field) combination is entirely unique for all galaxies in the catalogue.\\ 

\textbf{\emph{Note B}}: $z_{\rm conf}$ measurements are classified as follows:
\begin{itemize}
        \item Confidence 1: the redshift is based on a single ambiguous or low-SNR emission line, or several low SNR absorption features.\smallskip 
        
        \item Confidence 2: the redshift is based on a single emission line without additional information, several moderate S/N absorption features, or a Confidence 1 detection whose redshift confidence is increased by the identification of a multiply-imaged system.\smallskip
        
        \item Confidence 3: the redshift is based on multiple clear spectral features, or on a single high S/N emission line with additional information (e.g., an obvious asymmetry in the line profile or a clear non-detection in {\it HST} bands bluewards of the line).
    \end{itemize}

%

\begin{table*}
  \centering
  \caption{Galaxy Catalogue}
  \label{tbl:redshifts}
  \begin{tabular}{lllllllllllll}
    \hline
    ID & Source & Field & RA & Dec & $z$ & $z_{\rm conf}$ & m$_{\rm F606W}$  & m$_{\rm F606W_{err}}$ & m$_{\rm F814W}$  & m$_{\rm F814W_{err}}$ & [...] & Mult ID \\
  & &  & [deg] & [deg] & & & [ABmag] & [ABmag] & [ABmag] & [ABmag] & [...] & \\
    \hline
    451 &MUSELET &CORE &39.9621409 &-1.5644924 &0.4345 &3 &99.999 &9.999 &99.999 &9.999 &[...] &--\\
    3421 &PRIOR &CORE &39.9705140 &-1.5944456 &1.3399 &3 &23.261 &0.001 &22.863 &0.001 &[...] &--\\
    8620 &PRIOR &CORE &39.9671636 &-1.5768729 &0.8041 &3 &23.782 &0.002 &22.970 &0.001 &[...] &1.1\\
    10196 &PRIOR &CORE &39.9622333 &-1.5720594 &4.9160 &2 &27.574 &0.018 &27.003 &0.018 &[...] &24.2\\
    68 &MUSELET &P01 &40.0065392 &-1.5619935 &3.7215 &2 &99.999 &9.999 &99.999 &9.999 &[...] &--\\
    682 &PRIOR &P04 &39.9612627 &-1.5431322 &1.5487 &2 &23.320 &0.005 &23.012 &0.006 &[...] &--\\
    1107 &PRIOR &P07 &39.9893987 &-1.5905450 &0.2304 &3 &25.478 &0.008 &25.152 &0.010 &[...] &--\\
    52 &MUSELET &P10 &39.9608187 &-1.6082485 &1.2075 &2 &99.999 &9.999 &99.999 &9.999 &[...] &--\\
    612 &PRIOR &P10 &39.9604783 &-1.5997095 &0.3740 &1 &25.399 &0.011 &24.579 &0.009 &[...] &--\\
    \hline
  \end{tabular}
\end{table*}


\bsp	
\label{lastpage}
\end{document}